\documentclass[a4paper,11pt]{article}
\pdfoutput=1 % if your are submitting a pdflatex (i.e. if you have
\usepackage{jheppub} % for details on the use of the package, please
\usepackage[T1]{fontenc} % if needed

\usepackage{color}
\usepackage{amsmath}
\font\mybb=msbm10 at 12pt

\font\mybbsub=msbm10 at 8pt
\font\mybbsmall=msbm10 at 10pt

\def\bb#1{\hbox{\mybb#1}}

\def\bbsub#1{\hbox{\mybbsub#1}}
\def\bbsmall#1{\hbox{\mybbsmall#1}}

\def\FFsub{\bbsub{F}}
\def\FFsmall {\bbsmall{F}}
\def\ZZ {\bb{Z}}

\def\ZZsmall {\bbsmall{Z}}

\def\CCsub {\bbsub{C}}
\def\CCsmall {\bbsmall{C}}
\def\PP {\bb{P}}
\def\PPsub{\bbsub{P}}
\def\PPsmall {\bbsmall{P}}

\def\ms{\hspace{-0.05cm}}
\newcommand\beqa{\begin{eqnarray}}
\newcommand\eeqa{\end{eqnarray}}
\newcommand\n{\nonumber\\}

\def\beq#1\eeq{\begin{equation}#1\end{equation}}
\def\bes#1\ees{\begin{equation}\begin{split}#1
               \end{split}\end{equation}}
\def\bea#1\eea{\begin{align}#1\end{align}}

\begin{document}

{~}

\title{Half-hypermultiplets and incomplete/complete resolutions in F-theory }

\author[a]{Naoto Kan,}
\author[a,b]{Shun'ya Mizoguchi,}
\author[c]{Taro Tani}

\affiliation[a]{Graduate University for Advanced Studies (Sokendai)\\
Tsukuba, Ibaraki, 305-0801, Japan}
\affiliation[b]{Theory Center, Institute of Particle and Nuclear Studies, KEK\\
Tsukuba, Ibaraki, 305-0801, Japan}
\affiliation[c]{National Institute of Technology, Kurume College, \\Kurume, Fukuoka, 830-8555, Japan}
% e-mail addresses: one for each author, in the same order as the authors
\emailAdd{naotok@post.kek.jp}
\emailAdd{mizoguch@post.kek.jp}
\emailAdd{tani@kurume-nct.ac.jp}

\abstract{
We consider resolutions of 
codimension-two enhanced singularities 
from $SO(12)$ to $E_7$ 
and from $E_7$ to $E_8$ in six-dimensional F-theory,  
where a half-hypermultiplet arises for
generic complex structures achieving them. 
The exceptional fibers at the enhanced point 
exhibit different structures depending on how 
the colliding 7-brane approaches the stack of 
gauge 7-branes, as previously 
observed by Morrison and Taylor 
in the case of the enhancement from $SU(6)$ to $E_6$.
When the colliding brane 
approaches them as $O(s)$, 
where $s$ is the coordinate of the base space 
along the gauge 7-branes,
the resolution process 
ends up with fewer exceptional fibers than naively 
expected from the Kodaira classification,  
with a non-Dynkin intersection matrix  
including half-integral 
intersection numbers. 
We confirm that the exceptional fibers at the enhanced point 
form extremal rays of the cone of the positive weights of the 
relevant pseudo-real representation,
explaining why a half-hypermultiplet arises there. 
By altering the ordering of the singularities blown up 
in the process, 
we obtain, for both $SO(12)\rightarrow E_7$ 
and $E_7\rightarrow E_8$, the intersection diagram 
on every other row of the corresponding box graphs.
We  present detailed derivations of 
the intersection diagrams of the exceptional fibers at the 
singularity enhanced points by examining 
how an exceptional curve is lifted up on the chart arising 
due to the subsequent blowing-up process. 
When the colliding brane 
approaches the stack of branes as $O(s^2)$, we 
obtain additional conifold singularity at the enhanced point, 
which completes 
the full Dynkin diagram of the enhanced group 
as was found 
previously.
}

\preprint{KEK-TH-2196}
%\pacs{???}
%\date{\today}
\date{February 26, 2020}
\maketitle

\section{Introduction}
The significance of F-theory \cite{Vafa} 
in modern particle physics model 
building cannot be overestimated. The characteristic features 
of the Standard Model
%---the complicated-looking 
%assignment of hypercharges, a generation consisting of 16 
%quarks and leptons, the absence of gauge anomalies---they all 
can be naturally explained by the $SU(5)$ and $SO(10)$ grand 
unified theories, which are engineered in F-theory. 
It can achieve matter fields in the spinor representation 
as well as the exceptional group gauge symmetry. 
F-theory has an advantage over the $E_8\times E_8$ 
heterotic string theory in that it can address the issue 
of the gauge/gravity coupling correlation in the latter \cite{Witten96}.
F-theory can also generate up-type Yukawa couplings 
perturbatively forbidden in D-brane models \cite{Dbranemodels1,Dbranemodels2}.
 
In F-theory,  matter typically 
arises\footnote{aside from adjoint matter in four dimensions 
arising from 7-branes wrapped over a four-cycle with 
nontrivial relevant cohomologies.} 
at codimension-two singularities \cite{MV1,MV2,BIKMSV,KatzVafa,Tani}
in the base space of the elliptic fibration.
To be specific, consider a six-dimensional F-theory 
compactified on an elliptically fibered three-fold 
defined by a hypersurface 
$\Phi(x,y,z,s)=0$ in $\CCsmall^4\ni (x,y,z,s)$
over a complex two-fold with local coordinates $(z,s)\in\CCsmall^2$.
Over a generic point with fixed $(z,s)$, this equation describes a genus-one curve.
We assume that this hypersurface %$\Phi(x,y,z,s)=0$
has a codimension-one singularity of some compact 
group $H$ along $z=0$ for generic $s\neq 0$, which is 
enhanced to some 
compact group $G\supset H$ at a codimension-two particular 
locus along the line $z=0$, say, at $s=0$.
%Here by ``enhanced to $G$'' 
%we mean that the unbroken gauge symmetry one would 
%obtain is $G$ when the equation $\Phi(x,y,z,s=0)=0$ 
%is regarded as an elliptic fibration over a complex 
%one-dimensional base.
%
%
%Thus this $G$ can be determined by the information on
%the slice $s=0$ only, according to the Kodaira classification
%of singular fibers. As was pointed out in \cite{MT}, 
%however, in the codimension-two case 
%the set of resulting exceptional curves at $s=0$ 
%does not necessarily coincide with the naive set of Kodaira 
%fibers. 
%

In this set-up, one typically obtains 
chiral matter hypermultiplets 
corresponding to the homogeneous K\"ahler 
manifold $G/(H\times U(1))$ \cite{MV1,MV2,BIKMSV,KatzVafa,Tani,FFamilyUnification}
at the enhanced  
singularity.  
However, if $G$ and $H$ are such that
\beqa
G&\supset&H\times SU(2)\n
\mbox{\bf dim\boldmath $G$}&=&(\mbox{\bf dim\boldmath $H$},{\bf 1})
\oplus({\bf 2n},{\bf 2})
\oplus({\bf 1},{\bf 3}),
\label{eq:GHdecomp}
\eeqa
where $n=10$, $16$, $28$ for $(G,H)=(E_6,SU(6))$, 
$(E_7,SO(12))$, $(E_8,E_7)$, respectively,\footnote{They are also 
the pairs (though different real forms) 
appearing in ``magical'' supergravity theories \cite{magical1,magical2}.}
the matter %multiplets 
arising 
%at the codimension-two singularities 
there
are not the full hypermultiplets but {\em half-}hypermultiplets 
for generic complex structures achieving such singularities.
%In order to have full hypermultiplets, one needs to tune 
%the complex structure appropriately \cite{MT,BoxGraphs,MizoguchiTanianomaly}. 

In all the cases above, ${\bf 2n}$ is a pseudo-real 
representation of $H$, of which the representation matrix can be 
written as an $Sp(2n)$ matrix. 
For these representations, 
one can impose the symplectic Majorana 
condition on the $2n$ complex spinors. 
One can also define a similar relation among 
$2n$ pairs of the complex scalars by using a $U(1)_R$ 
rotation. In this way, the degrees of freedom of 
hypermultiplets are halved, 
yielding half-hypermultiplets in these representations.
%See Appendix \ref{Appendixhalfhyper} for more detail.

In \cite{MT}, the resolution of a codimension-two 
enhancement from $SU(6)$ to $E_6$ was studied 
in six dimensions. 
It was shown there that, 
%how the tuning of the complex structure affected the result of 
%the resolution 
%
%There it was found that, depending on the ``speed'' 
%of the colliding brane,  
%the number of exceptional curves changed. 
%
%That is,
%
if the colliding brane approaches $z=0$ like 
$O(s)$ %for $s$ defined in the text,
which is the generic case, 
only {\em five} exceptional curves appear through the blowing-up
process even at the singularity enhanced point.
This type of resolution was called the {\em incomplete 
resolution} in \cite{MT}. 
Note that, by resolving a codimension-{\em one} $E_6$ 
singularity, one obtains {\em six} exceptional fibers 
consisting the $IV^*$ fiber type of Kodaira.
%This type of resolution was called the {\em incomplete 
%resolution} in \cite{MT}. 

This means that there are not enough new degrees of 
freedom arising at the enhanced point 
to generate matter in a full hypermultiplet. % as usual. 
The five exceptional fibers form a curious non-Dynkin 
intersection diagram, however. In particular, some of the 
exceptional curves turn out to have 
self-intersection number $-\frac32$. This 
is equal to the minus of the length squared of 
a {\em weight} of the {\bf 20} representation. 
One can verify that these five exceptional curves 
form extremal rays of the cone of the positive 
weights of the {\bf 20} representation. 
(Actually, one can show that there are just 20 integer linear combinations 
of these exceptional curves which have self-intersection number $-\frac{3}{2}$;
half of them have positive coefficients and the other half have negative 
coefficients, giving a whole set of the weights of ${\bf 20}$.)
%
%Therefore, 
%although the degrees of freedom are less than usual, 
%the matter can form a {\bf 20} representation of $SU(6)$.
Thus the matter forms a {\em single} ${\bf 20}$ representation of $SU(6)$.
This shows the mechanism of how the {\em half}-hypermultiplet  
appears in this codimension-two singularity enhancement.
(Note that for a full hypermultiplet one needs two ${\bf 20}$ representations,
either one of which survives as massless matter in six dimensions.)

On the other hand, 
if the brane collides like $O(s^2)$, an extra 
singularity arises at the intersection. 
This is a conifold singularity. Resolving it 
% Blowing up this singularity 
yields another exceptional curve, 
completing the proper Dynkin intersection 
diagram expected from the ordinary Kodaira classification. 
This was called the {\em complete 
resolution} \cite{MT}.

Higher codimension singularities were also studied in \cite{BoxGraphs}
by exploring the phases of three-dimensional gauge theory 
arising from the compactification of M-theory on a Calabi-Yau 
four-fold (see  \cite{EsoleYau,MarsanoSSNameki,EsoleShaoYau,BraunNameki}). 
%(For earlier discussions, see \cite{EsoleYau,MarsanoSSNameki}.
%See also \cite{EsoleShaoYau,BraunNameki})
%In the M-theory Coulomb branch approach, this phenomenon 
In this approach, the generation of a half-hypermultiplet 
was explained as a result of the reduction of the massless states 
occurring due to the monodromy among the fibers \cite{BoxGraphs}. 
It was noted there that if the complex structure was 
tuned so that there were extra sections, the monodromy was 
reduced and a full hypermultiplet appeared. 
%
%See Appendix \ref{AppendixSU(6)E6}
%for a summary of the incomplete and 
%complete resolutions for the enhancement 
%$SU(6)\rightarrow E_6$.

In this paper, we study the resolutions of 
the other two cases, 
$SO(12)\rightarrow E_7$ and $E_7\rightarrow E_8$,
of the codimension-two singularity enhancement
in which half-hypermultiplets appear as massless states 
at the singularity.
%by a step-by-step blowing up process.
%
The strategy is the same as that used in \cite{MT}.
To blow up this singularity along the line $z=0$ 
we replace %, as a first step, 
the local ambient space 
$\CCsmall^4$ with
\beqa
{\widehat\CCsmall}^4&=&\left\{\left.((x,y,z,s);(\xi:\eta:\zeta))\in\CCsmall^4\times\PPsmall^2
\right|
x\eta-y\xi=y\zeta-z\eta=z\xi-x\zeta=0
\right\}
\label{C4hat}
\eeqa
and consider the hypersurface in it. 
(\ref{C4hat}) inserts a continuous family of $\PPsmall^2$ along 
the complex line $z=0$ for arbitrary $s\in\CCsmall$. 
Then we find that there still are several, (again) codimension-one 
singularities on the intersection of ${\widehat\CCsmall}^4$ 
(\ref{C4hat}) and the hypersurface $\Phi=0$. 
To resolve these singularities, we further 
make a  replacement similar to 
$\CCsmall^4\rightarrow{\widehat\CCsmall}^4$ along 
each singular line and consider the hypersurface in this  
blown-up ambient space. 
Then if we still find some singularities of codimension one, 
we perform a codimension-one blow up along each of them.
Repeating these steps we 
end up with either of the two cases:\\
(1) The case where there are no more singularities of any kind 
on the final hypersurface.\\
(2) The case where there are no more codimension-one singularities, but 
still there is an isolated codimension-two singularity at 
$s=0$ on the final hypersurface.

This is the phenomenon known as the partial simultaneous 
resolution of singularities \cite{PSR1,PSR2,KatzMorrison,BoxGraphs},
and which case we will end up with depends on the vanishing order 
of the relevant section of the projective-space bundle for the 
respective singularity enhancement. For the enhancement 
$SU(6)\rightarrow E_6$ studied in \cite{MT}, the relevant 
section is the polynomial %$h_{n+2}(s)$\footnote{Or, 
%more precisely, 
$t_r(s)$ in eq.(\ref{hHqfgA5}) (see below),
%} in the
%notation of \cite{BIKMSV}, 
for $SO(12)\rightarrow E_7$ it is $H_{n+4}(s)$, 
and for $E_7\rightarrow E_8$ it is $f_{n+8}(s)$.
%\footnote{In 
%Tate's form, $h_{n+2},H_{n+4},f_{n+8}$ correspond to
%$a_{1,0},a_{2,1},a_{4,3}$, respectively, where 
%${a}_{p,m}$ is the coefficient of the degree-$m$ term in $z$
%of the section $a_p$.} 
We will show that, also for the cases $SO(12)\rightarrow E_7$
and $E_7\rightarrow E_8$ we will consider in this paper, 
we are led to the result (1)
if these sections 
vanish like $O(s)$ as $s\rightarrow 0$.  
The codimension-two singularity is then resolved only 
by the codimension-one blow-ups in the ambient space, 
and therefore the number of 
times of the blowing-up operations is the same as that needed 
to resolve a Kodaira $H$ singularity 
%(that is, a codimension-one 
%singularity in an elliptic fibration over a complex plane) 
for fixed generic $s\neq 0$. Thus there are not enough 
%singular fibers 
exceptional curves 
to form the proper Dynkin diagram of $G$ as their 
intersection matrix. %This is the incomplete resolution.
Even so, we will see that 
the intersection diagram of the exceptional curves 
obtained by the $s\rightarrow 0$ limit of the 
exceptional surfaces is different from 
the Dynkin diagram of $H$. Rather, it turns out that 
some of the curves have their self-intersection number $-\frac32$, 
as was observed in \cite{MT} for the 
$SU(6)\rightarrow E_6$ enhancement.

%The intersection diagram of the fibers at the codimension-two 
%point $s=0$ changes depending on the order of 
%the codimension-one singularities we blow up at generic $s\neq 0$. 
%In this paper, 
We perform blowing-ups for all possible inequivalent 
orderings of blowing up the singularities in both the enhancements 
$SO(12)\rightarrow E_7$ 
and $E_7 \rightarrow E_8$. We find that the 
intersection diagram on {\em every other} row in Figures 33 and 44 of \cite{BoxGraphs} can be   
obtained, but not all of them.\footnote{In the enhancement $SU(6)\rightarrow E_6$
studied in \cite{MT}, there is no such option since only one 
singularity appears at each step of blowing up.}
Although the intersection diagrams are different, the fibers obtained 
in the limit $s\rightarrow 0$ always form extremal rays of 
the cone of the positive weights of the relevant pseudo-real representation.

On the other hand, if they vanish like $O(s^2)$, 
we are led to the result (2), where we need a further resolution 
of the isolated singularity. This is the complete resolution; 
we have one more additional node to the incomplete 
intersection diagram obtained in the incomplete resolution, 
obtaining the full Dynkin diagram of the group $G$.
We will find that, as was observed in \cite{MT}, 
this final singularity appearing at codimension two 
is a conifold singularity for all the cases we examine in this paper.

The plan of this paper is as follows. 
In section 2, we review in what circumstances
massless matter fields appear as half-hypermultiplets 
in the global six-dimensional F-theory compactification 
on an elliptic Calabi-Yau three-fold over a Hirzebruch surface.  
In section 3, we present the detailed process of blow-ups for 
the codimension-two singularity 
enhancement from $SO(12)$ to $E_7$. 
In section 4, we turn to the resolution of the enhancement 
from $E_7$ to $E_8$. Section 5 summarizes the conclusions. 
In Appendix A, we explain the relations between 
symplectic Majorana-Weyl spinors, 
pseudo-real representations and 
half-hypermultiplets. In Appendix B, 
we summarize the results of \cite{MT} on the 
resolution of the enhancement $SU(6)\rightarrow E_6$. 
Finally, in Appendix C,  we present a basic explanation 
of the small resolution of a conifold.

\section{Half-hypermultiplets in six-dimensional F-theory}
Half-hypermultiplets arise when the unbroken gauge group is $SU(6)$, 
$SO(12)$ or $E_7$ \cite{BIKMSV}. These models can be systematically 
obtained by tuning the complex structure of the $SU(5)$ model.

We start with the six-dimensional compactification on F-theory 
on an elliptically fibered 
Calabi-Yau three-fold over a Hirzebruch surface 
$\FFsmall_n$ \cite{MV1,MV2}. Let $z$, $z'$ be affine coordinates 
of the fiber and base $\PPsmall^1$'s, respectively. The Weierstrass model 
\beqa
y^2=x^3+ f(z,z') x + g(z,z')
\label{Weierstrass}
\eeqa
develops an $SU(5)$ singularity if \cite{BIKMSV}
\beqa
f(z,z')&=&
-3 h_{n+2}^4%
+12  h_{n+2}^2 H_{n+4}z
-12  \left(H_{n+4}^2-h_{n+2} q_{n+6}\right)z^2
+ f_{n+8}z^3
+f_8 z^4,\n
g(z,z')&=&
 2h_{n+2}^6
    -12  h_{n+2}^4 H_{n+4}z
  +   \left(24 h_{n+2}^2 H_{n+4}^2-12 h_{n+2}^3 q_{n+6}\right)z^2\n
&&+ \left(-f_{n+8}
   h_{n+2}^2+24 h_{n+2} H_{n+4} q_{n+6}-16 H_{n+4}^3\right) z^3\n
&&+ \left(-f_8 h_{n+2}^2+2 f_{n+8} H_{n+4}+12 q_{n+6}^2\right)  z^4
 + g_{n+12}z^5
   +g_{12} z^6,
 \label{SU(5)fg}
\eeqa
where $h_{n+2}$,$H_{n+4}$,$q_{n+6}$,$f_{n+8}$ and $g_{n+12}$ are 
polynomials of $z'$ of degrees specified by the subscripts. They are 
sections of Looijenga's weighted projective space bundle \cite{FMW,DonagiWijnholt,MizoguchiTaniLooijenga}
characterizing the $SU(5)_{\rm instanton}$ vector bundle of the 
dual heterotic string theory. This Calabi-Yau three-fold 
admits a $K3$ fibration, and we work with one of the 
rational elliptic surfaces in the stable degeneration limit of the $K3$ 
so that 
the orders of the polynomials $f(z,z')$ and $g(z,z')$ are truncated 
at $z^4$ and $z^6$, respectively. 
This suffices since the anomalies cancel for each $E_8$ gauge 
group, and also we are interested in the local structure of the 
singularity.
$x$ and $y$ are then taken 
to be sections of ${\cal O}(2(-K_{\FFsub_n}-C_0))$ and
${\cal O}(3(-K_{\FFsub_n}-C_0))$, 
where $C_0$ is a divisor class with $C_0^2=-n$,  
satisfying $-K_{\FFsub_n}=2C_0+(2+n)f$ with the fiber class $f$. 
Similar modifications are necessary for $f(z,z')$ and $g(z,z')$. 
This deviation from the anti-canonical class (and hence from a Calabi-Yau) 
is because we consider a rational-elliptic-surface fibration.
% F-theory IV clover 1

The Weierstrass equation (\ref{Weierstrass}) with (\ref{SU(5)fg}) can be 
written in Tate's form as
\beqa
y'^2+x'^3+\alpha_4 z^4 x' + \alpha_6 z^6 +
a_0 z^5 + a_2 z^3 x' + a_3 z^2 y' + a_4 z x'^2 + a_5 x' y' =0 
\label{Tate's}
\eeqa
with
%\bes
\beqa
&a_5=&2\sqrt{3} i h_{n+2},  \n %\\
&a_4=&-6 H_{n+4},  \n %\\
&a_3=&4\sqrt{3} i q_{n+6}, \n %\\
&a_2=&f_{n+8},  \n  %\\
&a_0=&g_{n+12}-2H_{n+4}f_8, \n %\\
&\alpha_4=&f_8, \n %\\
&\alpha_6=&g_{12}.
\eeqa
%\ees
For completeness we write $x$, $y$ in (\ref{Weierstrass}) 
in terms of $x'$, $y'$ in (\ref{Tate's}):
\beqa
x&=&x'+\frac13\left(a_4 z - \frac14 a_5^2\right),\n
y&=&i\left(y'+\frac12\left(
a_5 x' + a_3 z^2
\right)
\right).
\eeqa

\newpage
\noindent
\underline{\em $SU(6)$}  \\

To obtain an equation for $SU(6)$ gauge group, which yields 
half-hypermultiplets, we set \cite{BIKMSV}
\beqa
h_{n+2}&=&t_r h_{n+2-r},\n
H_{n+4}&=&t_r H_{n+4-r},\n
q_{n+6}&=&u_{r+4} h_{n+2-r},\n
f_{n+8}&=&t_r f_{n+8-r}-12 u_{r+4}H_{n+4-r} ,\n
g_{n+12}&=&2(u_{r+4} f_{n-r+8}+ f_8 t_r H_{n-r+4}).
\label{hHqfgA5}
\eeqa
Then the spectral cover factorizes as
\beqa
0&=&a_0 z^5 + a_2 z^3 x' + a_3 z^2 y' + a_4 z x'^2 + a_5 x' y' \n 
&=&\left(x' t_r+2 z^2 u_{r+4}\right) \left(z^3 f_{n-r+8}+2 i \sqrt{3} y' h_{n-r+2}-6 z x'
   H_{n-r+4}\right),
\eeqa
indicating that the $SU(5)$ instanton is reduced to an $SU(3)\times SU(2)$
instanton in the heterotic dual, at the same time 
the Mordell-Weil rank of the rational elliptic surface is reduced. 
The Mordell-Weil lattice 
is No.15 in the Oguiso-Shioda classification \cite{OguisoShioda}.
In this specification $f(z,z')$ and $g(z,z')$ become
\beqa
f_{SU(6)}(z,z')&=&-3 t_r^4
   h_{n-r+2}^4
   +12 z t_r^3 h_{n-r+2}^2 H_{n-r+4}
   +z^2 \left(12 t_r u_{r+4}
   h_{n-r+2}^2-12 t_r^2 H_{n-r+4}^2\right)
   \n
   &&
   +z^3 \left(t_r f_{n-r+8}-12 u_{r+4} H_{n-r+4}\right)+f_8 z^4,\label{fSU(6)}
   \\
g_{SU(6)}(z,z')&=&
   2 t_r^6 h_{n-r+2}^6-12 z t_r^5 h_{n-r+2}^4 H_{n-r+4}\n
   &&
   +z^2 \left(24 t_r^4 h_{n-r+2}^2 H_{n-r+4}^2-12
   t_r^3 u_{r+4} h_{n-r+2}^4\right)\n
   &&+z^3 \left(-t_r^3 f_{n-r+8} h_{n-r+2}^2+36 t_r^2 u_{r+4}
   h_{n-r+2}^2 H_{n-r+4}-16 t_r^3 H_{n-r+4}^3\right)\n
&& 
   +z^4 \left(-f_8 t_r^2 h_{n-r+2}^2+2 t_r^2 f_{n-r+8} H_{n-r+4}+12 u_{r+4}^2 h_{n-r+2}^2-24
   t_r u_{r+4} H_{n-r+4}^2\right)\n
   &&
     +z^5 \left(2 f_8 t_r H_{n-r+4}+2
   u_{r+4} f_{n-r+8}\right)
   +g_{12} z^6.\label{gSU(6)}
\eeqa
The discriminant reads
\beqa
\Delta_{SU(6)}&=&9  z^6 t_r^3 h_{n-r+2}^4\left[
t_r^3 \left(12 g_{12} h_{n-r+2}^2-f_{n-r+8}^2\right)\right.\n
&&~~~~~~~~~~~~~~~~+t_r^2 \left(-24 f_8 u_{r+4}
   h_{n-r+2}^2-24 u_{r+4} f_{n-r+8} H_{n-r+4}\right)\n
   &&~~~~~~~~~~~~~~~~\left.-144 t_r
   u_{r+4}^2 H_{n-r+4}^2
-96 u_{r+4}^3 h_{n-r+2}^2\right]+O(z^7).
\eeqa
Thus the Weierstrass model with (\ref{fSU(6)}),(\ref{gSU(6)}) 
indeed has a codimension-one $SU(6)$ singularity along $z=0$.

The zero loci of $t_r$ are the points where the $SU(6)$ singularity 
is enhanced to $E_6$, those of $h_{n-r+2}$ are the ones to $D_6$,
and those of the remaining factor of degree $2n+r+16$ are the ones 
to $A_6$. They respectively yield $r$ {\em half-hypermultiplets} in {\bf 20}, 
$n-r+2$ hypermultiplets in {\bf 15} and $2n+r+16$ hypermultiplets in {\bf 6}.

The number of the complex structure moduli is $3n-r+21$,
which satisfies the anomaly-free constraint for one of the $E_8$ factors \cite{BIKMSV}
\beqa
n_H-n_V&=&20\cdot\frac r2+15(n-r+2)+6(2n+r+16)+3n-r+21~-35\n
&=&30n+112.
\eeqa
Note that this condition does not hold if the multiplets in {\bf 20} are 
ordinary hypermultiplets.\\

\noindent
\underline{\em $SO(12)$}\\

To further obtain an equation for $SO(12)$ gauge group, 
one only needs to set $h_{n+2-r}=0$ in (\ref{hHqfgA5}).
%\beqa
%h_{n+2}&=&0,\n
%H_{n+4}&=&t_r H_{n+4-r},\n
%q_{n+6}&=&0,\n
%f_{n+8}&=&t_r f_{n+8-r}-12 u_{r+4}H_{n+4-r} ,\n
%g_{n+12}&=&2(u_{r+4} f_{n-r+8}+ f_8 t_r H_{n-r+4}).
%\label{hHqfgD6}
%\eeqa
%
The spectral cover is now
\beqa
%0&=&a_0 z^5 + a_2 z^3 x' + a_3 z^2 y' + a_4 z x'^2 + a_5 x' y' \n 
%&=&-6 z
\left(x' t_r+2 z^2 u_{r+4}\right) \left(x' H_{n-r+4}-\frac16 z^2 f_{n-r+8}\right)
&=&0.
\eeqa
These factors are in the same form, corresponding to two $SU(2)$'s 
of the instanton gauge group of the heterotic theory. The Mordell-Weil lattice 
is No.26 in \cite{OguisoShioda}.

Then $f(z,z')$ and $g(z,z')$ are 
\beqa
f_{SO(12)}(z,z')&=&
    -12 z^2 t_r^2 H_{n-r+4}^2
   +z^3 \left(t_r f_{n-r+8}-12 u_{r+4} H_{n-r+4}\right)+f_8 z^4,\label{fSO(12)}
\label{eq:so12f}
   \\
g_{SO(12)}(z,z')&=&
     -16 z^3 t_r^3 H_{n-r+4}^3
        +2 z^4 \left(
    t_r^2 f_{n-r+8} H_{n-r+4}
   -12 t_r u_{r+4} H_{n-r+4}^2\right)\n
   &&
     +2 z^5 \left( f_8 t_r H_{n-r+4}+
   u_{r+4} f_{n-r+8}\right)
   +g_{12} z^6.\label{gSO(12)}
\label{eq:so12g}
\eeqa
The discriminant is
\beqa
\Delta_{SO(12)}&=&
-36 z^8 t_r^2 H_{n-r+4}^2 \left(t_r f_{n-r+8}+12 u_{r+4}
   H_{n-r+4}\right){}^2+O(z^9).
\eeqa

The zero loci of both $t_r$ and $H_{n-r+4}$ give rise to $E_7$ 
singularities to yield $n+4$ half-hypermultiplets.
The loci of the remaining factor are $A_7$ singularities, giving 
$n+8$ hypers in {\bf 12}. 
With additional neutral hypermultiplets from the $2n+18$ complex structure moduli, we have
\beqa
n_H-n_V&=&32\cdot \frac {n+4}2+12(n+8)+2n+18~-66\n
&=&30n+112
\eeqa
as it should be. Again, if {\bf 32} is not a half-hyper, the anomaly does not cancel.
\\

\noindent
\underline{\em $E_7$}\\

Finally, the $E_7$ model can be obtained by 
setting 
%$H_{n-r+4}=0$ in the $SO(12)$ model. 
%This amounts to set 
$h_{n+2}=H_{n+4}=q_{n+6}=0$ in the $SU(5)$ model. The gauge group of the heterotic 
vector bundle is $SU(2)$. 
The Mordell-Weil lattice 
is No.43 in \cite{OguisoShioda}.
$f(z,z')$ and $g(z,z')$ are simply given by
\beqa
f_{E_7}(z,z')&=&   f_{n+8 }z^3 +f_8 z^4,\label{fE7}
   \\
g_{E_7}(z,z')&=&
        g_{n+12}z^5  
   +g_{12} z^6.\label{gE7}
\eeqa
The discriminant
\beqa
\Delta_{E_7}&=&4f_{n+8 }^3 z^9 + O(z^{10})
\eeqa
implies that $n+8$ half-hypermultiplets in {\bf 56} of $E_7$ 
arise. Again they must be half-hyper as
\beqa
n_H-n_V&=&56\cdot \frac {n+8}2+2n+21~-133\n
&=&30n+112.
\eeqa 

\section{$SO(12)\rightarrow E_7$}

%As in \cite{MT}, 
We consider a Weierstrass model 
on a base two-fold $B_2$ with local coordinates $\{z,s\}$,
where 
the codimension-one singularity arises along $z=0$ and the 
codimension-two singularity arises at $s=0$ on the $z=0$ complex line. 
$s$ corresponds to $z'$
 in the previous section.

\subsection{Incomplete resolution: Blowing up $p_1$ first}
\label{p1first}
\subsubsection{Blowing up process}
\label{p1firstprocess}
%$SU(5)$ ???? 
%$f_8=g_{12}=g_{n+12}=q_{n+6}=h_{n+2}=0$, $f_{n+8}=1$, $H_{n+4}=s$ ??????????????B
We consider a concrete Weierstrass model of incomplete resolution by
%The model for incomplete resolution is obtained by 
setting~\footnote{The same model is obtained by setting
$t_r=s$, $H_{n-r+4}=-\frac{1}{2}$, $u_{r+4}=\frac{1}{6}$ and $f_{n-r+8}=f_8=g_{12}=0$.}
$H_{n-r+4}=s$, $t_r=-\frac{1}{2}$, $f_{n-r+8}=-2$ and $u_{r+4}=f_8=g_{12}=0$
in \eqref{eq:so12f} and \eqref{eq:so12g}:
\beqa
\Phi(x,y,z,s)&=&-y^2 + x^3 +f(z,s) x + g(z,s)=0,
\label{eq:Phi0}
\eeqa
where
\beqa
f(z,s)&=&-3 s^2 z^2+z^3,\n
g(z,s)&=&2 s^3 z^3-s z^4.
\label{eq:fgD6E70}
\eeqa 
At $s\neq 0$, the orders of $f$, $g$ and  the discriminant $\Delta$ in $z$ are $(2,3,8)$,
while at $s=0$, they satisfy $(3,\ge \hspace{-0.05cm}5,9)$. Therefore 
\eqref{eq:fgD6E70} describes the enhancement $I_2^*\rightarrow III^*$ ($SO(12)\rightarrow E_7$)
of the Kodaira type, satisfying the requirement.

%???????????????}???????????????? $z'$ ??? $s$ ??????????????B$s=0$ ???? $g$ ??????P??????I??????O???????????? 
%$\rm{ord}(g)=\infty$ ??????????A$E_7$??t??@??C??o??[???????????????????????????? $\rm{ord}(g)\geq 5$
%???????????????}????????????????????????????B $g_{n+12}=1$ ???????????????? $g(t)$ ???? $t^5$????
%??c??????A???????????}??????????l??????u??????[??A??b??v??????????????????A??r??????A???}??????????_????????????????????????
%??????????????????????????????????????????????_??????V??t??g???????????}??????????????????????????????B???????}???????? $g_{n+12}=1$
%?????????????????A??V??t??g???????????????????????????????$t=0$ ??????T?????l???????? $t$???????????????????????????
%??V??t??g??????????????????A??????????????????????????? $g_{n+12}=0$ ???????????????????????????÷??????????????B
%???}???}?????????????????????????@$g_{n+12}=0$ ??????????????u??????[??A??b??v?????s????Ê??B
%
\paragraph{1st blow up}
With \eqref{eq:fgD6E70}, the equation \eqref{eq:Phi0}  reads
\beqa
\Phi(x,y,z,s)=x z^2 (z - 3 s^2) + s z^3 (2 s^2 - z) + x^3 - y^2&=&0. \label{PhiD6E7}
\eeqa
(\ref{PhiD6E7}) has a codimension-one singularity at $p_0 \equiv (0,0,0,s)$. 
We blow up this by replacing the complex line $(x,y,z)=(0,0,0)$ 
with $\PPsmall^2\times\CCsmall$ in $\CCsmall^4$  by passing to the following charts 
corresponding to three affine patches of $\PPsmall^2$ for fixed $s$:
\\
\noindent
\underline{Chart $1_x$}
\beqa
\Phi(x, x y_1, x z_1,s)&=&x^2\Phi_x(x,y_1,z_1,s),\n
\Phi_x(x,y_1,z_1,s)&=&
x^2 \left(z_1^3-s z_1^4\right)+x \left(s z_1-1\right)^2 \left(2 s
   z_1+1\right)-y_1^2.\n
   \mbox{$\cal C$$_{p_0}$ in $1_x$}&:&x=0,~~y_1=0.\n
   \mbox{Singularities}&:&(x,y_1,z_1,s)=  (0,0,\frac{1}{s},s), (0,0,-\frac{1}{2 s},s).
\label{eq:1xsing}
\eeqa
These singularities are of codimension one, which we refer to as $p_1$ and $q_1$,
respectively.
 \mbox{$\cal C$$_{p_0}$} is the exceptional curve at fixed $s$.
\bigskip 

\noindent
\underline{Chart $1_y$}
\beqa
\Phi(x_1 y, y, y z_1,s)&=&y^2\Phi_y(x_1, y, z_1,s),\n
\Phi_y(x_1,y,z_1,s)&=&
2 s^3 y z_1^3+x_1 y z_1^2 \left(y z_1-3 s^2\right)-s y^2 z_1^4+x_1^3 y-1.
\n
   \mbox{$\cal C$$_{p_0}$ in $1_y$}&:&\mbox{Invisible in this patch.}\n
   \mbox{Singularities}&:&\mbox{None.}
\eeqa
In chart $1_y$, the exceptional curve cannot be seen and there
is no singularity. \bigskip
 
\noindent
\underline{Chart $1_z$}
\beqa
\Phi(x_1 z, y_1 z, z,s)&=&z^2\Phi_z(x_1,y_1,z,s),\n
\Phi_z(x_1,y_1,z,s)&=&z \left(2 s^3-3 s^2 x_1-s z+x_1^3+x_1 z\right)-y_1^2.
\n
   \mbox{$\cal C$$_{p_0}$ in $1_z$}&:&z=0,~~y_1=0.\n
   \mbox{Singularities}&:&(x_1,y_1,z,s)=(s,0,0,s), (-2s,0,0,s).
\label{eq:1zsing}
\eeqa
The first singularity is $p_1$, while the second is $q_1$.

\paragraph{2nd blow up}\label{2ndp1}
By the 1st blow up we found two singularities. There are 
two ways to resolve them; either we blow up  $p_1$ first, 
or $q_1$ first. In this section we blow up $p_1$ first.

In order to blow up the singularity $p_1$ in $\Phi_z(x_1,y_1,z,s)=0$, 
we shift the coordinate $x_1$ so that the position of the singularity 
becomes $(0,0,0,s)$:
\beqa
\Psi_z(\tilde x_1, y_1, z,s)&\equiv&\Phi_z(\tilde x_1+s, y_1, z,s).
\label{eq:1z1}
\eeqa
The singularities of 
$\Psi_z(\tilde x_1, y_1, z,s)=0$ are now at 
$(0,0,0,s)$ $(=p_1)$ and $(-3s,0,0,s)$ $(=q_1)$. 
We blow up the singularity of 
$\Psi_z(\tilde x_1, y_1, z,s)=0$ at $(0,0,0,s)$.

\noindent
\underline{Chart $2_{zx}$}
\beqa
\Psi_z(\tilde x_1, \tilde x_1 y_2,\tilde x_1 z_2,s)&=&
\tilde x_1^2\Psi_{zx}(\tilde x_1,y_2,z_2,s),\n
\Psi_{zx}(\tilde x_1,y_2,z_2,s)&=&\tilde x_1 z_2 (3 s+\tilde x_1+z_2)-y_2^2
.\n
   \mbox{$\cal C$$_{p_1}$ in $2_{zx}$}&:&\tilde x_1=0,~~y_2=0.\n
   \mbox{Singularities}&:&(\tilde x_1,y_2,z_2,s)=
   (0,0,0,s) (=p_2), (0,0,-3s,s) (=r_2), \n
   &&(-3s, 0,0,s) (=q_1).
\label{eq:2zxsing}
\eeqa
We find three singularities in this chart and name them as shown 
in the parentheses.  \bigskip

\noindent
\underline{Chart $2_{zy}$}
\quad In this chart, we find no singularity so we omit the details of $\Psi_{zy}$.
\bigskip

\noindent
\underline{Chart $2_{zz}$}
\beqa
\Psi_z(x_2 z,y_2 z ,z,s)&=&
z^2\Psi_{zz}(x_2,y_2,z,s),\n
\Psi_{zz}(x_2,y_2,z,s)&=&x_2 z \left(3 s x_2+x_2^2 z+1\right)-y_2^2
.\n
   \mbox{$\cal C$$_{p_1}$ in $2_{zz}$}&:&z=0,~~y_2=0.\n
   \mbox{Singularities}&:&(x_2,y_2,z,s)=
   (0,0,0,s) (=q_2), (-\frac1{3s},0,0, s) (=r_2). 
\label{eq:2zzsing}
\eeqa
We observe two singularities. The former ($q_2$) is one which can only be 
seen in this chart, while the latter ($r_2$) is already seen in chart $2_{zx}$. 

Here the process branches off in three ways depending on 
which of the three singularities  $p_2$, $r_2$ and $q_1$ in chart $2_{zx}$ is 
blown up next. 
Although they are separated for $s \neq0$, they coincide with each other 
at $s=0$ (see \eqref{eq:2zxsing}). For this reason, changing the order of blowing up these 
three singularities changes the subsequent geometries. 
In this section, we consider the case $p_2$ is
blown up next.
On the other hand, $q_2$ in chart $2_{zz}$ is separated with the other three singularities 
even at $s=0$, and hence it can be independently blown up.
We leave the blow-up of $q_2$ until later and proceed with blowing up $p_2$.

\paragraph{3rd blow up at $p_2$}
We blow up $p_2$ in chart $2_{zx}$:

\noindent
\underline{Chart $3_{zxx}$}
\beqa
\Psi_{zx}(\tilde x_1,\tilde x_1 y_3,\tilde x_1 z_3,s)&=&
\tilde x_1^2\Psi_{zxx}(\tilde x_1,y_3,z_3,s),\n
\Psi_{zxx}(\tilde x_1,y_3,z_3,s)&=&z_3 (3 s+\tilde x_1 z_3+\tilde x_1)-y_3^2
.\n
   \mbox{$\cal C$$_{p_2}$ in $3_{zxx}$}&:&\tilde x_1=0,~~y_3^2=3s z_3.\n
   \mbox{Singularities}&:&(\tilde x_1,y_3,z_3,s)=
  (-3s, 0,0,s) (=q_1).
\label{eq:3zxxsing}
\eeqa
%This singularity is $q_1$, which we have already 
%seen in charts $1_x$ and $1_z$.
%If $s\neq 0$, this singularity is not on the exceptional 
%curve \mbox{$\cal C$$_3$}.
%\bigskip

\noindent
\underline{Chart $3_{zxy}$}
\quad Regular.
\bigskip

\noindent
\underline{Chart $3_{zxz}$}
\beqa
\Psi_{zx}(x_3 z_2,y_3 z_2 ,z_2,s)&=&
z_2^2\Psi_{zxz}(x_3,y_3,z_2,s),\n
\Psi_{zxz}(x_3,y_3,z_2,s)&=&3 s x_3+x_3 (x_3+1) z_2-y_3^2
.\n
   \mbox{$\cal C$$_{p_2}$ in $3_{zxz}$}&:&z_2=0,~~y_3^2=3s x_3.\n
   \mbox{Singularities}&:&(x_3,y_3,z_2,s)=
   (0,0,-3s, s) (=r_2).
\label{eq:3zxzsing} 
\eeqa
%This singularity is also not on \mbox{$\cal C$$_3$} when $s\neq 0$.

%There are no singularities on \mbox{$\cal C$$_3$} when $s\neq0$.
%This is the reason why the resolution is incomplete; as we will see
%in the later section,   
%there appears in the complete case 
%another codimension-two singularity on \mbox{$\cal C$$_3$} 
%so that the intersection diagram acquires an additional node to 
%comprise the $E_7$ Dynkin diagram.

The remaining singularities are resolved by blowing up $q_1$, $r_2$ and $q_2$,
which are all codimension one.
% Since $r_2$ and $q_1$ are different points on \mbox{$\cal C$$_3$}, while $q_2$ 
% is not on 
% \mbox{$\cal C$$_3$} but on \mbox{$\cal C$$_2$}, 
% they can be independently blown up.%at $r_2$, $q_1$, $q_2$,
$q_1$ and $r_2$, which were overlapping at $s=0$ in chart $2_{zx}$ before blowing up $p_2$, 
are now contained in different charts  $3_{zxx}$ and $3_{zxz}$, respectively, and are separated even at $s=0$.
Also, $q_2$ stays in chart $2_{zz}$ and never coincides with them.
Therefore, the remaining three singularities are all separated with each other for any $s$
and can be independently blown up.  
%These blow-ups do not cause any new singularities.
%
The procedure is similar as before and is easily done,
but for later use, we complete the process here and present the relevant results.
After blowing up $q_1$ in chart $3_{zxx}$ and $r_2$ in chart $3_{zxz}$,
we return to chart $2_{zz}$ and blow up $q_2$.  
These blow-ups do not cause any new singularities.

\paragraph{4th blow up at $q_1$}
To blow up $q_1$ in chart $3_{zxx}$, we  
shift the $\tilde{x}_1$ coordinate so that $q_1$ is at the origin:
\beq
   \Sigma_{zxx} (\tilde{\tilde{x}}_1,y_3,z_3,s) \equiv \Psi_{zxx}(\tilde{\tilde{x}}_1-3s,y_3,z_3,s).
\label{eq:q1shift}
\eeq 

\noindent
\underline{Chart $4_{zxxx}$}
\beqa
\Sigma_{zxx}(\tilde{\tilde{x}}_1, \tilde{\tilde{x}}_1 y_4, \tilde{\tilde{x}}_1 z_4,s)&=&
\tilde{\tilde{x}}_1^2\Sigma_{zxxx}(\tilde{\tilde{x}}_1,y_4,z_4,s),\n
\Sigma_{zxxx}(\tilde{\tilde{x}}_1,y_4,z_4,s)&=&-y_4^2 + z_4 (1 + \tilde{\tilde{x}}_1 z_4 - 3 z_4 s) .\n
   \mbox{$\cal C$$_{q_1}$ in $4_{zxxx}$}&:&\tilde{\tilde{x}}_1=0,~~y_4^2=  z_4(1-3z_4 s). %\n
%   \mbox{Singularities}&:& \mbox{none}.
\label{eq:4zxxxsing}
\eeqa

\noindent
\underline{Chart $4_{zxxy}$} 
\quad We omit the details.
\bigskip

%\beqa
%\Sigma_{zxx}(x_4  y_3, y_3, y_3 z_4,s)&=&
%y_3^2\Sigma_{zxxy}(x_4,y_3,z_4,s).\n
%\Sigma_{zxxy}(x_4,y_3,z_4,s)&=&-1 + x_4 z_4 (1 + y_3 z_4) - 3 z_4^2 s. \n
%   \mbox{$\cal C$$_{q_1}$ in $4_{zxxx}$}&:& y_3=0,~~-1 + x_4 z_4 - 3 z_4^2 s=0.\n
%   \mbox{Singularities}&:& \mbox{none}.
%\label{eq:4zxxysing}
%\eeqa

\noindent
\underline{Chart $4_{zxxz}$}
\beqa
\Sigma_{zxx}(x_4 z_3, y_4 z_3, z_3,s)&=&
z_3^2\Sigma_{zxxz}(x_4,y_4,z_3,s),\n
\Sigma_{zxxz}(x_4,y_4,z_3,s)&=& - y_4^2 + x_4(1+ z_3) - 3 s .\n
   \mbox{$\cal C$$_{q_1}$ in $4_{zxxz}$}&:&z_3=0,~~y_4^2 = x_4  - 3 s. %\n
%   \mbox{Singularities}&:& \mbox{none}.
\label{eq:4zxxzsing}
\eeqa

\paragraph{4th blow up at $r_2$}
To blow up $r_2$ in chart $3_{zxz}$, we  
shift the $z_2$ coordinate so that $r_2$ is at the origin:
\beq
   \Sigma_{zxz} (x_3,y_3, \tilde{z}_2,s) \equiv \Psi_{zxz}(x_3,y_3,\tilde{z}_2-3s,s).
\eeq 

\noindent
\underline{Chart $4_{zxzx}$}
\beqa
\Sigma_{zxz}(x_3, x_3 y_4, x_3 z_4,s)&=&
x_3^2\Sigma_{zxzx}(x_3,y_4,z_4,s),\n
\Sigma_{zxzx}(x_3,y_4,z_4,s)&=&-y_4^2 + z_4(1 + x_3) - 3 s .\n
   \mbox{$\cal C$$_{r_2}$ in $4_{zxzx}$}&:&x_3=0,~~y_4^2=  z_4 - 3s. %\n
%   \mbox{Singularities}&:& \mbox{none}.
\label{eq:4zxzxsing}
\eeqa

\noindent
\underline{Chart $4_{zxzy}$}
\quad We omit the details.
\bigskip

%\beqa
%\Sigma_{zxz}(x_4 y_3, y_3, y_3 z_4,s)&=&
%y_3^2\Sigma_{zxzy}(x_4,y_3,z_4,s).\n
%\Sigma_{zxzy}(x_4,y_3,z_4,s)&=&-1 + x_4 z_4 + x_4^2 (y_3 z_4 - 3 s) .\n
%   \mbox{$\cal C$$_{r_2}$ in $4_{zxzy}$}&:&y_3=0,~~-1 + x_4 z_4 -3 x_4^2 s=0 .\n
%   \mbox{Singularities}&:& \mbox{none}.
%\label{eq:4zxzysing}
%\eeqa

\noindent
\underline{Chart $4_{zxzz}$}
\beqa
\Sigma_{zxz}(x_4 \tilde{z}_2, y_4 \tilde{z}_2, \tilde{z}_2,s)&=&
\tilde{z}_2^2\Sigma_{zxzz}(x_4,y_4,\tilde{z}_2,s),\n
\Sigma_{zxzz}(x_4,y_4,\tilde{z}_2,s)&=& - y_4^2 + x_4+ x_4^2 (\tilde{z}_2 - 3 s)   .\n
   \mbox{$\cal C$$_{r_2}$ in $4_{zxzz}$}&:&\tilde{z}_2=0,~~y_4^2 =  x_4 -3x_4^2 s. %\n
%   \mbox{Singularities}&:& \mbox{none}.
\label{eq:4zxzzsing}
\eeqa

\paragraph{3rd blow up at $q_2$}
Finally, we go back to chart $2_{zz}$ and blow up $q_2$ \eqref{eq:2zzsing}:

\noindent
\underline{Chart $3_{zzx}$}
\beqa
\Psi_{zz}(x_2, x_2 y_3, x_2 z_3,s)&=&
x_2^2\Psi_{zzx}(x_2,y_3,z_3,s),\n
\Psi_{zzx}(x_2,y_3,z_3,s)&=& -y_3^2 + z_3 (1 + x_2^3 z_3 + 3 x_2 s) .\n
   \mbox{$\cal C$$_{q_2}$ in $3_{zzx}$}&:& x_2=0,~~y_3^2=z_3. %\n
%   \mbox{Singularities}&:& \mbox{None}.
\label{eq:3zzxsing}
\eeqa

\noindent
\underline{Chart $3_{zzy}$}
\quad We omit the details.
\bigskip

\noindent
\underline{Chart $3_{zzz}$}
\beqa
\Psi_{zz}(x_3 z,y_3 z ,z,s)&=&
z^2\Psi_{zzz}(x_3,y_3,z,s),\n
\Psi_{zzz}(x_3,y_3,z,s)&=&  - y_3^2 + x_3 + x_3^3 z^3 + 3 x_3^2 z s . \n
   \mbox{$\cal C$$_{q_2}$ in $3_{zzz}$}&:&z=0,~~y_3^2= x_3. %\n
%   \mbox{Singularities}&:& \mbox{None}.
\label{eq:3zzzsing} 
\eeqa

The whole process of blowing up is summarized 
in Table \ref{SO(12)E7p1first}.

\newcommand{\ctext}[1]{\raise0ex\hbox{\textcircled{\normalsize{#1}}}}
\begin{table}[htp]
\caption{$SO(12)\rightarrow E_7$: Incomplete case when 
$p_1$ is blown up first and then $p_2$ is blown up ($p_0\to p_1 \to p_2$). 
The singularities appearing at each step of 
the process are shown with their homogeneous coordinates on $\PPsub^2$.
The ones marked by a circle
are those blown up at the subsequent processes. 
$p_0$ denotes the original 
singularity on the fiber. 
The notes in the parentheses (such as $\tilde x_1=-3s$ for $q_1$) 
imply that they are not generically ({\em i.e.} unless $s\neq0$) 
the points on the $\PPsub^2$ arising at 
the respective step of the blowing-up process. }
\begin{center}
\begin{tabular}{|l|l|l|l|l|}
\hline
&1st blow up&2nd blow up&3rd  blow up&4th blow up\\
\hline
\ctext{$p_0$}$\rightarrow$
&\ctext{$p_1(s:0:1)$}$\rightarrow$
&\ctext{$p_2(1:0:0)$}~$\mbox{(in $2_{zx}$)}\rightarrow$
&regular
&\\
&$q_1(-2s:0:1)$&$q_1(1:0:0)(\tilde x_1=-3s)$
&\ctext{$q_1(1:0:0)(\tilde x_1=-3s)$}$\rightarrow$
&regular\\
&&$r_2(1:0:-3s)$&\ctext{$r_2(0:0:1)(z_2=-3s)$}$\rightarrow$
&regular\\
&&\ctext{$q_2(0:0:1)$}~$\mbox{(in $2_{zz}$)}\rightarrow$
&regular
&\\
\hline
\end{tabular}
\end{center}
\label{SO(12)E7p1first}
\end{table}%

\subsubsection{Intersection patterns at $s \neq 0$ and $s=0$}

Through the blowing up process, we have obtained six exceptional curves 
${\cal C}_I$ ($I \in \{p_0$, \ms$p_1$, \ms$p_2$, \ms$q_1$, \ms$r_2$, \ms$q_2\}$) 
%${\cal C}_{p_0}$, ${\cal C}_{p_1}$, ${\cal C}_{p_2}$, ${\cal C}_{q_1}$, ${\cal C}_{r_2}$ and ${\cal C}_{q_2}$ 
at general $s$.
For given ${\cal C}$, we define the corresponding exceptional curve $\delta$  
as the $s \to 0$ limit of ${\cal C}$ {\it in the chart} where ${\cal C}$ originally is defined.
One can then show that the intersection pattern of $\{ {\cal C}_I \}$ at $s\neq 0$ and 
that of $\{ \delta_I \}$ at $s=0$ are $D_6$ and $A_6$, respectively. 
%(See Figure \ref{fig:SO(12)E7p1first}; the meaning of the triangular node in $A_6$
%will be clarified there.)
%
%

Let us present the detail of the derivation.
Suppose ${\cal C}$ and $\delta$ arise from some blow-up and are defined in chart $A$, 
while ${\cal C}'$ and $\delta'$ arise from a subsequent blow-up and are defined in chart $A'$.
In order to see how  ${\cal C}$ and  $\delta$ intersect with  ${\cal C}'$ and $\delta'$,
one has to locate their positions in the same chart.
This is done by lifting up ${\cal C}$ and $\delta$ from chart $A$ to chart $A'$.

Let us start from defining $\delta_{p_0}$ by ${\cal C}_{p_0}$ in chart 1 (take chart $1_z$: see \eqref{eq:1zsing}) 
as 
%\beq
\begin{flalign} & 
\mbox{\underline{Chart $1_z$}} 
\hspace{2.6cm}
 {\cal C}_{p_0} : z = 0, \,\, y_1 = 0 \quad, \,\, \delta_{p_0} : z=0, \,\, y_1 = 0. 
 &
\label{eq:1zC1}
\end{flalign}
%\eeq
Since ${\cal C}_{p_0}$ does not depend on $s$, $\delta_{p_0}$ has the same form as ${\cal C}_{p_0}$.

\begin{figure}[h]
  \begin{center}
         \includegraphics[clip, width=14.6cm]{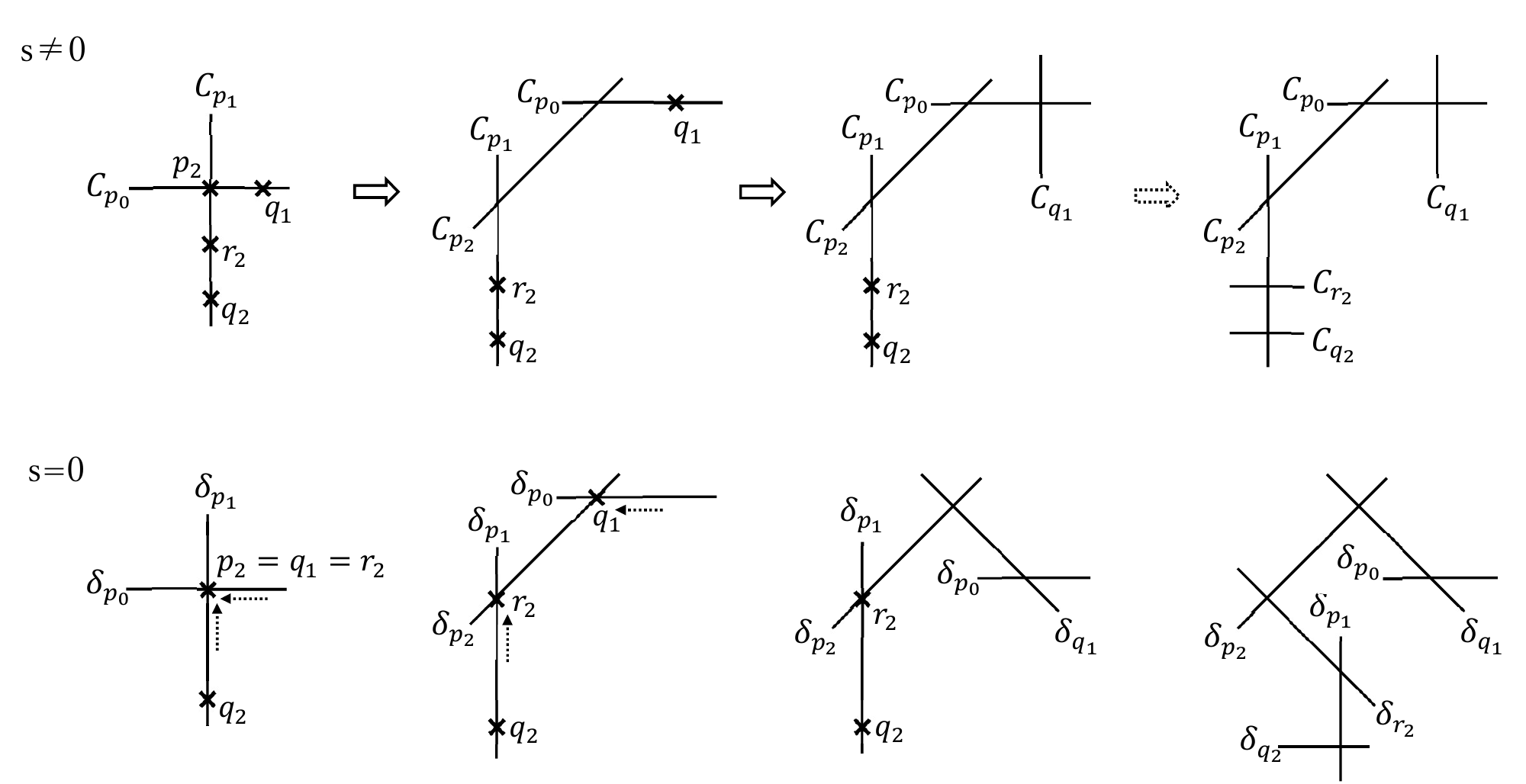}
                \caption{Exceptional curves and singularities of $SO(12)\rightarrow E_7$~:~$p_0\rightarrow
                             p_1\rightarrow p_2$ case.}
    \label{Fig:p1p2Cdelta}
  \end{center}
%\label{fig:p1p2Cdelta}
\end{figure}

Next, we lift them up in chart $2_{zx}$. 
Lifting up is done by transforming the coordinates from chart $A$ to chart $A'$:
in this case, from chart $1_z$ to chart $2_{zx}$.
The relation between these charts is
$(\tilde{x}_1,y_1,z) = (\tilde{x}_1,\tilde{x}_1y_2 , \tilde{x}_1z_2 )$ (see \eqref{eq:1z1} and \eqref{eq:2zxsing}).
Substituting it into \eqref{eq:1zC1}, we see that ${\cal C}_{p_0}$ and $\delta_{p_0}$ are written as
$\tilde{x}_1 z_2 =0\,, \, \tilde{x}_1 y_2= 0$. It is reduced to $z_2 =0\,, \, y_2 =0$,
because $\tilde{x}_1\, (=x_1-s)$ parameterizes ${\cal C}_{p_0}$ and $\delta_{p_0}$, and thus 
takes a non-zero value. 
Together with ${\cal C}_{p_1}$ and $\delta_{p_1}$ defined in chart $2_{zx}$ (see \eqref{eq:2zxsing}), we have
\begin{flalign} &
\mbox{\underline{Chart $2_{zx}$}}\hspace{2.6cm}
 {\cal C}_{p_1}  : \tilde{x}_1 = 0 \,,\, y_2 =0  \quad, \,\, \delta_{p_1} : \tilde{x}_1 = 0 \,,\, y_2 =0, & \nonumber \\
&\hspace{4.2cm}  {\cal C}_{p_0}   : z_2 = 0, \,\,\, y_2 = 0 \quad, \,\, \delta_{p_0} : z_2=0, \,\, y_2 = 0.&
\label{eq:2zxC}
\end{flalign}
Thus
 ${\cal C}_{p_0}$ ($\delta_{p_0}$) intersects with ${\cal C}_{p_1}$ ($\delta_{p_1}$) %perpendicularly 
 in this chart:
\beq
{\cal C}_{p_0}\cdot {\cal C}_{p_1} \neq 0 \quad , \quad \delta_{p_0}\cdot \delta_{p_1} \neq 0.
\eeq
Comparing the locus of the singularities \eqref{eq:2zxsing} with the positions of ${\cal C}_i$ \eqref{eq:2zxC},
we find that $p_2=(0,0,0,s)$ is located at the intersection point of 
${\cal C}_{p_0}$ and ${\cal C}_{p_1}$,
while $q_1=(-3s,0,0,s)$ and $r_2=(0,0,-3s,s)$ are on ${\cal C}_{p_0}$ and ${\cal C}_{p_1}$, respectively. 
As $s \to 0$, $q_1$ and $r_2$ approach $p_2$.
At $s=0$, they overlap with 
$p_2$ at the intersection point of $\delta_{p_0}$ and $\delta_{p_1}$.
Similarly, in chart $2_{zz}$, we have 
\begin{flalign}
& \mbox{\underline{Chart $2_{zz}$}}\hspace{2.6cm}
 {\cal C}_{p_1}  : z = 0 \,,\, y_2 =0  \quad, \,\, \delta_{p_1} : z = 0 \,,\, y_2 =0, & \nonumber \\
& \hspace{4.2cm} {\cal C}_{p_0} : \mbox{Invisible} \hspace{1.3cm}, \,\,  \delta_{p_0} : \mbox{Invisible}. &
\label{eq:2zzC}
\end{flalign}
The reason why ${\cal C}_{p_0}$ and $\delta_{p_0}$ are invisible in chart $2_{zz}$ is as follows. 
%In chart $1_z$, ${\cal C}_{p_0}$ and $\delta_{p_0}$ are given by $(x_1,y_1,z)=(x_1\neq0,0,0)$.
As seen from \eqref{eq:1z1} and \eqref{eq:2zzsing}, the coordinates of $1_z$ and $2_{zz}$ 
are related by $(\tilde{x}_1,y_1,z)=(x_2 z,y_2 z,z)$. %with $\tilde{x}_1=x_1-s$.
Then ${\cal C}_{p_0}$ and $\delta_{p_0}$ \eqref{eq:1zC1} are given by $(x_2 z,y_2 z,z) = (\tilde{x}_1\neq0,0,0)$.
It yields $x_2 = \infty$, which means that the lift-ups of ${\cal C}_{p_0}$ and $\delta_{p_0}$ cannot 
be seen in the finite region of chart $2_{zz}$.
The locus of the singularities $q_2$ and $r_2$ can be read from 
\eqref{eq:2zzsing} and \eqref{eq:2zzC}.
For $s \neq 0$, both of them are on ${\cal C}_{p_1}$.
At $s = 0$, $q_2$ stays on $\delta_{p_1}$, whereas $r_2$ has gone to infinity. 
The positions of the exceptional curves and the singularities in chart $2$ are 
schematically depicted in the leftmost column of Figure \ref{Fig:p1p2Cdelta}.

In the same way, lifting up from chart $2_{zx}$ to chart $3_{zxx}$ yields
\begin{flalign}
& \underline{\mbox{Chart $3_{zxx}$}} \hspace{2.4cm} 
   {\cal C}_{p_2}  : \tilde{x}_1=0 ,\, y_3^2 = 3s z_3 \hspace{0.3cm} ,\,\,\delta_{p_2} : \tilde{x}_1=0, \,\, y_3=0, & \n
& \hspace{4.2cm}  {\cal C}_{p_1}  :  \mbox{Invisible}  \,\hspace{1.75cm} ,\,\, \delta_{p_1} : \mbox{Invisible}, &\n
& \hspace{4.2cm} {\cal C}_{p_0}  : z_3 =0,\,  y_3 =0     \hspace{0.85cm} , \,\, \delta_{p_0}: z_3=0,\,  y_3 =0.&
\label{eq:3zxx}
\end{flalign}
This leads to  
\beq
   {\cal C}_{p_0}\cdot {\cal C}_{p_2} \neq0 \quad, \quad \delta_{p_0} \cdot \delta_{p_2} \neq0.
\eeq
The singularity $q_1 =(-3s,0,0,s)$ \eqref{eq:3zxxsing} is contained in ${\cal C}_{p_0}$ but 
not in ${\cal C}_{p_2}$ for $s\neq 0$, whereas it is located at the intersection point of $\delta_{p_0}$ and 
$\delta_{p_2}$ at $s=0$.
Lift-up from chart $2_{zx}$ to chart $3_{zxz}$ yields
\begin{flalign}
&\underline{\mbox{Chart $3_{zxz}$}} \hspace{2.4cm} 
       {\cal C}_{p_2}  : z_2=0 ,\, y_3^2 = 3s \tilde{x}_3 \hspace{0.4cm} ,\,\,  \delta_{p_2} : z_2=0, \,\, y_3=0, &  \n
&\hspace{4.2cm} {\cal C}_{p_1} : \tilde{x}_3 =0,\, y_3 =0 \hspace{0.95cm} , \,\, 
                         \delta_{p_1}: \tilde{x}_3=0,\, y_3 =0,   &\n       
&\hspace{4.2cm} {\cal C}_{p_0}  : \mbox{Invisible}  \hspace{1.95cm} ,\,\, \delta_{p_0} : \mbox{Invisible},  &
\label{eq:3zxz}
\end{flalign}
and hence
\beq
   {\cal C}_{p_1}\cdot {\cal C}_{p_2} \neq0 \quad, \quad \delta_{p_1} \cdot \delta_{p_2} \neq0.
\eeq
The singularity $r_2=(0,0,-3s,s)$ \eqref{eq:3zxzsing} is contained in ${\cal C}_{p_1}$ but 
not in ${\cal C}_{p_2}$ for $s\neq 0$, whereas it is located at the intersection point of $\delta_{p_1}$ and 
$\delta_{p_2}$ at $s=0$. The positions of these objects in chart $3$ (together with the objects 
in chart $2_{zz}$) are depicted 
in the second column of Figure \ref{Fig:p1p2Cdelta}.

As seen in the previous subsection, the remaining three singularities 
$q_1$ in chart $3_{zxx}$, $r_2$ in chart $3_{zxz}$ and $q_2$ in chart $2_{zz}$
are independently blown up.
Here we consider the blow-up of $q_1$ and lift all the information 
of chart $3_{zxx}$ \eqref{eq:3zxx} up in chart $4$. 
In chart $4_{zxxx}$ \eqref{eq:4zxxxsing}, the result is
%Next, we lift all of them up in chart 4. 
%Together with ${\cal C}_{q_1}, \delta_{q_1}$,
%${\cal C}_{r_2}$ and $\delta_{r_2}$, which arise in chart 4, we give their explicit forms.
%The blow-up of $q_1$ in chart $3_{zxx}$ is considered first. ${\cal C}_{q_1}$ and $\delta_{q_1}$
%arise in chart $4_{zxxx}$:
\begin{flalign}
& \underline{\mbox{Chart $4_{zxxx}$}} \hspace{2.05cm} 
         {\cal C}_{q_1} : \tilde{\tilde{x}}_1 = 0 ,\, y_4^2=z_4-3z_4^2 s 
                               \hspace{0.95cm} ,\,\,  \delta_{q_1}: \tilde{\tilde{x}}_1=0,y_4^2=z_4,  & \n
&\hspace{4.0cm}  {\cal C}_{p_2}  : \tilde{\tilde{x}}_1-3s=0 ,\, y_4^2 = z_4 \hspace{1.3cm} , \, \, 
                          \delta_{p_2} : \mbox{Invisible},  & \n             
&\hspace{4.0cm}  {\cal C}_{p_1}  :  \mbox{Invisible}  \hspace{3.3cm} ,\,\, \delta_{p_1} : \mbox{Invisible}, & \n
&\hspace{4.0cm}  {\cal C}_{p_0}  : z_4 =0,\,  y_4 =0     \hspace{2.4cm} , \,\, \delta_{p_0}: z_4=0,\,  y_4 =0. &   
\label{eq:4zxxx}
\end{flalign}
%${\cal C}_{q_1}$ is the exceptional curve arising from the blow-up of $q_1$ at general $s$.
%${\delta}_{q_1}$ is its $s \to 0$ limit.
We will briefly explain how the forms of ${\cal C}_{p_2}$ and $\delta_{p_2}$ are obtained.
The coordinates of chart $3_{zxx}$ and $4_{zxxx}$ are related by 
(see \eqref{eq:q1shift} and \eqref{eq:4zxxxsing})
\beq
(\tilde{x}_1+3s,y_3,z_3)=(\tilde{\tilde{x}}_1,  \tilde{\tilde{x}}_1 y_4, \tilde{\tilde{x}}_1 z_4).
\label{eq:3zxx4zxxx}
\eeq
Substituting it into \eqref{eq:3zxx}, we have
\beq
 {\cal C}_{p_2}: \tilde{\tilde{x}}_1 - 3s =0 \, ,\,  \tilde{\tilde{x}}_1^2 y_4^2 = 3s \tilde{\tilde{x}}_1 z_4 .
\eeq
The second equation is rewritten by using the first equation as
$\tilde{\tilde{x}}_1^2 y_4^2= \tilde{\tilde{x}}_1^2 z_4$.
Since $s\neq 0$, the first equation leads $\tilde{\tilde{x}}_1 \neq 0$.
Thus the second equation is reduced to $y_4^2= z_4$.
To see the form of $\delta_{p_2}$, we set $s=0$ in \eqref{eq:3zxx4zxxx}. 
Then, from \eqref{eq:3zxx}, $\delta_{p_2}$ is given by
$(\tilde{\tilde{x}}_1,  \tilde{\tilde{x}}_1 y_4,  \tilde{\tilde{x}}_1 z_4) =(0,0,z_3\neq 0)$.
It yields $z_4=\infty$ and is invisible in chart $4_{zxxx}$.
From \eqref{eq:4zxxx}, one can see the following intersections in this chart:
\beq
       {\cal C}_{p_0}\cdot {\cal C}_{p_2} \neq 0, \,\, {\cal C}_{p_0} \cdot {\cal C}_{q_1} \neq 0 \quad ,\quad 
      \delta_{p_0} \cdot \delta_{q_1} \neq 0.
\eeq
In chart $4_{zxxy}$,  one can show that ${\cal C}_{p_0}$ is invisible as well, and no intersection can be seen.
In chart $4_{zxxz}$ \eqref{eq:4zxxzsing}, we have
\begin{flalign}
& \underline{\mbox{Chart $4_{zxxz}$}} \hspace{2.0cm} 
  {\cal C}_{q_1} : z_3 = 0 ,\, y_4^2=x_4-3 s \hspace{0.55cm} ,\,\,  \delta_{q_1}: z_3=0,y_4^2=x_4, & \n
&\hspace{4.0cm}   {\cal C}_{p_2}  : x_4 z_3-3s=0 ,\, y_4^2 = x_4 \,\, , \, \, 
                           \delta_{p_2} : x_4=0 \, , \, y_4=0,  &  \n                       
&\hspace{4.0cm}   {\cal C}_{p_1}  :  \mbox{Invisible}  \hspace{2.5cm} ,\,\, \delta_{p_1} : \mbox{Invisible}, & \n
&\hspace{4.0cm}   {\cal C}_{p_0} : \mbox{Invisible}   \hspace{2.5cm} , \,\, \delta_{p_0}:  \mbox{Invisible}, &
\label{eq:4zxxz}
\end{flalign}
where $(\tilde{x}_1+3s,y_3,z_3) = (x_4 z_3,y_4 z_3, z_3)$.
${\cal C}_{q_1}$ does not intersect with ${\cal C}_{p_2}$, whereas  
$\delta_{q_1}$ intersects  with $\delta_{p_2}$:
\beq
   \delta_{p_2}\cdot{\delta}_{q_1} \neq 0.
\eeq
There is no singularity in these charts. 
The positions of the exceptional curves and the singularities after this blow-up 
are given in the third column in Figure \ref{Fig:p1p2Cdelta}.
Since $q_1$ was on ${\cal C}_{p_0}$ but not on ${\cal C}_{p_2}$ in chart $3$, 
${\cal C}_{q_1}$ intersects only with ${\cal C}_{p_0}$ in chart $4$. 
On the other hand, $q_1$ was on the intersection of $\delta_{p_0}$ and $\delta_{p_2}$,
and hence $\delta_{q_1}$ bridges them after the blow-up.

The intersections after blowing up the remaining two singularities
$r_2$ and $q_2$ are obtained in a similar manner.
The result is given in the rightmost column in Figure \ref{Fig:p1p2Cdelta}.
The final %forms of the 
intersection patterns for $s \neq 0$ and $s=0$ are the 
$D_6$ and $A_6$ Dynkin diagrams, respectively
(see also Figure \ref{fig:SO(12)E7p1first} in the next subsection; the meaning of the triangular node in 
$A_6$ will be clarified there).

\subsubsection{Intersection diagram at $s=0$ : Transmutation of a root into a weight}% and generation of a 
%half-hypermultiplet}
\label{sec:matrixp1}
%In this subsection, we explain the following phenomenon:
%one of the exceptional curves at $s\neq 0$ forming one of the simple roots
%transmutates into a weight in the limit $s \to 0$.

In this subsection, we examine how $D_6$ Dynkin diagram at $s\neq 0$ 
becomes $A_6$ at $s=0$; in other words, how the sets $\{ {\cal C}_I \}$ and 
$\{ \delta_I \}$ are related.
For this, as discovered in \cite{MT}, it is important to know the $s\to 0$ limit of ${\cal C}$, 
which we write $\lim_{s \to 0}{\cal C}$.
Here, we give a detailed explanation how this limit is explicitly calculated.
It is worth noting that $\lim_{s\to 0}{\cal C}$ does not necessarily coincide with $\delta$.
Suppose ${\cal C}$ and $\delta$ are defined in chart $A$,
while ${\cal C}'$ and $\delta'$ are defined in a ``deeper'' chart $A'$.
By definition, $\lim_{s\to 0}{\cal C}=\delta$ in chart $A$.
However, after ${\cal C}$ and $\delta$ are lifted up in chart $A'$, 
${\cal C}$ coexists with $\delta'$ and hence
$\lim_{s\to 0}{\cal C}$ may contain $\delta'$. 
%One can realize this possibility only after the lift-up.
This can happen only after the lift-up.
That is, lifting up and taking $s\to 0$ limit do not commute in general,
and we should take $s\to 0$ {\it after} the lift-up. 
 
Since the lift-ups have been completed in the previous subsection, 
all that is left is letting $s \to 0$.
Suppose the limit is taken in a chart.
As seen in the previous subsection, some of the lift-ups of ${\cal C}$'s and/or $\delta$'s
may be invisible; that is, the limit consists only of the components visible in that chart.
Thus the limit should be taken in every chart
and the final form of $\lim_{s\to 0}{\cal C}$ is obtained as their union.

In chart $1$ and chart $2$, we can see from \eqref{eq:1zC1}, \eqref{eq:2zxC} and 
\eqref{eq:2zzC} that $\lim_{s \to 0} {\cal C}_i  = \delta_i$ for $i=p_0, p_1$.
In chart $3$, we encounter the first nontrivial result.
From \eqref{eq:3zxx}, we find in chart $3_{zxx}$ as
\beq
\lim_{s \to 0} {\cal C}_{p_0}  = \delta_{p_0}, \quad
\lim_{s \to 0} {\cal C}_{p_2}  =  \{ \tilde{x}_1=0, \, y_3^2=0 \}  = 2 \delta_{p_2}.
\label{eq:3C3zxx}
\eeq
${\cal C}_{p_2}$ has only one component for $s\neq 0$, but after $s\rightarrow 0$, it 
splits into 
two overlapping (multiplicity two) $\delta_{p_2}$'s. 
Similarly, from \eqref{eq:3zxz}, we find in chart $3_{zxz}$ as
\beq
\lim_{s \to 0} {\cal C}_{p_1}  = \delta_{p_1}, \quad
\lim_{s \to 0} {\cal C}_{p_2}  =  \{ z_2=0, \, y_3^2=0 \}  = 2 \delta_{p_2}.
\label{eq:3C3zxz}
\eeq
In chart $3_{zxy}$, $\lim_{s\to 0} {\cal C}_{p_2}$ is invisible.

Next we consider charts $4_{zxx*}$ arising from the blow-up of $q_1$.
In chart $4_{zxxx}$, ${\cal C}_{p_2}$ is visible, but $\delta_{p_2}$ is 
invisible (see \eqref{eq:4zxxx}).
Nevertheless, $\lim_{s \to 0}{\cal C}_{p_2}$ does exist.
Actually, one can see from \eqref{eq:4zxxx} that
\beq
  \lim_{s \to 0} {\cal C}_{p_2} = \delta_{q_1}.
\label{eq:4zxxxC3}
\eeq
Also, we find 
$
\lim_{s \to 0} {\cal C}_{p_0} = \delta_{p_0} 
\, , \, \lim_{s \to 0} {\cal C}_{q_1} = \delta_{q_1}.
$
One can show that the same results are obtained in chart $4_{zxxy}$ except that ${\cal C}_{p_0}$ is invisible.
In chart $4_{zxxz}$, one can see from \eqref{eq:4zxxz} that ${\cal C}_{p_2}$ splits into two components as
follows:
%\bes
\beqa
 \lim_{s \to 0} {\cal C}_{p_2} & =& \{ x_4 z_3 = 0, y_4^2 = x_4 \}   \n %\\
                    &=& \{ x_4=0,y_4^2=0\} \cup \{z_3=0, y_4^2 = x_4\}  \n %\\
                    &=& 2\delta_{p_2}+\delta_{q_1}.
\label{eq:4zxxzC3}
\eeqa
%\ees
Also, $\lim_{s \to 0} {\cal C}_{q_1} = \delta_{q_1}$ is satisfied.

%There are two singularities $r_2$ and $q_2$ left to be blown up. 
In charts $4_{zxz*}$, which arise from the blow-up of $r_2$, one can show 
by repeating the same argument that 
%\bes
\beqa
& &\lim_{s \to 0} {\cal C}_{p_2} = 2 \delta_{p_2} + \delta_{r_2} \quad  (\mbox{chart $4_{zxzx}$}),
                                      %  \quad,\,\,   \lim_{s \to 0} {\cal C}_{r_2}=\delta_{r_2}  
                                     \n %  \\
& &\lim_{s \to 0} {\cal C}_{p_2} = \delta_{r_2}, 
                                      % \hspace{1.4cm} ,\,\,  \lim_{s \to 0} {\cal C}_{r_2}=\delta_{r_2} 
                                      % \hspace{1.4cm} ,\,\,  \lim_{s \to 0} {\cal C}_{r_2}=\delta_{r_2} 
                                      % \,\, ,\,\, \lim_{s \to 0} {\cal C}_{2}=\delta_{2} 
                                  \hspace{1.35cm}  (\mbox{chart $4_{zxzy}$, $4_{zxzz}$}) .  
\label{eq:4zxz*C3}
\eeqa
%\ees
All the other curves satisfy $\lim_{s \to 0} {\cal C}_{i} = \delta_i$.

Finally, after blowing up $q_2$, charts $3_{zz*}$ contain ${\cal C}_{q_2}$ as well as the 
lift-ups of ${\cal C}_{p_0}$ and ${\cal C}_{p_1}$.
A similar analysis shows that $\lim_{s \to 0} {\cal C}_{i} = \delta_i$ for $i =p_0, p_1$ and $q_2$.
%\bes
% \underline{\mbox{Chart $3_{zzx}$}} & \hspace{1cm} \lim_{s \to 0} {\cal C}_{q_2} = \delta_{q_2}
%                                       \quad,\,\,   \lim_{s \to 0} {\cal C}_{2}=\delta_{2}   \\
% \underline{\mbox{Chart $3_{zzy}$}} & \hspace{1cm} \lim_{s \to 0} {\cal C}_{q_2} = \delta_{q_2}  \\
% \underline{\mbox{Chart $3_{zzz}$}} & \hspace{1cm} \lim_{s \to 0} {\cal C}_{q_2} = \delta_{q_2}.  
%\label{eq:3zz*C3}
%\ees

Collecting all the above results, we find only ${\cal C}_{p_2}$ has a non-trivial limit.
The final form of $\lim_{s \to 0} {\cal C}_{p_2}$ is given by the union of 
the components that are visible in each chart, and hence from
\eqref{eq:3C3zxx}, \eqref{eq:3C3zxz}, \eqref{eq:4zxxxC3}, \eqref{eq:4zxxzC3} and
\eqref{eq:4zxz*C3},
\beq
 \lim_{s \to 0} {\cal C}_{p_2} = 2\delta_{p_2}+\delta_{q_1}+\delta_{r_2}.
\eeq
In conclusion, we find (Hereafter, $\lim_{s \to 0}$ will be omitted.)
%\bes
\beqa
 & {\cal C}_{p_0} =& \delta_{p_0}, \n %\\
 & {\cal C}_{p_1} =& \delta_{p_1}, \n %\\
 & {\cal C}_{p_2} =& 2\delta_{p_2}+\delta_{q_1}+\delta_{r_2}, \n %\\
 & {\cal C}_{q_1} =& \delta_{q_1}, \n %\\
 & {\cal C}_{r_2} =& \delta_{r_2}, \n %\\
 & {\cal C}_{q_2} =& \delta_{q_2}.
\label{eq:so12e7ica6}
\eeqa
The intersection matrix of ${\cal C}$'s is 
the minus of the $D_6=SO(12)$ Cartan matrix:
\beqa
-{\cal C}_I\cdot {\cal C}_J &=&
\left(
\begin{array}{cccccc}
 2 & -1 & 0 & 0 & 0 & 0 \\
 -1 & 2 & -1 & 0 & 0 & 0 \\
 0 & -1 & 2 & -1& 0  & 0 \\
 0 & 0 & -1 & 2 & -1 & -1 \\
 0 & 0 & 0 & -1 & 2 & 0 \\
 0 & 0 & 0 & -1 & 0 & 2 
\end{array}
\right),
 \label{SO12Cartan}
\eeqa
where $I,J=q_1,p_0,p_2,p_1,r_2,q_2$ in this order.
Note that this is different from the order of the blow-ups
(see the upper rightmost diagram in Figure \ref{Fig:p1p2Cdelta}, or equivalently, the upper diagram in 
Figure \ref{fig:SO(12)E7p1first} below).
From \eqref{eq:so12e7ica6}, $\delta$'s are expressed in terms of ${\cal C}$'s. 
Then one can compute the intersection matrix of $\delta$'s,
which is found to be 
\beqa
-\delta_I\cdot \delta_J &=&
\left(
\begin{array}{cccccc}
 2 & -1 & -1 & 0 & 0 & 0 \\
 -1 & 2 & 0 & 0 & 0 & 0 \\
 -1 & 0 & \frac{3}{2} & 0 & -1 & 0 \\
 0 & 0 & 0 & 2 & -1 & -1 \\
 0 & 0 & -1 & -1 & 2 & 0 \\
 0 & 0 & 0 & -1 & 0 & 2 
\end{array}
\right).
\label{eq:deltaintersectionsp1first}
\eeqa
%This agrees with the diagram in the previous section. 
This is the minus of the $A_6=SU(7)$ Cartan matrix, except that the self-intersection 
number of $\delta_{p_2}$ is not $-2$, but $-\frac{3}{2}$. 
Namely, it is true that the intersection pattern of $\delta$'s 
is given by the $A_6$ Dynkin diagram, 
but their intersection ``numbers'' 
are slightly different from the corresponding $A_6$ Cartan matrix.
This difference is expressed in the lower diagram in Figure \ref{fig:SO(12)E7p1first} 
as the triangular (not circular) node. 
In summary,
%\bes
\beqa
& &\mbox{intersection pattern \, $\Rightarrow$ $A_6$ Dynkin diagram} ,  \n % \\
& &\mbox{intersection numbers $\Rightarrow$ $A_6$ Dynkin diagram with a triangular node}.
\eeqa
%\ees
Let us call the latter diagram ``intersection diagram''.
Since it contains a $-\frac{3}{2}$ node, it is {\it not} a usual $A_6$ Dynkin diagram.
In this sense, we conclude in the present case that the intersection diagram  
is an $A_6$ ``non-Dynkin'' diagram.

\begin{figure}[h]
  \begin{center}
         \includegraphics[clip, width=7.6cm]{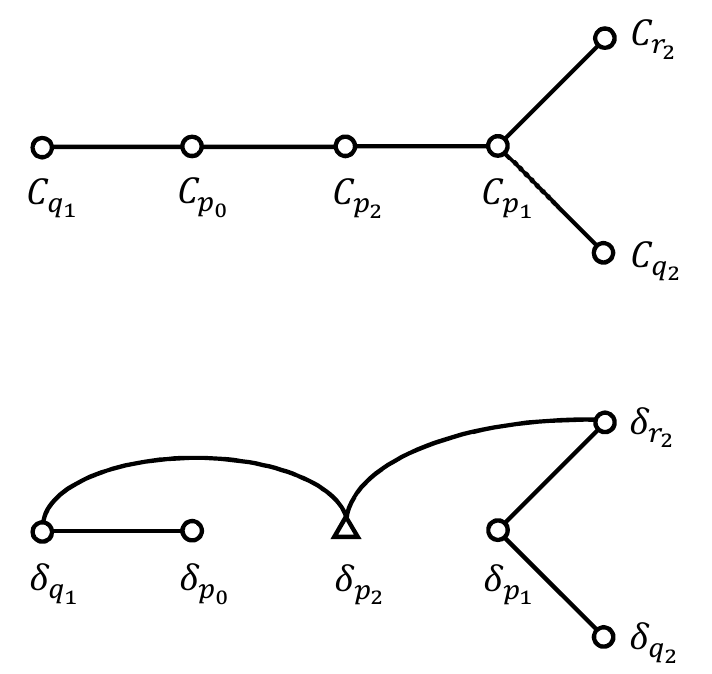}
   
                \caption{Generic $SO(12)$ intersection diagram at $s \neq 0$ (upper) and
                             incomplete intersection diagram at $s=0$ (lower) of 
                             $SO(12)\rightarrow E_7$~:~$p_1$-first ($p_0\rightarrow p_1 \rightarrow p_2$) 
                            case. }
    \label{fig:SO(12)E7p1first}
  \end{center}
%\label{fig:SO(12)E7p1first}
\end{figure}

In terms of ${\cal C}$'s, $\delta_{p_2}$ is written as
$\delta_{p_2} = \frac{1}{2}({\cal C}_{p_2}-{\cal C}_{q_1}-{\cal C}_{r_2})$.
This is one of the weights in the spinor representation of $SO(12)$.
Thus we see that at a generic $s\neq 0$ codimension-one locus 
of the singularity the exceptional fibers after the resolutions form 
a root system of $SO(12)$, but at $s=0$ one of the simple roots (${\cal C}_{p_2}$)
is transmuted to a weight in the spinor representation ($\delta_{p_2}$). 
%A similar but slightly different observation was made in \cite{BoxGraphs}.

These $\delta$'s form a basis of
the two-cycles appearing at the codimension-two singularity 
after the resolution. 
Let us consider the lattice spanned by $\delta$'s~:
\beq
  J = \sum_{I = q_1, p_0,p_2,p_1,r_2,q_2} n_I \delta_I  \quad (n_I \in {\boldmath Z}),
\label{eq:J}
\eeq
where each lattice point expresses a two-cycle at $s=0$.
One can show by using \eqref{eq:deltaintersectionsp1first} that
%\bes
\beqa
&  &\sharp \, (J\cdot J=-2\,)  \,\,\, = 60,  \n %\\
&  &\sharp \, \Big(J\cdot J=-\frac{3}{2} \,\Big)  = 32.
\eeqa
%\ees
They respectively correspond to the adjoint representation (except the Cartan part) of $SO(12)$
and the spinor representation ${\bf 32}$ of $SO(12)$.
The latter representation consists of 16 states with $n_I\geq 0$ for all $I$ and 
16 states with $n_I\leq 0$ for all $I$.
Note that, unlike in the cases of the ordinary 
(the complete) resolutions \eqref{eq:GHdecomp}, there appears only {\em one} irreducible 
representation ($={\bf 32}$) in the integer span of the two-cycles 
at the singularity, indicating that it is a half-hypermultiplet.

%%%%%%%%%%%%
\if0
Interestingly, as was observed in \cite{MT}, 
the self-intersection of one of the exceptional curves ($\delta_3$) is 
$-\frac32$,
which is the minus of the length squared of a weight 
in the spinor representation of $SO(12)$.
Thus we see that atMT a generic $s\neq 0$ codimension-one locus 
of the singularity the exceptional fibers after the resolutions form 
a root system of $SO(12)$, but at $s=0$ one of the simple roots 
is transmuted to a weight in the spinor representation. A similar 
but slightly different observation was made in \cite{BoxGraphs}.

These $\delta$'s form a basis of
the two-cycles appearing at the codimension-two singularity 
after the resolution. On the lattice spanned by these $\delta$'s, there
are precisely 32 points of length squared $\frac32$.
They are of the form
$
\sum_{I=q_1,1,3,2,r_2,q_2}n_I \delta_I
$
with either $n_I\geq 0$ for all $I$, or $n_I\leq 0$ for all $I$.
Note that, unlike in the the cases of the ordinary 
or the complete resolutions, there appears only {\em one} irreducible 
representation ($={\bf 32}$) in the integer span of the two-cycles 
at the singularity.

%The reason why this ``incomplete'' set of exceptional fibers 
%give rise to a half-hypermultiplet can be understood as follows:

%
%\begin{figure}[h]%
%\centerline{
%\includegraphics[width=0.5\textwidth]{Fig1.pdf}}
%\caption{\label{}
%}
%\end{figure}
%\begin{figure}[h]%
%\centerline{
%\includegraphics[sidth=0.5\textsidth]{Fig2.pdf}
%}
%\caption{\label{Fig2}
%}
%\end{figure}
%

\fi
%%%%%%%%%%%%%

\subsection{Complete resolution: Blowing up $p_1$ first}

\subsubsection{Blowing up process  \if0 : existence of a conifold singularity \fi }

The geometry of complete resolution is given by setting $H_{n-r+4} = s^2$ 
instead of $H_{n-r+4} = s$ in \eqref{eq:so12f} and \eqref{eq:so12g}
(the %forms of the 
other polynomials are unchanged), 
we have
\beqa
\Phi(x,y,z,s)&=&-y^2 + x^3 +f(z,s) x + g(z,s)=0,
\label{Phi}
\eeqa
where
\beqa
f(z,s)&=&-3 s^4 z^2+z^3,\n
g(z,s)&=&2 s^6 z^3-s^2 z^4.
\label{fgD6E7}
\eeqa 

Singularities can be blown up in the same way as in section \ref{p1firstprocess}.
As a result, a new isolated (codimension-two) singularity $p_3$ arises in chart 3 after 
$p_2$ is blown up in chart $2_{zx}$. 
Regarding the structure of singularities, this is the only difference between complete and incomplete cases
(see Table \ref{SO(12)E7completep1first} below).
%As we will see below, $p_3$ is the conifold singularity.
%
\begin{table}[htp]
\caption{$SO(12)\rightarrow E_7$: Complete case when 
$p_1$ is blown up first ($p_0\to p_1 \to p_2$). The new isolated codimension-two 
conifold singularity is shown in red. 
%$\delta_3$ is now 
%an ordinary node represented by a circle (cf. Figure 1).
}
\begin{center}
\begin{tabular}{|l|l|l|l|l|}
\hline
&1st blow up&2nd blow up&3rd  blow up&4th blow up\\
\hline
\ctext{$p_0$}$\rightarrow$
&\ctext{$p_1(s^2:0:1)$}$\rightarrow$
&\ctext{$p_2(1:0:0)$}~$\mbox{(in $2_{zx}$)}\rightarrow$
&{\color{red}\ctext{$p_3(1:0:0;s=0)$}$\mbox{(codim.2)}\rightarrow$}
&{\color{red}regular}\\
&$q_1(-2s^2:0:1)$&$q_1(1:0:0)(\tilde x_1=-3s^2)$
&\ctext{$q_1(1:0:0)(\tilde x_1=-3s^2)$}$\rightarrow$
&regular\\
&&$r_2(1:0:-3s^2)$&\ctext{$r_2(0:0:1)(z_2=-3s^2)$}$\rightarrow$
&regular\\
&&\ctext{$q_2(0:0:1)$}~$\mbox{(in $2_{zz}$)}\rightarrow$
&regular
&\\
\hline
\end{tabular}
\end{center}
\label{SO(12)E7completep1first}
\end{table}%

Explicitly, the 3rd blow up \eqref{eq:3zxxsing} and \eqref{eq:3zxzsing} are replaced as follows:

\paragraph{3rd blow up}  \hspace{5cm}

\noindent
\underline{Chart $3_{zxx}$}
\beqa
\Psi_{zx}(\tilde x_1,\tilde x_1 y_3,\tilde x_1 z_3,s)&=&
\tilde x_1^2\Psi_{zxx}(\tilde x_1,y_3,z_3,s),\n
\Psi_{zxx}(\tilde x_1,y_3,z_3,s)&=&z_3 (3 s^2+\tilde x_1 z_3+\tilde x_1)-y_3^2
.\n
   \mbox{$\cal C$$_{p_2}$ in $3_{zxx}$}&:&\tilde x_1=0,~~y_3^2=3s^2 z_3.\n
   \mbox{Singularities}&:&(\tilde x_1,y_3,z_3,s)=
  (-3s^2, 0,0,s) (=q_1), (0,0,-1,0) (=p_3).
\label{eq:3zxxsingcomp}
\eeqa
%This singularity is $q_1$, which we have already 
%seen in charts $1_x$ and $1_z$.
%If $s\neq 0$, this singularity is not on the exceptional 
%curve \mbox{$\cal C$$_3$}.

\noindent
\underline{Chart $3_{zxy}$}
\quad Regular.
\bigskip

\noindent
\underline{Chart $3_{zxz}$}
\beqa
\Psi_{zx}(x_3 z_2,y_3 z_2 ,z_2,s)&=&
z_2^2\Psi_{zxz}(x_3,y_3,z_2,s),\n
\Psi_{zxz}(x_3,y_3,z_2,s)&=&3 s^2 x_3+x_3 (x_3+1) z_2-y_3^2
.\n
   \mbox{$\cal C$$_{p_2}$ in $3_{zxz}$}&:&z_2=0,~~y_3^2=3s^2 x_3.\n
   \mbox{Singularities}&:&(x_3,y_3,z_2,s)=
   (0,0,-3s^2, s) (=r_2), (-1,0,0,0) (=p_3).
\label{eq:3zxzsingcomp} 
\eeqa
The new isolated singularity $p_3$ is the conifold singularity.
To see this, consider chart $3_{zxx}$ \eqref{eq:3zxxsingcomp} and shift $p_3$ to the origin via
\beq
\tilde{z}_3 \equiv z_3+1.
\label{eq:3zxxshift}
\eeq
The defining equation then reads
\beq
 \Upsilon_{zxx}(\tilde{x}_1,y_3,\tilde{z}_3,s) \equiv \Psi_{zxx}(\tilde{x}_1,y_3,\tilde{z}_3-1,s).
\eeq 
The explicit form is
\beq
 \Upsilon_{zxx}(\tilde{x}_1,y_3,\tilde{z}_3,s) = -y_3^2 + (-1 + \tilde{z}_3) (\tilde{x}_1 \tilde{z}_3 + 3 s^2),
\label{eq:conifold0}
\eeq
where $p_3$ is located at $(\tilde{x}_1,y_3,\tilde{z}_3,s)=(0,0,0,0)$.
The leading terms of $\Upsilon_{zxx}$  can be written as 
$
  \Upsilon_{zxx} \sim \tilde{z}_3(-\tilde{x}_1)-(y_3+i\sqrt{3}s)(y_3-i\sqrt{3}s)
                      \sim  X_1 X_4-X_2 X_3,
$
and hence $ \Upsilon_{zxx}=0$ is a conifold.
Including the subleading terms, we have
%\bes
\beqa
  \Upsilon_{zxx}(\tilde{x}_1,y_3,\tilde{z}_3,s) 
           & =& \tilde{z}_3 (-\tilde{x}_1+\tilde{x}_1 \tilde{z}_3 + 3 s^2)-(y_3 + i \sqrt{3} s)(y_3 - i \sqrt{3} s)  \n %\\
           & =& X_1 X_4-X_2 X_3,
\eeqa
%\ees
where $X_i$ are given by~\footnote{
One may exchange the definitions of $X_1$ and $X_4$, or $X_2$ and $X_3$, but 
it does not change the result (\eqref{eq:SO(12)E7Cdeltacomp} below). 
It also holds for all the other cases discussed in this paper. 
\label{foot:Xxechange1}}
%\bes
\beqa
  X_1 & =& \tilde{z}_3,  \n %\\
  X_4 & =& -\tilde{x}_1+\tilde{x}_1 \tilde{z}_3 + 3 s^2,  \n %\\
  X_2 & =& y_3 + i \sqrt{3} s,  \n %\\
  X_3 & =& y_3 - i \sqrt{3} s.
\label{eq:X}
\eeqa
%\ees
$p_3$ is located at the origin $(X_1,X_2,X_3,X_4)=(0,0,0,0)$, which is the conifold singularity.

The exceptional curves existing in chart $3_{zxx}$ are given by replacing $s$ to $s^2$ in 
\eqref{eq:3zxx}.
After the coordinate changes \eqref{eq:3zxxshift} and \eqref{eq:X}, their positions read

\smallskip
\noindent
\underline{\mbox{Chart $3_{zxx}$}}
%\bes
\beqa
& &\hspace{-0.0cm} {\cal C}_{p_2}  : X_4=-3s^2 ,\, 
                                            \frac{(X_2+X_3)^2}{4} = 3s^2 (X_1-1) \hspace{0.1cm}
                                           ,\, \, \delta_{p_2} : X_4=0, \, X_2=X_3=0,  \n %\\
& &\hspace{-0.0cm} {\cal C}_{p_1}  : \mbox{Invisible}  \,\hspace{5.2cm} ,\, \,\delta_{p_1} : \mbox{Invisible},  \n %\\
& &\hspace{-0.0cm} {\cal C}_{p_0}  : X_1 =1,\, X_2+X_3 =0   \hspace{3.1cm} , \, \,\delta_{p_0}: X_1=1,\, X_2=X_3 =0,  
\label{eq:3zxxX}
\eeqa
%\ees
where $s =-\frac{i}{2\sqrt{3}}(X_2-X_3)$.
The conifold singularity $p_3$ is contained in $\delta_{p_2}$, but not in $\delta_{p_0}$.
Similarly, in chart $3_{zxz}$, 
$p_3$ is on $\delta_{p_2}$, but not on $\delta_{p_1}$.
As for Figure \ref{Fig:p1p2Cdelta}, one new point $p_3$ is added on $\delta_{p_2}$ 
in the second figure of the bottom line.

\subsubsection{Intersection diagram at $s=0$}
\label{sec:conifoldlimit}

Since $p_3$ is a codimension-two singularity, it does not change the intersection 
diagram for generic $s \neq 0$ (ordinary $SO(12)$ Dynkin diagram in Figure \ref{fig:SO(12)E7p1first}).
Here we examine how the intersection diagram at $s=0$ is modified via the resolution of $p_3$.

The conifold singularity $p_3$ is resolved by inserting  an exceptional curve $\PPsmall^1$ at the origin.
This process is called the small resolution.
We write the inserted $\PPsmall^1$ as $\delta_{\mbox{\scriptsize complete}}$. 
Since $p_3$ is contained only in $\delta_{p_2}$, 
$\delta_{\mbox{\scriptsize complete}}$ intersects
only with $\delta_{p_2}$.
Adding this node to Figure \ref{fig:SO(12)E7p1first},
we find that the intersection pattern becomes $E_7$ (see Figure \ref{Fig:SO(12)E7completep1first} below).
At this stage, however, it is not clear what intersection matrix is associated with 
this diagram. To clarify it, we need a lift-up.

The smooth geometry after the small resolution is covered by two local coordinate 
patches $H_+$ and $H_-$ (see Appendix \ref{sec:smallresolution}).
By lifting up the relevant objects from chart $3$ into these patches,
one can examine how the limit $s\to 0$ is modified from the 
incomplete case and what intersection matrix is obtained.
%One can also explicitly check how $\delta_3$ and 
%$\delta_{\mbox{\scriptsize complete}}$ are intersecting.
Since $p_3$ is separated from $\delta_{p_0}$ and $\delta_{p_1}$, 
lifting up ${\cal C}_{p_2}$ and $\delta_{p_2}$ is sufficient. We consider chart $3_{zxx}$
\eqref{eq:3zxxX}.

The local coordinates of $H_+$ are $(X'_2,X'_4,\lambda)$ and the resolved geometry 
is obtained by the replacement \eqref{eq:H+}:
\beq
 (X_1,X_2,X_3,X_4) = (-\lambda X'_2, X'_2,-\lambda X'_4,X'_4).
\label{eq:H+d}
\eeq  
Here we put $'$ for the coordinates after the small resolution.
The explicit form of $\delta_{\mbox{\scriptsize complete}}$ in chart $H_+$ is given by \eqref{eq:H+delta}.
The lift-ups of ${\cal C}_{p_2}$ and $\delta_{p_2}$ are given by substituting \eqref{eq:H+d} into \eqref{eq:3zxxX}.
Then we have

\bigskip
\noindent 
$\underline{\mbox{Chart $H_+$}}$
%\bes
\beqa
  & &\hspace{7.5cm} \,  \delta_{\mbox{\scriptsize complete}}:(X'_2,X'_4,\lambda)=(0,0,\lambda), \n %\\
                   & &\hspace{-0.8cm} {\cal C}_{p_2}  : X'_4=-3s^2, 
                        \frac{(X'_2-\lambda X'_4)^2}{4}= 3s^2 (-\lambda X'_2-1) \hspace{0.1cm}
                                           ,\,\delta_{p_2} : \mbox{Invisible} , 
%\underline{\mbox{Chart $H^+$}} & \\
%                      & \hspace{-1.6cm} {\cal C}_1  : -\lambda X'_2 =1,\, X'_2=\lambda X'_4  
%                                                      \hspace{3.1cm} , \, \delta_1: \mbox{Invisible}  \\
%                      & \hspace{-1.6cm} {\cal C}_2  : \mbox{Invisible}  \,\hspace{5.3cm} ,\, \delta_2 : \mbox{Invisible} \\ 
\label{eq:3zxxXH+}
\eeqa
%\ees
where
\beq  
s=-\frac{i}{2\sqrt{3}}(X'_2+\lambda X'_4).
\eeq
$\delta_{p_2}$ is invisible, since
$\delta_{p_2}$ \eqref{eq:3zxxX} is given by $(-\lambda X'_2, X'_2,-\lambda X'_4,X'_4)=(X_1,0,0,0)$
with $X_1 \neq 0$, which is impossible. 
In this patch, $\delta_{p_2}$ and $\delta_{\mbox{\scriptsize complete}}$ do not intersect.
The $s \to 0$ limit is given by the replacement 
$X'_2 = -\lambda X'_4$, and hence 
%\bes
\beqa
 \lim_{s \to 0}{\cal C}_{p_2} & =& \{X'_4 =0, \lambda^2 X'^2_4  =0 \} \quad \mbox{with $X'_2 =-\lambda X'_4$}   \n %\\
                          & =& \{(X'_2,X'_4,\lambda) = (0,0,\lambda)\}  \n %\\
                          & =& \delta_{\mbox{\scriptsize complete}}. 
\label{eq:H+limC3}
\eeqa
%\ees

In the other patch $H_-$, the local coordinates are $(X'_1,X'_3,\mu)$ and the resolution is given by \eqref{eq:H-}:
\beq
 (X_1,X_2,X_3,X_4) = (X'_1,-\mu X'_1, X'_3,-\mu X'_3).
\label{eq:H-d}
\eeq
In this patch, $\delta_{\mbox{\scriptsize complete}}$ \eqref{eq:H-delta} as well as
${\cal C}_{p_2}$ and $\delta_{p_2}$ \eqref{eq:3zxxX} 
take the following forms: 

\bigskip
\noindent
$\underline{\mbox{Chart $H_-$}}$
%\bes
\beqa
                  &    & \hspace{7.4cm} \,       \delta_{\mbox{\scriptsize complete}}:(X'_1,X'_3,\mu)=(0,0,\mu),  \n %\\
                  &    & \hspace{-0.4cm} {\cal C}_{p_2}  : \mu X'_3=3s^2 ,\, 
                                          \frac{(X'_3-\mu X'_1)^2}{4} = 3s^2 (X'_1-1) \hspace{0.1cm}
                                           ,\,\delta_{p_2} : (X'_1,X'_3,\mu)=(X'_1,0,0) ,
%\underline{\mbox{Chart $H^-$}} & \\
%                      & \hspace{-1.6cm} {\cal C}_1  : X'_1 =1,\, X'_3=\mu X'_1  
%                                                      \hspace{3.0cm} , \, \delta_1: (X'_1,X'_3,\mu)=(1,0,0) \\
%                      & \hspace{-1.6cm} {\cal C}_2  : \mbox{Invisible}  \,\hspace{4.7cm} ,\, \delta_2 : \mbox{Invisible}, \\
\label{eq:3zxxXH-}
\eeqa
%\ees
where 
\beq
 s=\frac{i}{2\sqrt{3}}(X'_3+\mu X'_1).
\eeq
This time, $\delta_{p_2}$ 
%is given by $(X'_1,-\mu X'_1, X'_3,-\mu X'_3)=(X_1,0,0,0)$, which 
is visible.
$\delta_{p_2}$ and $\delta_{\mbox{\scriptsize complete}}$ are intersecting 
%perpendicularly 
at $(X'_1,X'_3,\mu)=(0,0,0)$:
\beq
  \delta_{p_2} \cdot \delta_{\mbox{\scriptsize complete}} \neq 0.
\eeq
The $s \to 0$ limit is given by the replacement $X'_3 = -\mu X'_1$:
%\bes
\beqa
 \lim_{s \to 0}{\cal C}_{p_2} & =& \{-\mu^2 X'_1 =0, \mu^2 X'^2_1  =0 \} \quad \mbox{with $X'_3 =-\mu X'_1$}  \n % \\
                          & =& \{X'_1=0\} \cup \{\mu^2=0\} \quad \mbox{with $X'_3 =-\mu X'_1$}  \n %\\
                          & =& \{(X'_1,X'_3,\mu)=(0,0,\mu)\} \cup \{(X'_1,X'_3,\mu)=(X'_1,0,0)\}^{\otimes 2}  \n %\\
                          & =& \delta_{\mbox{\scriptsize complete}}+2\,\delta_{p_2}. 
\label{eq:H+limC3}
\eeqa
%\ees
In conclusion, for the complete resolution, 
the new exceptional curve $\delta_{\mbox{\scriptsize complete}}$ is contained in $\lim_{s\to 0}{\cal C}_{p_2}$
with multiplicity one,
and \eqref{eq:so12e7ica6} is modified as 
%\bes
\beqa
 & {\cal C}_{p_0} =& \delta_{p_0},   \n %\\
 & {\cal C}_{p_1} =& \delta_{p_1},  \n %\\
 & {\cal C}_{p_2} =& 2\delta_{p_2}+\delta_{q_1}+\delta_{r_2}+\delta_{\mbox{\scriptsize complete}},  \n %\\
 & {\cal C}_{q_1} =& \delta_{q_1},  \n %\\
 & {\cal C}_{r_2} =& \delta_{r_2},  \n %\\
 & {\cal C}_{q_2} =& \delta_{q_2}.
\label{eq:SO(12)E7Cdeltacomp}
\eeqa
%\ees
%
Then, assuming that the intersection matrix of these seven $\delta$'s 
is just the minus of ordinary $E_7$ Cartan matrix, 
we find that the intersection matrix of the six ${\cal C}$'s computed by 
\eqref{eq:SO(12)E7Cdeltacomp} is precisely 
the minus of the $SO(12)$ Cartan matrix.
It means that, in contrast to the incomplete case, $\delta$'s have no node with self-intersection 
number $-\frac{3}{2}$ and the intersection diagram is the ordinary $E_7$ Dynkin diagram.
That is, $\delta_{p_2}$, whose self-intersection number is $-\frac{3}{2}$ (triangular node) in the 
incomplete case, becomes the ordinary node with self-intersection $-2$ (circular node)
by virtue of the existence of $\delta_{\mbox{\scriptsize complete}}$.
The result is summarized in Figure \ref{Fig:SO(12)E7completep1first}.
As usual, two cycles $J$ spanned by the seven $\delta$'s with 
$J\cdot J = -2$ contain two ${\bf 32}$'s as in \eqref{eq:GHdecomp}.

%Let us note that there is a relation between the intersection diagrams for
%incomplete (and complete) resolutions and the box graphs:
%Figure \ref{fig:SO(12)E7p1first} (with Figure \ref{Fig:SO(12)E7completep1first})
%corresponds to the box graph in the right column of Figure 33 in \cite{BoxGraphs}. 

\begin{figure}[htbp]
  \begin{center}
\includegraphics[clip, width=7.6cm]{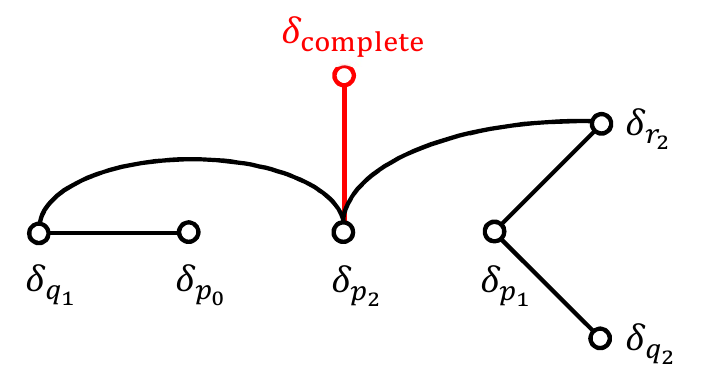}
         
\caption{Complete $E_7$ intersection diagram of 
$SO(12)\rightarrow E_7$~:~$p_1$-first ($p_0\rightarrow p_1 \rightarrow p_2$) case.
%The $E_7$ Dynkin diagram obtained by 
%a complete resolution with $p_1$ blown up first. 
}

    \label{Fig:SO(12)E7completep1first}
  \end{center}
\end{figure}

\subsection{Incomplete resolution: Blowing up $q_1$ first}
\label{q1first}
\subsubsection{Blowing up process}
In section \ref{p1first},
between the two singularities $p_1$ and $q_1$ arising 
from the first blow up in chart $1_z$
\eqref{eq:1zsing}, $p_1$ was blown up first.
In this section, let us blow up $q_1$ first and see the differences.
This time we make a shift of the coordinate $x_1$ 
so that $q_1$ comes to $(0,0,0,s)$:
We define 
\beqa
\Sigma_z(\tilde x_1, y_1, z,s)&\equiv&\Phi_z(\tilde x_1-2s, y_1, z,s).
\label{Sigmatxcoordinateshift}
\eeqa
$\Sigma_z(\tilde x_1, y_1, z,s)=0$ has singularities 
$(3s,0,0,s)$ $(=p_1)$ and $(0,0,0,s)$ $(=q_1)$. 
We blow up the latter singularity.
The process is completely parallel to that in 
the previous section so we will only describe 
the relevant charts and show the main differences 
from the previous case. 
For later use, we also present the definition of $\delta$ and 
the lift-ups of ${\cal C}$ and $\delta$.
For the $p_1$-first case studied in the previous sections, 
the exceptional curves arising from the blow-ups of the singularities 
$p_0$, $p_1$, $p_2$, $q_1$, $r_2$ and $q_2$ are respectively expressed by
${\cal C}_{p_0}$, ${\cal C}_{p_1}$, ${\cal C}_{p_2}$, ${\cal C}_{q_1}$, ${\cal C}_{r_2}$ and ${\cal C}_{q_2}$.
We use the same notation for the $q_1$-first case as well.

\begin{figure}[h]
  \begin{center}
         \includegraphics[clip, width=11.0cm]{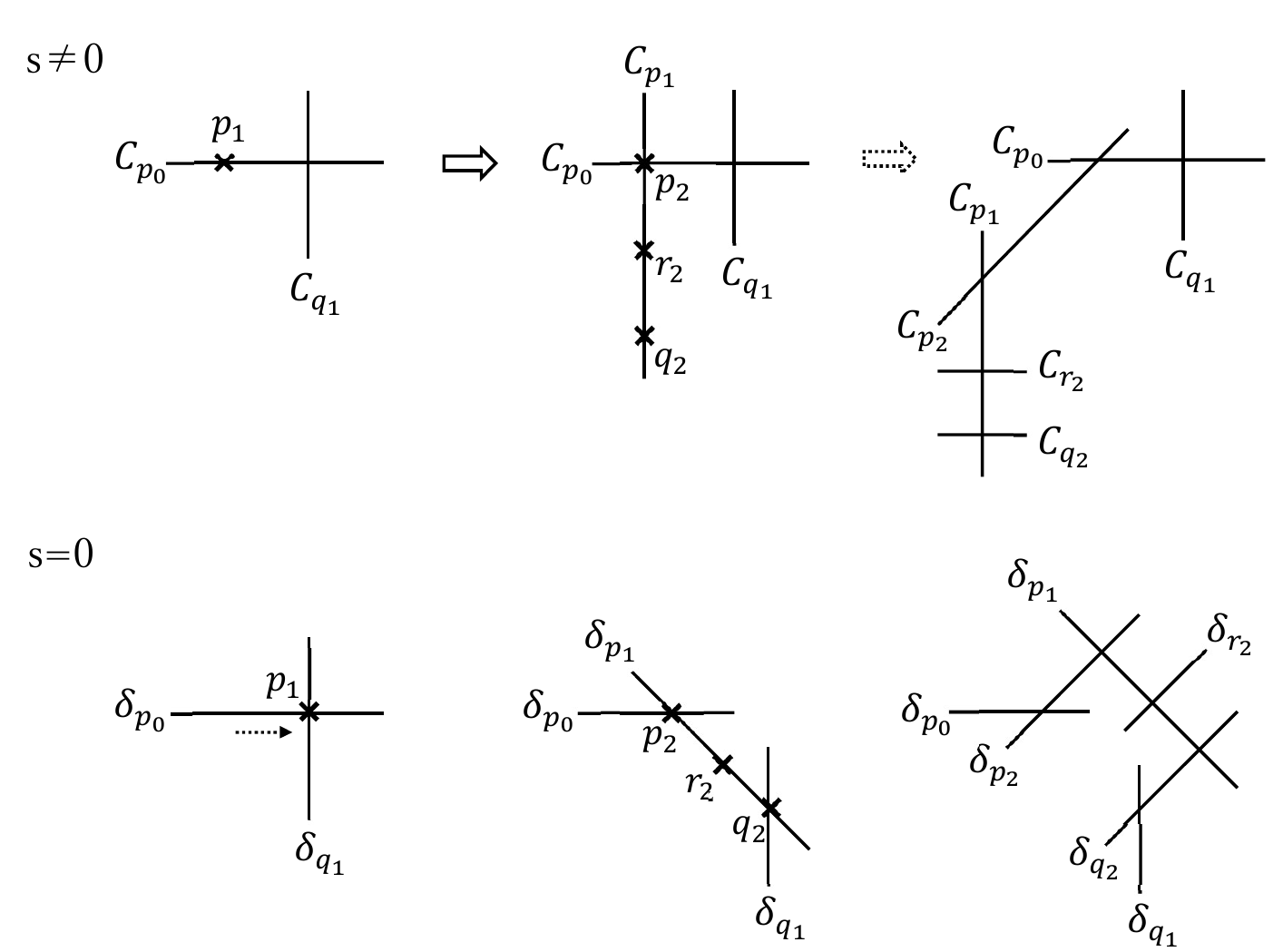}
   
                \caption{Exceptional curves and singularities of $SO(12)\rightarrow E_7$~:~$p_0\rightarrow
                             q_1$ case.}
    \label{Fig:q1Cdelta}
  \end{center}
%\label{fig:p1p2Cdelta}
\end{figure}

\paragraph{2nd blow up} From the blow-up of $q_1$, ${\cal C}_{q_1}$ arises.
${\cal C}_{p_0}$ and $\delta_{p_0}$ are lifted up from chart $1_{z}$.
\label{2ndq1} 
%
%\hskip -1em\noindent
\noindent
\underline{Chart $2_{zx}$}
\beqa
\Sigma_z(\tilde x_1, \tilde x_1 y_2,\tilde x_1 z_2,s)&=&
\tilde x_1^2\Sigma_{zx}(\tilde x_1,y_2,z_2,s),\n
\Sigma_{zx}(\tilde x_1,y_2,z_2,s)&=&
z_2 (3 s-\tilde x_1) (3 s-\tilde x_1-z_2)-y_2^2.\n
   \mbox{$\cal C$$_{q_1}$}&:&
   \tilde x_1=0,~~3 s z_2 (3 s-z_2)-y_2^2=0,
   \quad \delta_{q_1}\,:\,\tilde x_1=0,~y_2=0, \n
   \mbox{$\cal C$$_{p_0}$}&:&
   z_2=0,~~y_2=0
   \hspace{2.8cm} , \quad \delta_{p_0}\,:\, z_2=0,~y_2=0. \n
   \mbox{Singularities}&:&(\tilde x_1,y_2,z_2,s)=
 (3s, 0,0,s) (=p_1).
 \label{chart2zx}
\eeqa
The positions of these objects are depicted in the leftmost column of Figure \ref{Fig:q1Cdelta}. 
There are no other singularities in chart $2_{zy}$ or $2_{zz}$, 
so we blow up $p_1$ in chart $2_{zx}$. Again, we need to shift 
the coordinate so that the singularity we now blow up comes 
to the origin:
\beqa
\Psi_{zx}(\!\tilde{\tilde {~x_1}}, y_2, z_2,s)&\equiv&\Sigma_{zx}(\!\tilde{\tilde {~x_1}}+3s,y_2,z_2,s).
\label{Sigmazxcoordinateshift}
\eeqa

\paragraph{3rd blow up}
The relevant charts are $3_{zxx}$ and $3_{zxz}$.
From the blow-up of $p_1$, ${\cal C}_{p_1}$ arises.

\noindent
\underline{Chart $3_{zxx}$}
\beqa
\Psi_{zx}(\!\tilde{\tilde {~x_1}},\!\tilde{\tilde {~x_1}} y_3,\!\tilde{\tilde {~x_1}} z_3,s)&=&
\!\tilde{\tilde {~x_1}}^2\Psi_{zxx}(\!\tilde{\tilde {~x_1}},y_3,z_3,s),\n
\Psi_{zxx}(\!\tilde{\tilde {~x_1}},y_3,z_3,s)&=&
\!\tilde{\tilde {~x_1}} z_3 (z_3 + 1) - y_3^2.
\n
   \mbox{$\cal C$$_{p_1}$}&:&\!\tilde{\tilde {~x_1}}=0,~~y_3=0
   \hspace{3.2cm} , \quad \delta_{p_1}\,\,:\, \tilde{\tilde {~x_1}}=0,~y_3=0,\n
    \mbox{$\cal C$$_{q_1}$}&:&
   \tilde{\tilde{~x_1}}=-3s,~~\tilde{\tilde{~x_1}} z_3 (1+z_3)-y_3^2=0,
   \quad \delta_{q_1}\,:\, \mbox{Invisible} , \n
   \mbox{$\cal C$$_{p_0}$}&:&
   z_3=0,~~y_3=0
   \hspace{3.3cm} , \quad \delta_{p_0}\,\,:\, z_3=0,~y_3=0. \n   
   \mbox{Singularities}&:&(\!\tilde{\tilde {~x_1}},y_3,z_3,s)=
  (0,0,-1,s) (=r_2), (0,0,0,s) (=p_2).
\label{eq:3zxxq1sing}
\eeqa

\noindent
\underline{Chart $3_{zxz}$}
\beqa
\Psi_{zx}(x_3 z_2, y_3 z_2,z_2,s)&=&
z_2^2\Psi_{zxz}(x_3,y_3,z_2,s),\n
\Psi_{zxx}(x_3,y_3,z_2,s)&=&x_3 (x_3+1) z_2-y_3^2.
\n
   \mbox{$\cal C$$_{p_1}$} &:&z_2=0,~~y_3=0
   \hspace{3.4cm} , \quad \delta_{p_1}\,\,:\, z_2=0,~y_3=0 ,\n
    \mbox{$\cal C$$_{q_1}$}&:&
   x_3 z_2=-3s,~~x_3 z_2 (x_3+1)-y_3^2=0,
   \quad \delta_{q_1}\,:\, x_3=0 ,\, y_3=0, \n
   \mbox{$\cal C$$_{p_0}$}&:&
   \mbox{Invisible}
   \hspace{4.6cm} , \quad \delta_{p_0}\,\,:\,\mbox{Invisible} . \n
  \mbox{Singularities}&:&(x_3,y_3,z_2,s)=
  (-1,0,0,s) (=r_2), (0,0,0,s) (=q_2).
\label{eq:3zxzq1sing}
\eeqa
Note that ${\cal C}_{q_1}$ and $\delta_{q_1}$ are independently lifted up from (\ref{chart2zx}).
The positions of these objects are depicted in the middle column of Figure \ref{Fig:q1Cdelta}. 
One can see from \eqref{eq:3zxxq1sing} and \eqref{eq:3zxzq1sing} that the remaining three 
singularities $p_2$, $r_2$ and $q_2$ are separated for all $s$ (not only for $s\neq 0$ but also for $s=0$), 
and hence they are independently blown up.
The whole blowing up process is summarized in Table \ref{SO(12)E7q1first}.
The resulting intersection patterns are depicted in the rightmost column of  Figure \ref{Fig:q1Cdelta}.
For $s\neq 0$, the intersection pattern is the $D_6$ Dynkin diagram,
which is identical to the one in the $p_1$-first case (see the upper rightmost diagram  
in Figure \ref{Fig:p1p2Cdelta}, or the upper diagram in Figure \ref{fig:SO(12)E7p1first}). 
The orders of the nodes are the same as well.
On the other hand, the intersection pattern at $s=0$ is the $E_6$ Dynkin diagram,
which is different from the one ($A_6$) in the $p_1$-first case.
% (see Figure \ref{Fig:SO(12)E7q1first} below).

\begin{table}[h]
\caption{$SO(12)\rightarrow E_7$: Incomplete case when 
$q_1$ is blown up first ($p_0\to q_1$).  }
\begin{center}
\begin{tabular}{|l|l|l|l|l|}
\hline
&1st blow up&2nd blow up&3rd blow up&4th blow up\\
\hline
\ctext{$p_0$}$\rightarrow$
&\ctext{$q_1(-2s:0:1)$}$\rightarrow$
&regular
&
&
\\
% 2??s????
&$p_1(s:0:1)$
&\ctext{$p_1(1:0:0)$}$(\tilde x_1=3s)\rightarrow$
&\ctext{$p_2(1:0:0)$}$\rightarrow$
&regular
\\
% 3??s????
&
&
&\ctext{$q_2(0:0:1)$}$\rightarrow$
&regular
\\
% 4??s????
&
&
&\ctext{$r_2(1:0:-1)$}$\rightarrow$
&regular
\\
\hline
\end{tabular}
\end{center}
\label{SO(12)E7q1first}
\end{table}%

\subsubsection{Intersection diagram at $s=0$: Differences from the $p_1$-first case}%

Following the same procedure given in section \ref{sec:matrixp1}, we find the
$s \to 0$ limit of ${\cal C}$ for the $q_1$-first case as
%\bes
\beqa
& {\cal C}_{p_0}=& \delta_{p_0},  \n %\\
& {\cal C}_{q_1}=& 2\delta_{q_1}+2\delta_{p_1}+\delta_{p_2}+2\delta_{q_2}+\delta_{r_2},  \n %\\
& {\cal C}_{p_1}=& \delta_{p_1}, \n %\\
& {\cal C}_{p_2}=& \delta_{p_2},  \n %\\
& {\cal C}_{r_2}=& \delta_{r_2}, \n %\\
& {\cal C}_{q_2}=& \delta_{q_2}. 
\label{Cswithds_q1first}
\eeqa
%\ees
%%%%%%%
\if0
\beqa
{\cal C}_2&=&2\delta_2+\delta_{p_2}+2\delta_3
+2\delta_{q_2}+\delta_{r_2},\n
{\cal C}_1&=&\delta_1,\n
{\cal C}_{p_2}&=&\delta_{p_2},\n
{\cal C}_3&=&\delta_3,\n
{\cal C}_{q_2}&=&\delta_{q_2},\n
{\cal C}_{r_2}&=&\delta_{r_2}.
\label{Cswithds_q1first}
\eeqa
\fi
%%%%%%
The first two terms of ${\cal C}_{q_1}$ are readily observed in chart $3$ (see \eqref{eq:3zxxq1sing} 
and \eqref{eq:3zxzq1sing}).
The other terms of ${\cal C}_{q_1}$ can only be seen in deeper charts. 
We skip the detail of the derivation.

\begin{figure}[htbp]
  \begin{center}
          
         \includegraphics[clip, width=7.6cm]{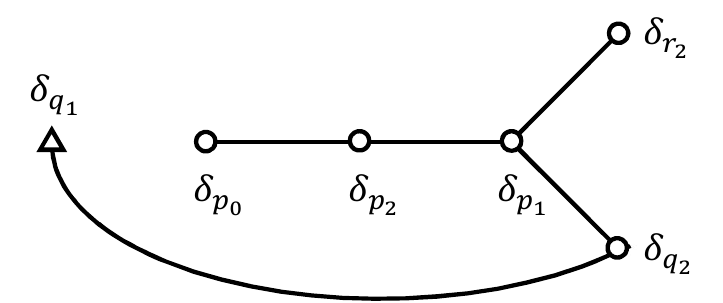}
       \caption{Incomplete intersection diagram of 
                   $SO(12)\rightarrow E_7$~:~$q_1$-first ($p_0\rightarrow q_1$) case.}
    \label{Fig:SO(12)E7q1first}
  \end{center}
\end{figure}

The intersection matrix among ${\cal C}$'s is the same one with the $p_1$-first case
and is given by (\ref{SO12Cartan}) with the same order 
$I,J=q_1,p_0,p_2,p_1,r_2,q_2$. 
Then (\ref{Cswithds_q1first}) 
yields the intersection matrix of $\delta$'s as
\beqa
-\delta_I\cdot \delta_J &=&
\left(
\begin{array}{cccccc}
 \frac{3}{2} & 0 & 0 & 0 & 0 & -1 \\
 0 & 2 & -1 & 0 & 0 & 0 \\
 0 & -1 & 2 & -1 & 0 & 0 \\
 0 & 0 & -1 & 2 & -1 & -1 \\
 0 & 0 & 0 & -1 & 2 & 0 \\
 -1 & 0 & 0 & -1 & 0 & 2 
\end{array}
\right).
\label{deltaintersectionsq1first}
\eeqa
%where $I,J=2,1,p_2,3,q_2,r_2$.
%In this case, we obtain an $E_6$-like diagram as one 
%representing the intersections of the exceptional curves 
%(\ref{deltaintersectionsq1first})
%at the codimension-two singularity. (\ref{deltaintersectionsq1first}) 
%is not, however, the $E_6$ Cartan matrix itself, as the 
%self-intersection of $\delta_1$ is $-\frac32$. 
This is almost the $E_6$ Cartan matrix except that the self-intersection 
number of $\delta_{q_1}$ is $-\frac{3}{2}$, which is expressed as a triangular node 
in Figure \ref{Fig:SO(12)E7q1first}.
Thus the intersection diagram for this case is an $E_6$ non-Dynkin diagram. 

We can search for the elements of the form 
$
\sum_{I=q_1,p_0,p_2,p_1,r_2,q_2}n_I \delta_I
$
whose square is $-\frac32$ to find, again, that there are $16+16$
such elements, the former of which have $n_I\geq 0$ for all $I$, and 
the latter of which have $n_I\leq 0$ for all $I$.
Thus, in this case as well, there is only {\em one} irreducible 
representation ($={\bf 32}$)  
at the singularity, showing that it is a half-hypermultiplet.

\subsection{Complete resolutions: Blowing up $q_1$ first}\label{completeresolutionSO(12)E7}

%To achieve the complete resolution, we have only to set 
% $H_{n+4}=s^2$ with other parameters being the same as  
%the previous section.
%The process of the blow-ups is almost the same as that in 
%the incomplete resolutions,
%except for the replacement $s\rightarrow s^2$.

% \subsubsection{Blowing up process}

\begin{figure}[htbp]
  \begin{center}

\includegraphics[clip, width=9.5cm %10.4cm
]{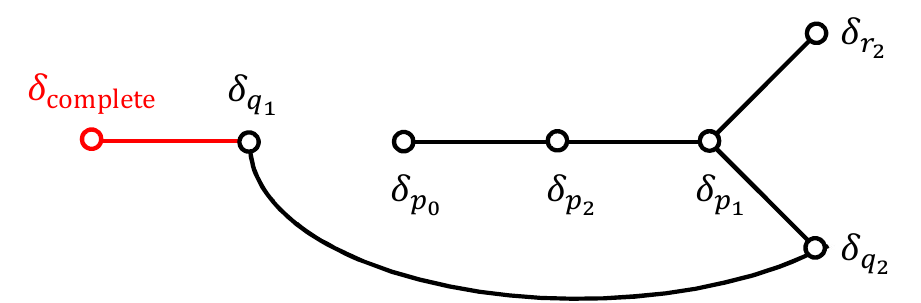} %{SO(12)E7completeq1first.pdf}       
         
\caption{Complete $E_7$ intersection diagram of 
            $SO(12)\rightarrow E_7$~:~$q_1$-first ($p_0\rightarrow q_1$) case.}
    \label{Fig:SO(12)E7completeq1first}
  \end{center}
\end{figure}

The process of the blow-ups is almost the same as that in 
the incomplete resolutions.
A difference arises in chart $2_{zz}$, where 
a conifold singularity is developed at 
$(x_2,y_2,z,s)=(0,0,0,0)$, which we denote by $q_3$ 
(shown in red in Table \ref{SO(12)E7completeq1first}), 
where the relation to the 
coordinates in chart $1_{z}$ is 
$(\tilde x_1,y_1,z,s)=(x_2 z,y_2 z,z,s)$. 
This is an isolated codimension-two singularity developed 
only at $s=0$. Since this is in chart $2_{zz}$, 
this singularity is located at $(0:0:1)$ on $\PPsmall^2$ 
emerged by the blow up at $s=0$. Therefore, 
it is not visible in chart $2_{zx}$ or $3_{zx*}$. 
Moreover, after the coordinate shift similar to 
(\ref{Sigmatxcoordinateshift}),
$\Psi_{zx}$ becomes identical to the incomplete case.
Thus the process is the same as the incomplete case 
afterwards. 
Therefore, the only extra exceptional curve is 
the one arising from the small resolution of the 
isolated conifold singularity on $\delta_{q_1}$. This 
adds an extra node to the diagram in the 
Figure \ref{Fig:SO(12)E7q1first}, as we show in Figure \ref{Fig:SO(12)E7completeq1first}. 
We denote this new curve as $\delta_{\rm complete}$ here.
This is $E_7$, 
and the extra node again extends from $\delta_{q_1}$
that was the ``weight'' node represented by the triangle 
in the incomplete case. 
How $\delta_{\rm complete}$ modifies the $\lim_{s\to 0}{\cal C}$ can be seen in the same way
as explained in section \ref{sec:conifoldlimit}.
%In the complete resolution, 
%it becomes an ordinary node with 
%self-intersection $-2$, being consistent with 
%the modified relation:
The result is
%\bes
\beqa
& {\cal C}_{p_0}=& \delta_{p_0},  \n %\\
& {\cal C}_{q_1}=& 2\delta_{q_1}+2\delta_{p_1}+\delta_{p_2}+2\delta_{q_2}+\delta_{r_2}+\delta_{\rm complete},  \n %\\
& {\cal C}_{p_1}=& \delta_{p_1},  \n %\\
& {\cal C}_{p_2}=& \delta_{p_2},  \n %\\
& {\cal C}_{r_2}=& \delta_{r_2},  \n %\\
& {\cal C}_{q_2}=& \delta_{q_2}. 
\label{Cswithds_q1firstmodified}
\eeqa
%\ees
%%%%%%%%%%%%%%
\if0
\beqa
{\cal C}_2&=&2\delta_2+\delta_{p_2}+2\delta_3
+2\delta_{q_2}+\delta_{r_2}+\delta_{\rm complete},\n
{\cal C}_1&=&\delta_1,\n
{\cal C}_{p_2}&=&\delta_{p_2},\n
{\cal C}_3&=&\delta_3,\n
{\cal C}_{q_2}&=&\delta_{q_2},\n
{\cal C}_{r_2}&=&\delta_{r_2}.
\label{Cswithds_q1firstmodified}
\eeqa
\fi
%%%%%%%%%%%%%%
This reproduces the minus of the $D_6$ 
Cartan matrix as the intersection matrix of ${\cal C}$'s
if the intersections among $\delta$'s are given by the minus 
of the proper $E_7$ Cartan matrix.
Therefore, in the complete case, $\delta_{q_1}$ becomes an ordinary
node with self-intersection $-2$ as shown in Figure  \ref{Fig:SO(12)E7completeq1first}.
%Figure \ref{Fig:SO(12)E7q1first} and \ref{Fig:SO(12)E7completeq1first}  form 
%the bottom-most box graph in the left column of Figure 33 in \cite{BoxGraphs}.

\begin{table}[h]
\caption{$SO(12)\rightarrow E_7$: Complete case when 
$q_1$ is blown up first ($p_0\to q_1$).  }
\begin{center}
\begin{tabular}{|l|l|l|l|l|}
\hline
&1st blow up&2nd blow up&3rd blow up&4th blow up\\
\hline
\ctext{$p_0$}$\rightarrow$
&\ctext{$q_1(-2s:0:1)$}$\rightarrow$
&{\color{red}\ctext{$q_3(0:0:1;s=0)$}$\mbox{(codim.2)}\rightarrow$}
&{\color{red}regular}
&
\\
% 2??s????
&$p_1(s:0:1)$
&\ctext{$p_1(1:0:0)$}$(\tilde x_1=3s)\rightarrow$
&\ctext{$p_2(1:0:0)$}$\rightarrow$
&regular
\\
% 3??s????
&
&
&\ctext{$q_2(0:0:1)$}$\rightarrow$
&regular
\\
% 4??s????
&
&
&\ctext{$r_2(1:0:-1)$}$\rightarrow$
&regular
\\
\hline
\end{tabular}
\end{center}
\label{SO(12)E7completeq1first}
\end{table}%

%
%\begin{figure}[h]%
%\centerline{
%\includegraphics[sidth=0.5\textsidth]{Fig3.pdf}}
%
%\caption{\label{Fig3}
%}
%\end{figure}
%

\subsection{Other inequivalent orderings}
So far we have considered incomplete and complete
resolutions where
the order of blowing up singularities is 
$p_0\rightarrow p_1\rightarrow p_2$ and
$p_0\rightarrow q_1$.
In the former case, we have also two codimension-one 
singularities $q_1$ and $r_2$, 
besides $p_2$ and $q_2$, after the 2nd blow up, as shown 
in Table \ref{SO(12)E7p1first}.
Since $q_2$ never coincides with the other three 
for any value of $s$, it can be blown up independently 
at any stage. On the other hand, $p_2$, $q_1$ and $r_2$ 
become the same point $(1:0:0)$ on the $\PPsmall^2$ at $s=0$, 
so a different  intersection diagram arises if we blow up  
$q_1$ or $r_2$ instead of $p_2$ after blowing up $p_1$. 
The concrete process is similar to the previous cases 
so we only describe the results. \bigskip

%%%%%%%%%%%%%%%%%%%%
\if0
\begin{figure}[h]
  \begin{center}
         \includegraphics[clip, width=8.6cm]{SO(12)E7p1q1.pdf}
   
                \caption{}
    \label{}
\label{Fig:SO(12)E7p1q1}
  \end{center}
%\label{Fig:SO(12)E7p1q1}
\end{figure}
\begin{figure}[h]
  \begin{center}
         \includegraphics[clip, width=8.6cm]{SO(12)E7p1r2.pdf}
   
                \caption{}
    \label{}
\label{Fig:SO(12)E7p1r2}
  \end{center}
%\label{Fig:SO(12)E7p1r2}
\end{figure}
\fi
%%%%%%%%%%%%%%%%%%%%
%
\noindent 
\underline{$p_0 \rightarrow p_1 \rightarrow q_1$ case}
\medskip 

If $q_1$ is blown up after $p_1$, 
the relations among 
${\cal C}$'s and $\delta$'s are given by
(the modifications via $\delta_{\scriptsize \mbox{complete}}$ in the complete case 
are shown in the parentheses)
%\bes
\beqa
& {\cal C}_{p_0}=& \delta_{p_0},  \n %\\
& {\cal C}_{p_1}=& \delta_{p_1}+\delta_{q_1}\,(+\delta_{\scriptsize \mbox{complete}}),  \n %\\
& {\cal C}_{q_1}=& 2\delta_{q_1}+\delta_{p_2}+\delta_{r_2}\,(+\delta_{\scriptsize \mbox{complete}}), \n %\\
& {\cal C}_{p_2}=& \delta_{p_2}, \n %\\
& {\cal C}_{r_2}=& \delta_{r_2}, \n %\\
& {\cal C}_{q_2}=& \delta_{q_2}. 
\label{CswithdsSO(12)E7p1q1}
\eeqa
%\ees
%
%%%%%%%%%%%%%%%%%%%%%
\if0
%
\beqa
{\cal C}_1&=&\delta_1,\n
{\cal C}_2&=&\delta_2+\delta_3,\n
{\cal C}_3&=&2\delta_3+\delta_{p_2}+\delta_{r_2},\n
{\cal C}_{p_2}&=&\delta_{p_2},\n
{\cal C}_{r_2}&=&\delta_{r_2},\n
{\cal C}_{q_2}&=&\delta_{q_2}.
\label{CswithdsSO(12)E7p1q1}
\eeqa
%
\fi
%%%%%%%%%%%%%%%%%%%%%
The intersection matrix of ${\cal C}$'s is the same one as before.
The intersection diagrams of  $\delta$'s for incomplete\,/\,complete cases are 
shown in Figure \ref{Fig:SO(12)E7p1q1}. 
%This is the third box graph 
%from the bottom in the left column of Figure 33 in \cite{BoxGraphs}.
The diagram for the incomplete case is an $E_6$ non-Dynkin diagram,
which is similar to the $p_0 \rightarrow q_1$ case.
But this time there are two $-\frac{3}{2}$ nodes.
The intersection matrix is given by 
\beqa
-\delta_I\cdot \delta_J &=&
\left(
\begin{array}{cccccc}
 \frac{3}{2} & 0 & -1 & -\frac{1}{2} & -1 & 0 \\
 0 & 2 & -1 & 0 & 0 & 0 \\
 -1 & -1 & 2 & 0 & 0 & 0 \\
 -\frac{1}{2} & 0 & 0 & \frac{3}{2} & 0 & -1 \\
 -1 & 0 & 0 & 0 & 2 & 0 \\
 0 & 0 & 0 & -1 & 0 & 2 
\end{array}
\right),
\label{deltaintersectionsp0p1q1}
\eeqa
with the same order $I,J=q_1,p_0,p_2,p_1,r_2,q_2$ as before.
Note that the intersection of the two $-\frac{3}{2}$ nodes is 
$\frac{1}{2}$.

Again, there is only one ${\bf 32}$ at $s=0$ as two-cycles $J$ \eqref{eq:J}
satisfying $J\cdot J=-\frac{3}{2}$.
For the complete case, $\delta_{\scriptsize \mbox{complete}}$
bridges the two $-\frac{3}{2}$ nodes and forms the ordinary $E_7$ intersection diagram.

\begin{figure}[h]
  \begin{center}
\vspace{0.2cm}
         \includegraphics[clip, width=7.4cm]{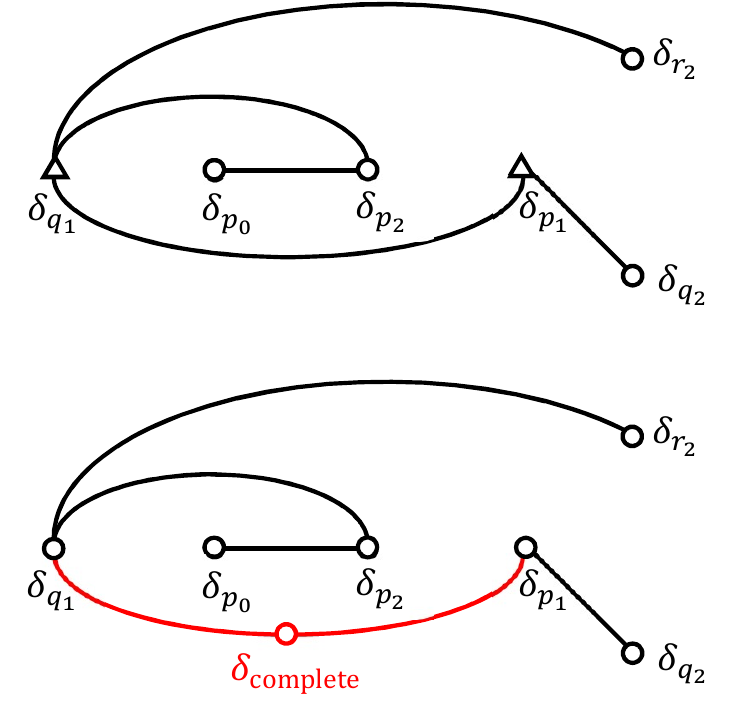}
   
                \caption{Incomplete/complete intersection 
                    diagrams of $SO(12)\rightarrow E_7$~:~$p_1$-first ($p_0\rightarrow p_1 \rightarrow q_1$) case.}
    \label{}
\label{Fig:SO(12)E7p1q1}
  \end{center}
%\label{Fig:SO(12)E7p1q1}
\end{figure}

\medskip
\noindent 
\underline{$p_0 \rightarrow p_1 \rightarrow r_2$ case}
\medskip 

If $r_2$ is blown up after $p_1$, 
the relations among 
${\cal C}$'s and $\delta$'s are found to be
%
%\bes
\beqa
& {\cal C}_{p_0}=& \delta_{p_0}+\delta_{r_2}\,(+\delta_{\scriptsize \mbox{complete}}), \n %\\
& {\cal C}_{p_1}=& \delta_{p_1}, \n %\\
& {\cal C}_{r_2}=& 2\delta_{r_2}+\delta_{p_2}+\delta_{q_1}\,(+\delta_{\scriptsize \mbox{complete}}), \n %\\
& {\cal C}_{q_1}=& \delta_{q_1}, \n %\\
& {\cal C}_{p_2}=& \delta_{p_2}, \n %\\
& {\cal C}_{q_2}=& \delta_{q_2},
\label{CswithdsSO(12)E7p1r2}
\eeqa
%\ees
%%%%%%%%%%%%%%%%%%%%
\if0
\beqa
{\cal C}_{q_1}&=&\delta_{q_1},\n
{\cal C}_1&=&\delta_1+\delta_3,\n
{\cal C}_{p_2}&=&\delta_{p_2},\n
{\cal C}_2&=&\delta_2,\n
{\cal C}_3&=&2\delta_3+\delta_{q_1}+\delta_{p_2},\n
{\cal C}_{q_2}&=&\delta_{q_2},
\label{CswithdsSO(12)E7p1r2}
\eeqa
\fi
%%%%%%%%%%%%%%%%%%%%
and the intersection diagrams of  $\delta$'s are 
as shown in Figure \ref{Fig:SO(12)E7p1r2}. 
%This is the fifth box graph 
%from the bottom in the left column of Figure 33 in \cite{BoxGraphs}.
This time, the diagram for the incomplete case is a $D_6$ non-Dynkin diagram. 
There are two nodes with self intersections $-\frac{3}{2}$ and their mutual 
intersection is $\frac{1}{2}$.
As before, only one ${\bf 32}$ appears at $s=0$ as $J\cdot J=-\frac{3}{2}$ states.
For the complete case, $\delta_{\scriptsize \mbox{complete}}$
bridges the two $-\frac{3}{2}$ nodes and forms the ordinary $E_7$ intersection diagram.

This exhausts all the possibilities of changing the order of 
the singularities we blow up.
\begin{figure}[h]
  \begin{center}
\vspace{0.3cm}
         \includegraphics[clip, width=7.4cm]{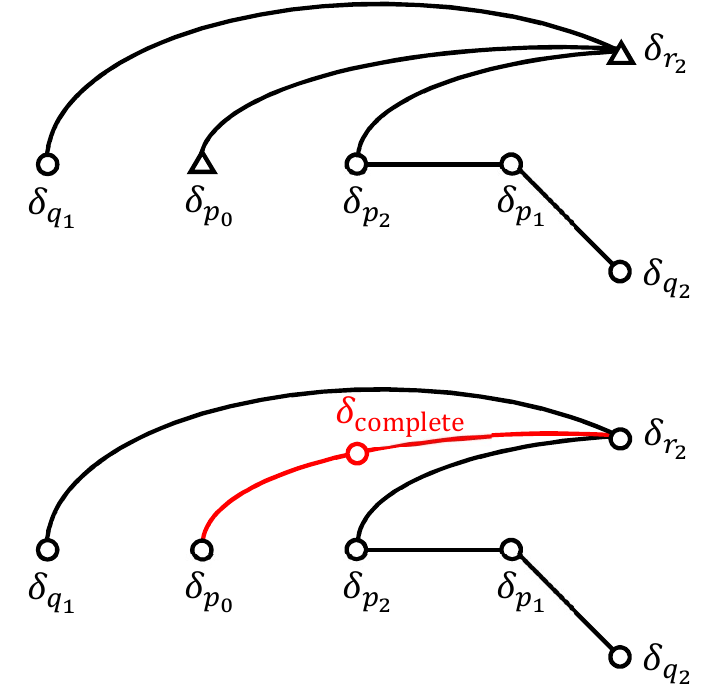}
   
                \caption{Incomplete/complete intersection diagrams of 
                            $SO(12)\rightarrow E_7$~:~$p_1$-first ($p_0\rightarrow p_1 \rightarrow r_2$) case.}
    \label{}
\label{Fig:SO(12)E7p1r2}
  \end{center}
%\label{Fig:SO(12)E7p1r2}
\end{figure}

% 0223 %
\subsection{Comparison with the results of the M-theory Coulomb branch analysis}
In the previous sections 
we have obtained {\em four} distinct incomplete intersection diagrams
of the fibers at the codimension-two singularity. 
Let us compare our results with those obtained by the 
M-theory Coulomb branch analysis \cite{BoxGraphs}.

In general, 
F-theory compactifications on Calabi-Yau four-folds are dual to M-theory compactifications on Calabi-Yau 
four-folds, which present three-dimensional ${\cal N}=2$ supersymmetric gauge theories.
The geometry of the Calabi-Yau four-fold determines the structure of the gauge theory.
In particular, the codimension-one singularity decides the gauge group, and the network of the resolution corresponds to the structure of the classical Coulomb phase since the resolution corresponds to the symmetry breaking.

We consider three-dimensional gauge theory with a gauge group $G$ and with $N_f$ chiral multiplets in a representation ${\bf R}_f$.
We set the  masses of the chiral multiplets to zero.
In addition, we assume that there is no classical Chern-Simons term.
The vector multiplet in the adjoint representation includes a real scalar field $\phi$.
In general, the gauge group $G$ breaks to $U(1)^r$ by the VEVs of the scalar, where $r={\rm rank}(G)$.
We choose the fundamental Weyl chamber as
\begin{align}
\label{eq:Weyl_chamber}
\vec{\alpha}_i\cdot\vec{\phi}>0,
\end{align}
where $\vec{\alpha}_i\ (i=1,2,\ldots,r)$ are the simple roots of $G$.
$\vec{\phi}=(\phi^1,\phi^2,\ldots,\phi^r)$ are the VEVs in the Cartan subalgebra of $G$.

Now we have the chiral multiplets, which make a substructure in the Coulomb branch.
The Lagrangian includes the mass terms of the chiral multiplets $Q^{(f)}$:
\begin{align}
{\cal L}_{\rm mass}=\sum_f\left|\phi Q^{(f)}\right|^2=\sum_f\left|\vec{\phi}\cdot\vec{\omega}_{f}\right|^2\left|Q^{(f)}\right|^2,
\end{align}
where $\vec{\omega}_{f}$ are weights of ${\bf R}_f$ representation.
Note that when $\vec{\phi}\cdot\vec{\omega}_{f}=0$, the corresponding matter becomes massless.

We can classify the regions of the Coulomb branch.
A region is bounded by the zero loci of $\vec{\phi}\cdot\vec{\omega}_{f}$, namely, it is characterized by $\vec{\phi}\cdot\vec{\omega}_{f}>0$ or $\vec{\phi}\cdot\vec{\omega}_{f}<0$.
However, not all regions are allowed 
since we are working on the fundamental Weyl chamber \eqref{eq:Weyl_chamber}.
The allowed regions of the Coulomb branch are completely classified by the decorated box graphs defined by a collection of  
boxes with signs (or colors) \cite{BoxGraphs}.

Although the analysis of \cite{BoxGraphs} is based on the 
Coulomb branch of the three-dimensional M-theory, 
which basically applies to 
a resolution of a Calabi-Yau four-fold such as \cite{EsoleYau,MarsanoSSNameki,EsoleShaoYau,BraunNameki},
it is interesting to compare our results with the corresponding
box graphs obtained in \cite{BoxGraphs} since the scalars of 
the five-dimensional M-theory, which are supposed to 
describe the resolution of a Calabi-Yau three-fold, partly 
comprise the three-dimensional scalars.

In \cite{BoxGraphs}, 
the intersection diagrams for the  
singularity enhancement $SO(12)\rightarrow E_7$ 
are given in Figure 33 of that paper.
We can see that the intersection diagrams for 
incomplete resolutions shown in Figures 5, 7 and 8 
of the present paper 
are the ones at the bottom, the third from the bottom and the 
second from the top of the left column of Figure 33 of \cite{BoxGraphs}. 
Also, the (lower) diagram in Figure 2 of this paper 
is the diagram in the right column of Figure 33 of \cite{BoxGraphs}.
Thus, all the four intersection diagrams we found here 
have corresponding box graphs obtained in \cite{BoxGraphs} 
describing different phases 
of the three-dimensional gauge theory.  Note that
any two of them are not adjacent to each other in 
the seven graphs 
of Figure 33 of \cite{BoxGraphs}. 
This is consistent, 
as the two adjacent graphs of \cite{BoxGraphs} are related by a flop, 
and the change of the order of the singularities 
is not a flop. Indeed, we do not have any conifold singularities 
until we consider a complete resolution. This is one of the 
differences between the resolutions in six and four dimensions.
% 0223 %

\section{$E_7\rightarrow E_8$}

\subsection{Incomplete resolution: Blowing up $p_2$ first}
\label{sec:p2firstincomp}
\subsubsection{Blowing up process}
\label{p2firstprocess}

In this case we take $f_{n+8}=s$ with setting $g_{n+12}=1$, $f_{8}=g_{12}=0$ 
in (\ref{fE7}) and (\ref{gE7}) :
\beqa
f(z,s)&=&s  z^3,\n
g(z,s)&=&z^5. \label{fandgE7E8}
\eeqa 
At $s\neq 0$, the orders of $f$, $g$ and $\Delta$ in $z$ are $(3,5,9)$, while at $s=0$,
they satisfy $(\ge \hspace{-0.05cm} 4,5,10)$. Hence it describes the enhancement $III^* \rightarrow II^*$ 
($E_7\rightarrow E_8$) of the Kodaira type. 

$\Phi(x,y,z,s)=0$ has a codimension-one singularity at $p_0=(0,0,0,s)$.  
The concrete process of the incomplete resolution
%of the codimension-two (<---one) singularity enhancement from 
%$E_7$ to $E_8$  
%where $p_3$ is blown up first, 
goes as follows (an exceptional curve arising 
from blowing up a singularity $p$ will be denoted by ${\cal C}_p$):
\paragraph{1st blow up\\}
%\vskip -10mm

\noindent
\hskip -1em\underline{Chart $1_x$}
\beqa
\Phi(x, x y_1,x z_1,s)&=&x^2\Phi_x(x,y_1,z_1,s),\n
\Phi_x(x,y_1,z_1,s)&=&
s x^2 z_1^3+x^3 z_1^5+x-y_1^2.\n
   \mbox{$\cal C$$_{p_0}$ in $1_x$}&:&x=0,~~y_1=0.\n
   \mbox{Singularities}&:&\mbox{None}
\eeqa

\noindent
\underline{Chart $1_y$} \quad
$\cal C$$_{p_0}$ is not visible in this chart.\bigskip

%\beqa
%\Phi(x_1 y, y, y z_1,s)&=&y^2\Phi_y(x_1, y, z_1,s),\n
%\Phi_y(x_1,y,z_1,s)&=&
%x_1 y^2 z_1^3 (s+y z_1)+x_1^3 y+y^3 z_1^5-1,
%\n
%   \mbox{$\cal C$$_1$ in $1_y$}&:&\mbox{Invisible in this patch}\n
%   \mbox{Singularities}&:&\mbox{None}
%\eeqa

\noindent
\underline{Chart $1_z$}
\beqa
\Phi(x_1 z, y_1 z, z,s)&=&z^2\Phi_z(x_1,y_1,z,s),\n
\Phi_z(x_1,y_1,z,s)&=&z \left(x_1 z s+x_1^3+z^2\right)-y_1^2.
\n
   \mbox{$\cal C$$_{p_0}$ in $1_z$}&:&z=0,~~y_1=0.\n
   \mbox{Singularities}&:&(x_1,y_1,z,s)=(0,0,0,s).
\eeqa
We refer to this singularity as $p_1$.

\paragraph{2nd blow up\\}
\noindent
\hskip -1em\underline{Chart $2_{zx}$}
\beqa
\Phi_z(x_1, x_1 y_2, x_1 z_2,s)&=&x_1^2\Phi_{zx}(x_1, y_2, z_2,s),\n
\Phi_{zx}(x_1, y_2, z_2,s)&=&z_2 x_1 \left(z_2
   (s+z_2)+ x_1\right)-y_2^2
.\n
   \mbox{$\cal C$$_{p_1}$ in $2_{zx}$}&:&x_1=0,~~y_2=0.\n
   \mbox{Singularities}&:&(x_1,y_2,z_2,s)=
   (0,0,-s,s) (=q_2), (0,0,0,s) (=p_2). 
\eeqa
Here we see two singularities on $\cal C$$_{p_1}$ which coincide with 
each other at $s=0$.
\bigskip

\noindent
\underline{Chart $2_{zy}$} \quad
$\cal C$$_{p_1}$ is not visible in this chart.\bigskip

%\beqa%x_1,y_1,z,s
%\Phi_z(x_2 y_1, y_1, y_1 z_2,s)&=&y_1^2\Phi_{zy}(x_2, y_1, z_2,s),\n
%\Phi_{zy}(x_2, y_1, z_2,s)&=&
%x_2 y_1 z_2^2 (s + y_1 z_2) + x_2^3 y_1^2 z_2 + y_1 z_2^3 - 1.\n
%   \mbox{$\cal C$$_2$ in $2_{zy}$}&:&\mbox{Invisible in this patch}\n
%   \mbox{Singularities}&:&\mbox{None}. 
%\eeqa

\noindent
\underline{Chart $2_{zz}$}
\beqa%x_1,y_1,z,s
\Phi_z(x_2 z, y_2 z,  z,s)&=&z^2\Phi_{zz}(x_2, y_2, z,s),\n
\Phi_{zz}(x_2, y_2, z,s)&=&z \left(s x_2+z x_2^3+1\right)-y_2^2
.\n
   \mbox{$\cal C$$_{p_1}$ in $2_{zz}$}&:&z=0,~~y_2=0.\n
   \mbox{Singularities}&:&(x_2,y_2,z,s)=
   (-\frac1s,0,0,s). 
\eeqa
This singularity is $q_2$, which is also seen in chart $2_{zx}$.
At this stage, we have two singularities $p_2$ and $q_2$. 
In this section we blow up $p_2$ first. We can see this singularity 
in chart $2_{zx}$ only, so we consider $\Phi_{zx}(x_1, y_2, z_2,s)$ 
in the next blow up.

\paragraph{3rd blow up\\}
\noindent%x_1, y_2, z_2,s
\hskip -1em
\underline{Chart $3_{zxx}$}
\beqa
\Phi_{zx}(x_1, x_1y_3, x_1z_3,s)&=&x_1^2\Phi_{zxx}(x_1, y_3, z_3,s),\n
\Phi_{zxx}(x_1, y_3, z_3,s)&=&
z_3 x_1 \left(s z_3+z_3^2 x_1+1\right)-y_3^2.\n
   \mbox{$\cal C$$_{p_2}$ in $3_{zxx}$}&:&x_1=0,~~y_3=0.\n
   \mbox{Singularities}&:&(x_1,y_3,z_3,s)=
   (0,0,-\frac1s,s) (=r_3), (0,0,0,s) (=p_3). 
\label{eq:E7E83zxxsing}
\eeqa
We name the first singularity $r_3$, and the second singularity $p_3$.
\bigskip

\noindent
\underline{Chart $3_{zxy}$} %x_1, y_2, z_2,s
\quad Regular.  \bigskip

\noindent
\underline{Chart $3_{zxz}$}%x_1, y_2, z_2,s
\beqa
\Phi_{zx}(x_3 z_2, y_3 z_2, z_2,s)&=&z_2^2\Phi_{zxz}(x_3, y_3, z_2,s),\n
\Phi_{zxz}(x_3, y_3, z_2,s)&=&
z_2 x_3 \left(s+z_2+x_3\right)-y_3^2
.\n
   \mbox{$\cal C$$_{p_2}$ in $3_{zxz}$}&:&z_2=0,~~y_3=0.\n
   \mbox{Singularities}&:&(x_3,y_3,z_2,s)=
   (0,0,-s,s) (=q_2), (0,0,0,s) (=q_3), \n
   &\,& \hspace{2.6cm}(-s,0,0,s) (=r_3). 
\label{eq:E7E83zxzsing}
\eeqa
The first singularity is not on $\cal C$$_{p_2}$ unless $s=0$; 
this is $q_2$. We name the second singularity $q_3$. The third one 
is $r_3$ already seen in chart $3_{zxx}$.

When $s \rightarrow 0$, the three singularities $q_2$, $q_3$ and $r_3$ in chart $3_{zxz}$ coincide 
with each other, and which one we blow up next affects the proceeding process.
In this section, we consider the case $q_3$ is blown up next. 
On the other hand, $p_3$ in chart $3_{zxx}$ is separated from these points even when $s\rightarrow 0$ and 
can be blown up independently.
We leave the blow-up of $p_3$ until later and work on the blow-up of $q_3$.

%%%%%%%%%
\if0

%On the other hand, $p_3$ is separated from these points even when $s\rightarrow 0$.
%Thus we can blow up $p_3$ and the other three singularities independently on charts $3_{zxx}$ and $3_{zxz}$,
%respectively. Let us first blow up $p_3$ using $\Phi_{zxx}(x_1, y_3, z_3,s)$:

\paragraph{4th blow up at $p_3$\\}
\noindent
\hskip -1em
\underline{Chart $4_{zxxx}$}%x_1, y_3, z_3,s
\beqa
\Phi_{zxx}(x_1, x_1y_4, x_1z_4,s)&=&x_1^2\Phi_{zxxx}(x_1, y_4, z_4,s),\n
\Phi_{zxxx}(x_1, y_4, z_4,s)&=&
s z_4^2 x_1+z_4^3 x_1^3+z_4-y_4^2.\n
   \mbox{$\cal C$$_{p_3}$ in $4_{zxxx}$}&:&x_1=0,~~y_4^2=z_4.\n
   \mbox{Singularities}&:&\mbox{None}. 
\eeqa

\noindent
\underline{Chart $4_{zxxy}$}\\%x_1, y_3, z_3,s
Regular.

\noindent
\underline{Chart $4_{zxxz}$}%x_1, y_3, z_3,s
\beqa
\Phi_{zxx}(x_4 z_3, y_4 z_3, z_3,s)&=&z_3^2\Phi_{zxxz}(x_4, y_4, z_3,s),\n
\Phi_{zxxz}(x_4, y_4, z_3,s)&=&
s z_3 x_4+z_3^3 x_4^2+x_4-y_4^2.\n
   \mbox{$\cal C$$_{p_3}$ in $4_{zxxz}$}&:&z_3=0,~~y_4^2=x_4.\n
   \mbox{Singularities}&:&(x_4,y_4,z_3,s)=
   (0,0,-\frac1s,s). 
\eeqa
This singularity is not on $\cal C$$_{p_3}$ even when $s=0$;
this is $r_3$. 

%There is no singularity any more on $\cal C$$_{p_3}$,
%so let us turn to the singularities observed in chart $3_{zxz}$:
%$(x_1,y_3,z_3,s)=
%   (0,0,-s,s) (=q_2), (0,0,0,s) (=q_3), (-s,0,0,s) (=r_3)$.
%There are three choices of blowing up depending on what singularity
%is blown up next. 
%In this section, we consider the case $q_3$ is blown up next.

\fi
%%%%%%%%%%%

\paragraph{4th blow up at $q_3$}
We next blow up $q_3$.
Using $\Phi_{zxz}(x_3, y_3, z_2,s)$, we find

\noindent
\underline{Chart $4_{zxzx}$}%x_3, y_3, z_2,s
\beqa
\Phi_{zxz}(x_3, x_3y_4, x_3z_4,s)&=&x_3^2\Phi_{zxzx}(x_3, y_4, z_4,s),\n
\Phi_{zxzx}(x_3, y_4, z_4,s)&=&
z_4 \left(s+z_4 x_3+x_3\right)-y_4^2.\n
   \mbox{$\cal C$$_{q_3}$ in $4_{zxzx}$}&:&x_3=0,~~y_4^2=s z_4.\n
   \mbox{Singularities}&:&(x_3,y_4,z_4,s)=
   (-s,0,0,s). 
\label{eq:E7E84zxzxsing}
\eeqa
This is $r_3$, which is not on ${\cal C}$$_{q_3}$ unless $s=0$.
\bigskip

\noindent
\underline{Chart $4_{zxzy}$} %x_3, y_3, z_2,s
\quad  Regular. \bigskip

\noindent
\underline{Chart $4_{zxzz}$}%x_3, y_3, z_2,s
\beqa
\Phi_{zxz}(x_4 z_2, y_4 z_2, z_2,s)&=&z_2^2\Phi_{zxzz}(x_4, y_4, z_2,s),\n
\Phi_{zxzz}(x_4, y_4, z_2,s)&=&
s x_4+z_2 x_4 \left(x_4+1\right)-y_4^2
.\n
   \mbox{$\cal C$$_{q_3}$ in $4_{zxzz}$}&:&z_2=0,~~y_4^2=s x_4.\n
   \mbox{Singularities}&:&(x_4,y_4,z_2,s)=
   (0,0,-s,s). 
\label{eq:E7E84zxzzsing}
\eeqa
This is $q_2$, which is not on ${\cal C}$$_{q_3}$ unless $s=0$, either.
$r_3$ and $q_2$ coincide with each other at $s=0$ before  blowing up $q_3$; 
but after the blow up,  they are never the same point even when $s=0$. 
Thus we can blow them up independently.

%So far, all the singularities except for $q_2$ and $r_3$ are 
%resolved. Since $q_2$ is located in the $(0:0:1)$ direction 
%on the $\PP^2$ blown up at $q_3$, whereas $r_3$ is 
%in the $(1:0:0)$ direction on the same $\PP^2$, 
%they are never the same point even when $s=0$. 
%Thus we can blow them up independently. 

\paragraph{5th blow up at $r_3$ %in chart $4_{zxzx}$
}
To blow up $r_3$ in chart $4_{zxzx}$ \eqref{eq:E7E84zxzxsing}, we shift the $x_3$ coordinate so that 
this singularity is represented 
as $(0,0,0,s)$ in the new coordinate $\tilde x_3$:
\beqa
\Psi_{zxzx}(\tilde x_3, y_4, z_4,s)&\equiv&\Phi_{zxzx}(\tilde x_3-s, y_4, z_4,s).
\eeqa
Then it can be verified that no singularity arises 
in charts $5_{zxzx*}$ below. %$\Psi_{zxzxx}$, $\Psi_{zxzxy}$ or $\Psi_{zxzxz}$ defined below.
The exceptional curves are:

\noindent
\underline{Chart $5_{zxzxx}$}%\tilde x_3, y_4, z_4,s
\beqa
\Psi_{zxzx}(\tilde x_3, \tilde x_3y_5, \tilde x_3z_5,s)&=&
\tilde x_3^2\Psi_{zxzxx}(\tilde x_3, y_5, z_5,s),\n
\Psi_{zxzxx}(\tilde x_3, y_5, z_5,s)&=&
z_5^2 (\tilde x_3-s)+z_5-y_5^2.\n
   \mbox{$\cal C$$_{r_3}$ in $5_{zxzxx}$}&:&\tilde x_3=0,~~y_5^2=-s z_5^2+z_5.
\label{eq:E7E85zxzxxsing}
\eeqa

\noindent
\underline{Chart $5_{zxzxy}$}%\tilde x_3, y_4, z_4,s
\quad Invisible.   \bigskip

\noindent
\underline{Chart $5_{zxzxz}$}%\tilde x_3, y_4, z_4,s
\beqa
\Psi_{zxzx}(x_5 z_4, y_5 z_4, z_4,s)&=&z_4^2\Psi_{zxzxz}(x_5, y_5, z_4,s),\n
\Psi_{zxzxz}(x_5, y_5, z_4,s)&=&
-s+z_4 x_5+x_5-y_5^2 .\n
   \mbox{$\cal C$$_{r_3}$ in $5_{zxzxz}$}&:&z_4=0,~~y_5^2=x_5-s.
\label{eq:E7E85zxzxzsing}
\eeqa

\paragraph{5th blow up at $q_2$ %in chart $4_{zxzz}$
}%x_4, y_4, z_2,s
Having resolved the singularity $r_3$, we turn to the resolution of 
$q_2$ in chart $4_{zxzz}$ \eqref{eq:E7E84zxzzsing}. For this we need a different coordinate shift:
\beqa
\Psi_{zxzz}(x_4, y_4, \tilde z_2,s)&\equiv&\Phi_{zxzz}(x_4, y_4, \tilde z_2-s,s).
\eeqa
Then $\Psi_{zxzz}$ has a singularity at $(x_4, y_4, \tilde z_2,s)=(0,0,0,s)$. 
Again, charts $5_{zxzz*}$ below %$\Psi_{zxzzx}$, $\Psi_{zxzzy}$ and $\Psi_{zxzzz}$ defined below 
have no singularity. The exceptional curves are:

\noindent
\underline{Chart $5_{zxzzx}$}%x_4, y_4, \tilde z_2,s
\beqa
\Psi_{zxzz}(x_4,  x_4y_5,  x_4z_5,s)&=&
x_4^2\Psi_{zxzzx}(x_4, y_5, z_5,s),\n
\Psi_{zxzzx}(x_4, y_5, z_5,s)&=&
-s+z_5 x_4+z_5-y_5^2 .\n
   \mbox{$\cal C$$_{q_2}$ in $5_{zxzzx}$}&:&x_4=0,~~y_5^2=z_5-s.
\label{eq:E7E85zxzzxsing}
\eeqa

\noindent
\underline{Chart $5_{zxzzy}$}  %x_4, y_4, \tilde z_2,s
\quad We omit the details.  \bigskip

\noindent
\underline{Chart $5_{zxzzz}$}%x_4, y_4, \tilde z_2,s
\beqa
\Psi_{zxzz}(x_5 \tilde z_2, y_5 \tilde z_2, \tilde z_2,s)&=&
\tilde z_2^2\Psi_{zxzzz}(x_5, y_5, \tilde z_2,s),\n
\Psi_{zxzzz}(x_5, y_5, \tilde z_2,s)&=&
x_5^2 \left(-s+\tilde z_2\right)+x_5-y_5^2
.\n
   \mbox{$\cal C$$_{q_2}$ in $5_{zxzzz}$}&:&\tilde z_2=0,~~y_5^2=-s x_5^2+x_5.
\label{eq:E7E85zxzzzsing}
\eeqa

\paragraph{4th blow up at $p_3$ %in chart $3_{zxx}$\\
}
Let us return to chart $3_{zxx}$ and blow up the remaining $p_3$ \eqref{eq:E7E83zxxsing}.

\noindent
%\hskip -1em
\underline{Chart $4_{zxxx}$}%x_1, y_3, z_3,s
\beqa
\Phi_{zxx}(x_1, x_1y_4, x_1z_4,s)&=&x_1^2\Phi_{zxxx}(x_1, y_4, z_4,s),\n
\Phi_{zxxx}(x_1, y_4, z_4,s)&=&
s z_4^2 x_1+z_4^3 x_1^3+z_4-y_4^2.\n
   \mbox{$\cal C$$_{p_3}$ in $4_{zxxx}$}&:&x_1=0,~~y_4^2=z_4.\n
   \mbox{Singularities}&:&\mbox{None}. 
\eeqa

\noindent
\underline{Chart $4_{zxxy}$} %x_1, y_3, z_3,s
\quad Regular . \bigskip

\noindent
\underline{Chart $4_{zxxz}$}%x_1, y_3, z_3,s
\beqa
\Phi_{zxx}(x_4 z_3, y_4 z_3, z_3,s)&=&z_3^2\Phi_{zxxz}(x_4, y_4, z_3,s),\n
\Phi_{zxxz}(x_4, y_4, z_3,s)&=&
s z_3 x_4+z_3^3 x_4^2+x_4-y_4^2.\n
   \mbox{$\cal C$$_{p_3}$ in $4_{zxxz}$}&:&z_3=0,~~y_4^2=x_4.\n
   \mbox{Singularities}&:&(x_4,y_4,z_3,s)=
   (0,0,-\frac1s,s). 
\eeqa
This singularity is not on $\cal C$$_{p_3}$ even when $s=0$;
this is $r_3$ (see  \eqref{eq:E7E83zxxsing}) and is blown up as in \eqref{eq:E7E85zxzxxsing}
and \eqref{eq:E7E85zxzxzsing}. 

The whole process of the incomplete resolution
of the codimension-two singularity enhancement from 
$E_7$ to $E_8$ is summarized in Table \ref{E7E8p1first}.

\begin{table}[htp]
\caption{$E_7\rightarrow E_8$: Incomplete case when $p_2$ is blown up first
($p_0 \to p_1 \to p_2 \to q_3$).}
\begin{center}
\begin{tabular}{|l|l|l|l|l|l|}
\hline
&1st blow up&2nd blow up&3rd  blow up&4th blow up&5th blow up\\
\hline
%??P??s????
\ctext{$p_0$}$\rightarrow$
&\ctext{$p_1(0:0:1)$}$\rightarrow$
&\ctext{$p_2(1:0:0)$}$\rightarrow$
&\ctext{$p_3(1:0:0)$}$\rightarrow$
&regular
&\\
%??Q??s????
&
&
&\ctext{$q_3(0:0:1)$}$\rightarrow$
&regular
&\\
%??R??s????
&
&$q_2(1:0:-s)$
&$q_2(0:0:1)(z_2=-s)$
&\ctext{$q_2(0:0:1)(z_2=-s)$}$\rightarrow$
&regular
\\
%??S??s????
&
&
&$r_3(-s:0:1)$
&\ctext{$r_3(1:0:0)(x_3=-s)$}$\rightarrow$
&regular
\\
\hline
\end{tabular}
\end{center}
\label{E7E8p1first}
\end{table}%

\subsubsection{Intersection diagram at $s=0$}

Exceptional curves $\delta_I$ at $s=0$ are defined from ${\cal C}_I$
($I \in \{ p_0,p_1,p_2,q_3,r_3,q_2,p_3\}$).
One can see from the explicit blowing up process 
that the intersection patterns of ${\cal C}$'s and $\delta$'s 
are $E_7$ and $A_7$, respectively (see Figure \ref{Fig:E7E8} below).
As explained in section \ref{sec:matrixp1},
the $s \to 0$ limit of ${\cal C}_I$ is derived 
through the careful lift-ups of ${\cal C}_I$ and $\delta_I$ from 
the chart where they are originally defined to the charts arise via
the subsequent blow-ups.
In this case, we have
%\bes
\beqa
& {\cal C}_{p_0}=& \delta_{p_0}, \n %\\
& {\cal C}_{p_1}=& \delta_{p_1}, \n %\\
& {\cal C}_{p_2}=& \delta_{p_2}, \n %\\
& {\cal C}_{q_3}=& 2\delta_{q_3}+\delta_{r_3}+\delta_{q_2}, \n %\\
& {\cal C}_{r_3}=& \delta_{r_3}, \n %\\
& {\cal C}_{q_2}=& \delta_{q_2}, \n %\\
& {\cal C}_{p_3}=& \delta_{p_3}.
\label{eq:CswithdsE7E8}
\eeqa
%\ees

The intersection matrix of ${\cal C}$'s is %assumed to be 
the minus of the ordinary $E_7$ Cartan matrix
\beqa
-{\cal C}_I\cdot {\cal C}_J &=&
\left(
\begin{array}{ccccccc}
 2 & -1 & 0 & 0 & 0 & 0 & 0\\
 -1 & 2 & -1 & 0 & 0 & 0 & 0\\
 0 & -1 & 2 & -1& 0 & 0 & -1\\
 0 & 0 & -1 & 2 & -1 & 0 & 0\\
 0 & 0 & 0 & -1 & 2 & -1 & 0\\
 0 & 0 & 0 &  0 & -1 & 2 & 0 \\
 0 & 0 & -1 &  0 & 0 & 0 & 2
\end{array}
\right),
 \label{eq:E7Cartan}
\eeqa
where $I,J=p_0,p_3,p_2,q_3,p_1,q_2,r_3$ in this order.
Then the relations \eqref{eq:CswithdsE7E8} imply that the 
intersection matrix of $\delta$'s is 
\beqa
-\delta_I\cdot \delta_J &=&
\left(
\begin{array}{ccccccc}
 2 & -1 & 0 & 0 & 0 & 0 & 0\\
 -1 & 2 & -1 & 0 & 0 & 0 & 0\\
 0 & -1 & 2 & 0& 0 & 0 & -1\\
 0 & 0 & 0 & \frac{3}{2} & 0 & -1 & -1\\
 0 & 0 & 0 & 0 & 2 & -1 & 0\\
 0 & 0 & 0 &  -1 & -1 & 2 & 0 \\
 0 & 0 & -1 &  -1 & 0 & 0 & 2
\end{array}\right).
\label{deltaintersectionsE7E8}
\eeqa
%
%%%%%%%%%%%%%%%%%%%%
\if0
\beqa
-{\cal C}_I\cdot {\cal C}_J &=&
\left(
\begin{array}{ccccccc}
 2 & 0 & 0 & -1 & 0 & 0 & 0\\
 0 & 2 & 0 & 0 & -1 & 0 & -1\\
 0 & 0 & 2 & -1& -1 & -1 & 0\\
 -1 & 0 & -1 & 2 & 0 & 0 & 0\\
 0 & -1 & -1 & 0 & 2 & 0 & 0\\
 0 & 0 & -1 &  0 & 0 & 2 & 0 \\
 0 & -1 & 0 &  0 & 0 & 0 & 2
\end{array}
\right),
 \label{eq:E7Cartan}
\eeqa
where $I,J=1,2,3,4,4',r_3,q_2$.
Then the relations \eqref{eq:CswithdsE7E8} imply that the 
intersection matrix among $\delta_I$'s is 
\beqa
-\delta_I\cdot \delta_I &=&
\left(
\begin{array}{ccccccc}
 2 & 0 & 0 & -1 & 0 & 0 & 0\\
 0 & 2 & 0 & 0 & 0 & 0 & -1\\
 0 & 0 & 2 & -1& 0 & -1 & 0\\
 -1 & 0 & -1 & 2 & 0 & 0 & 0\\
 0 & 0 & 0 & 0 & \frac32 & -1 & -1\\
 0 & 0 & -1 &  0 & -1 & 2 & 0 \\
 0 & -1 & 0 &  0 & -1 & 0 & 2
\end{array}\right).
\label{deltaintersectionsE7E8}
\eeqa
\fi
%%%%%%%%%%%%%%%%%%%%%%
%
This is the minus of the $A_7$ Cartan matrix except that 
one of the $\delta$'s ($=\delta_{q_3}$) has self-intersection $-\frac32$, 
which equals to the minus of the length squared of a weight 
in the {\bf 56} representation of $E_7$. 
The result is shown in Figure \ref{Fig:E7E8}.
For the two-cycles at $s=0$
\beq
 J \equiv \sum_{I = p_0,p_3,p_2,q_3,p_1,q_2,r_3} n_I \delta_I    \quad \quad (n_I \in \ZZ),
\eeq
one can show by using (\ref{deltaintersectionsE7E8}) that
%\bes
\beqa
 & & \sharp \, (J \cdot J = -2\, ) \,\,\, =126,  \n %\\
 & & \sharp \, \Big(J \cdot J = -\frac{3}{2}\,  \Big)  = 56. 
\eeqa
%\ees
They respectively are the adjoint (without Cartan part) and ${\bf 56}$ representations of $E_7$.
The latter consists of $28$  elements with $n_I\geq 0$ for all $I$ and 
$28$ elements with $n_I\leq 0$ for all $I$.
%It can be verified that there are precisely $28$ elements 
%of the form $\sum_{I=p_0,p_1,p_2,p_3,q_3,r_3,q_2} n_I \delta_I$ with non-negative 
%integer coefficients, $n_I\geq 0$ for all $I$, 
%such that the length squared is $-\frac32$, 
%and also there are the same number of elements 
%with non-positive  
%integer coefficients, $n_I\leq 0$ for all $I$.
%They all together form the whole weights 
%of the {\bf 56} representation. 
Again, there is only a single ${\bf 56}$ representation, 
indicating that it is a half-hypermultiplet.

\begin{figure}[h]
  \begin{center}
         \includegraphics[clip, width=8.6cm]{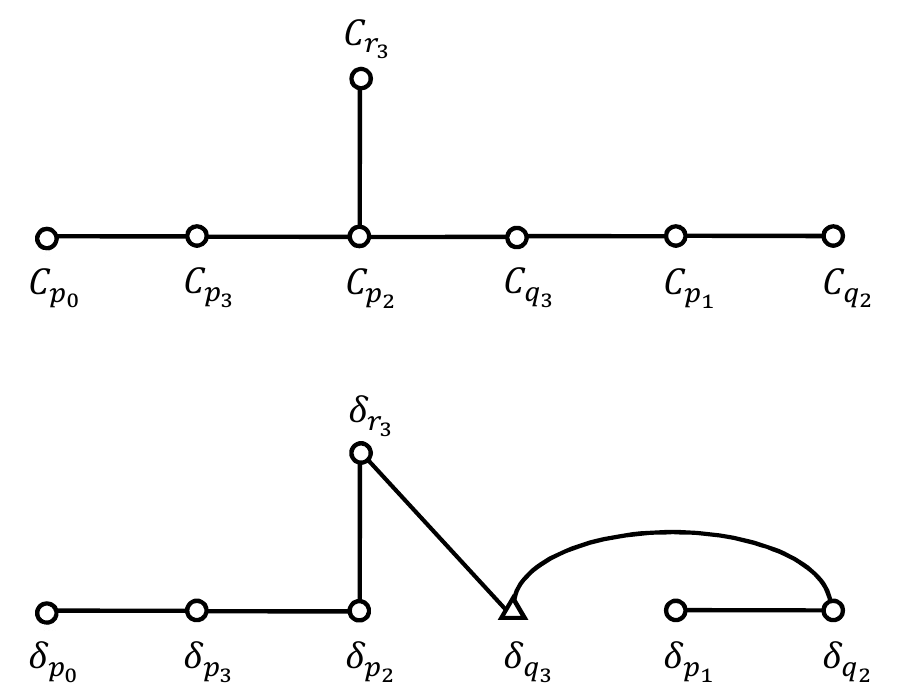}   
                \caption{Generic $E_7$ intersection diagram at $s\neq 0$ (upper) and
                             incomplete intersection diagram at $s=0$ (lower)
             of $E_7 \rightarrow E_8$~:~$p_2$-first ($p_0\rightarrow p_1 \rightarrow p_2 \rightarrow q_3$) case.}
    \label{Fig:E7E8}
  \end{center}
\end{figure}

%\newpage
\subsection{Complete resolution}
\label{sec:p2firstcomp}
We will now consider the complete resolution. 
This can be achieved by taking $f_{n+8}=s^2$ 
instead of $s$.
%similarly to what was done in 
%section \ref{completeresolutionSO(12)E7}.
This amounts to replacing $s$ in (\ref{fandgE7E8})  
with $s^2$. Similarly to the previous sections, 
we find an additional isolated codimension-two 
conifold singularity after we blow up $q_3$. 
As shown in red in Table \ref{E7E8completep1first}, this new singularity, 
which we denote by $r_4$, arises at $(1:0:-1)$ on 
the $\PPsmall^2$ particularly at $s=0$. This adds 
an extra node to the incomplete intersection diagram 
to form the correct $E_8$ Dynkin diagram as we 
show in Figure \ref{Fig:E7E8complete}. 
To see how the intersection matrix is modified,
we repeat the argument given in section \ref{sec:conifoldlimit}.
By carefully lifting up ${\cal C}$'s and ${\delta}$'s into the 
local coordinate system of the small resolution, we find
the modified relations
%\bes
\beqa
& {\cal C}_{p_0}=& \delta_{p_0}, \n %\\
& {\cal C}_{p_1}=& \delta_{p_1}, \n %\\
& {\cal C}_{p_2}=& \delta_{p_2}, \n %\\
& {\cal C}_{q_3}=& 2\delta_{q_3}+\delta_{r_3}+\delta_{q_2}+\delta_{\rm complete}, \n %\\
& {\cal C}_{r_3}=& \delta_{r_3}, \n %\\
& {\cal C}_{q_2}=& \delta_{q_2}, \n %\\
& {\cal C}_{p_3}=& \delta_{p_3}.
\label{CswithdsE7E8modified}
\eeqa
%\ees
One can then verify that: if the intersection matrix of these eight ${\delta}$'s   
is the minus of the ordinary $E_8$ Cartan matrix, the intersection matrix of
${\cal C}$'s \eqref{eq:E7Cartan} is reproduced.
Therefore, the node $\delta_{q_3}$, 
which was formerly represented by a triangle in 
Figure \ref{Fig:E7E8}, is now an ordinary node consisting of 
the root system of $E_8$ as in Figure \ref{Fig:E7E8complete}. 
\begin{table}[htp]
\caption{$E_7\rightarrow E_8$:  Complete case when $p_2$ is blown up first
($p_0 \to p_1 \to p_2 \to q_3$). }
\vspace{0.6cm}
\hspace{0.2cm}
%\begin{center}
\begin{tabular}{|l|l|l|l|}
\hline
&1st blow up&2nd blow up&3rd  blow up \\
\hline
%??P??s????
\ctext{$p_0$}$\rightarrow$
&\ctext{$p_1(0:0:1)$}$\rightarrow$
&\ctext{$p_2(1:0:0)$}$\rightarrow$
&\ctext{$p_3(1:0:0)$}$\rightarrow$
\\
%??Q??s????
&
&
&\ctext{$q_3(0:0:1)$}$\rightarrow$
\\
%??R??s????
&
&$q_2(1:0:-s^2)$
&$q_2(0:0:1)(z_2=-s^2)$
\\
%??S??s????
&
&
&$r_3(-s^2:0:1)$
\\
\hline
\end{tabular}

\hspace{6.8cm}
\begin{tabular}{|l|l|}
\hline  
4th blow up & 5th blow up  \\ \hline
regular
& \\
{\color{red}\ctext{$r_4(1:0:-1;s=0)$}$\mbox{(codim.2)}\rightarrow$}
&{\color{red}regular}\\
\ctext{$q_2(0:0:1)(z_2=-s^2)$}$\rightarrow$
&regular\\
\ctext{$r_3(1:0:0)(x_3=-s^2)$}$\rightarrow$
&regular
\\
\hline
\end{tabular}
%\end{center}
\label{E7E8completep1first}
\end{table}%

\begin{figure}[h]
\vspace{0.5cm}
  \begin{center}
         \includegraphics[clip, width=8.6cm]{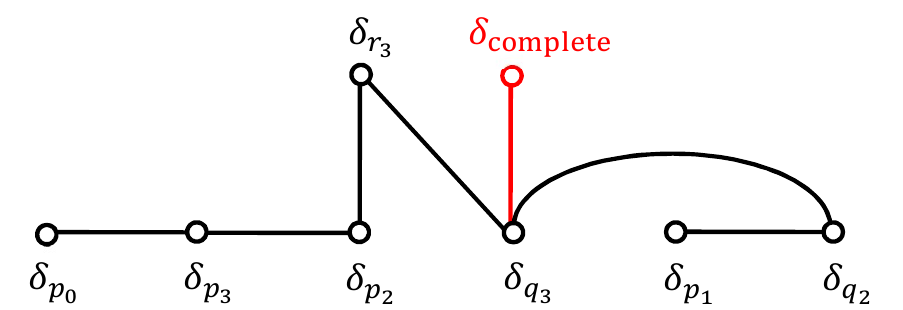}
   
                \caption{Complete $E_8$ intersection diagram of 
            $E_7 \rightarrow E_8$~:~$p_2$-first ($p_0\rightarrow p_1
                            \rightarrow p_2 \rightarrow q_3$) case.}
    \label{Fig:E7E8complete}
  \end{center}
%  \label{Fig:E7E8complete}
\end{figure}
%

%%%%%%%%%%%%%%%%%%%%%%%
\if0
%
\begin{table}[htp]
\caption{$E_7\rightarrow E_8$: Complete case.}
\begin{center}
\begin{tabular}{|l|l|l|l|l|l|}
\hline
&1st blow up&2nd blow up&3rd  blow up&4th blow up&5th blow up\\
\hline
%??P??s????
\ctext{$p_0$}$\rightarrow$
&\ctext{$p_1(0:0:1)$}$\rightarrow$
&\ctext{$p_2(1:0:0)$}$\rightarrow$
&\ctext{$p_3(1:0:0)$}$\rightarrow$
&regular
&\\
%??Q??s????
&
&
&\ctext{$q_3(0:0:1)$}$\rightarrow$
&{\color{red}\ctext{$r_4(1:0:-1;s=0)$}$\mbox{(codim.2)}\rightarrow$}
&{\color{red}regular}\\
%??R??s????
&
&$q_2(1:0:-s^2)$
&$q_2(0:0:1)(z_2=-s^2)$
&\ctext{$q_2(0:0:1)(z_2=-s^2)$}$\rightarrow$
&regular
\\
%??S??s????
&
&
&$r_3(-s^2:0:1)$
&\ctext{$r_3(1:0:0)(x_3=-s^2)$}$\rightarrow$
&regular
\\
\hline
\end{tabular}
\end{center}
\label{E7E8completep1first}
\end{table}%

\begin{figure}[h]
  \begin{center}
         \includegraphics[clip, width=9.6cm]{E7E8complete.pdf}
   
                \caption{}
    \label{Fig:E7E8complete}
  \end{center}
%  \label{Fig:E7E8complete}
\end{figure}
%
\fi
%%%%%%%%%%%%%%%%%%%%%%%%%

%\newpage
\subsection{Incomplete/complete resolutions: Blowing up $q_2$ first}
As we did for $SO(12)\rightarrow E_7$, we can change the order 
of the blow-ups to obtain a different intersection diagram. 
For instance, we can choose $q_2$ instead of $p_2$ for the 
3rd blow up. The procedures are analogous to the previous 
cases so we report only the results. 

For the incomplete resolution, the whole process of 
the blow-ups is as shown 
in Table \ref{E7E8q2first}. 
We use the same notation for ${\cal C}$'s (and hence for $\delta$'s) as was used in the $p_2$-first case.
%Namely, ${\cal C}_{1}$, ${\cal C}_{2}$, ${\cal C}_3$, ${\cal C}_{4}$, ${\cal C'}_4$,
%${\cal C}_{r_3}$ and ${\cal C}_{q_2}$ denote the exceptional sets
%arise from the blow ups of the singularities $p_0$, $p_1$, $p_2$, $p_3$, $q_3$, $r_3$ and $q_2$,
%respectively.
Then the intersection pattern of ${\cal C}$'s for the $q_2$-first case is the same $E_7$ Dynkin diagram 
with the $p_2$-first case given in Figure \ref{Fig:E7E8} and their intersection matrix is also the same with \eqref{eq:E7Cartan}.
The intersection pattern of $\delta$'s, however, is different from the one ($A_7$) 
given in Figure \ref{Fig:E7E8}.
This time, it is $E_7$ (see Figure \ref{Fig:E7E8q2first} below). 

One can verify the relations 
%
%\bes
\beqa
& {\cal C}_{p_0}=& \delta_{p_0}+\delta_{q_2}, \n %\\
& {\cal C}_{p_1}=& \delta_{p_1}, \n %\\
& {\cal C}_{q_2}=& 2\delta_{q_2}+2\delta_{p_2}+2\delta_{p_3}+\delta_{q_3}+\delta_{r_3}, \n %\\
& {\cal C}_{p_2}=& \delta_{p_2}, \n %\\
& {\cal C}_{p_3}=& \delta_{p_3}, \n %\\
& {\cal C}_{q_3}=& \delta_{q_3}, \n %\\
& {\cal C}_{r_3}=& \delta_{r_3}.
\label{CswithdsE7E8q2first}
\eeqa
%\ees
%
%%%%%%%%%%%%%%%%%%%%%
\if0
%
In this case 
we find two non-root nodes having the length squared $\frac32$. 
The relations between them are
%
\bes
& {\cal C}_1=\delta_1+\delta_3, \\
& {\cal C}_2=\delta_2, \\
& {\cal C}_3=2\delta_3+2\delta_4+2\delta_{p_3}+\delta_{q_3}+\delta_{r_3}, \\
& {\cal C}_4=\delta_4, \\
& {\cal C}_{p_3}=\delta_{p_3}, \\
& {\cal C}_{q_3}=\delta_{q_3}, \\
& {\cal C}_{r_3}=\delta_{r_3}.
\label{CswithdsE7E8q2first}
\ees
%
\fi
%%%%%%%%%%%%%%%%%%%%%
Then the intersection matrix among $\delta$'s has two $-\frac{3}{2}$ nodes as shown in 
Figure \ref{Fig:E7E8q2first}.  The intersection among these two nodes is $\frac{1}{2}$.
As is the same as the previous examples, one can form 
$28+28$ different linear combinations of $\delta$'s with non-negative and non-positive integer 
coefficients such that they have self-intersection $-\frac32$, giving a single ${\bf 56}$
representation.

\begin{table}[h]
\caption{$E_7\rightarrow E_8$: Incomplete case when $q_2$ is blown up first ($p_0\to p_1 \to q_2$).}
\begin{center}
\begin{tabular}{|l|l|l|l|l|l|}
\hline
&1st blow up&2nd blow up&3rd  blow up&4th blow up&5th blow up\\
\hline
%??P??s????
\ctext{$p_0$}$\rightarrow$
&\ctext{$p_1(0:0:1)$}$\rightarrow$
&\ctext{$q_2(1:0:-s)$}$\rightarrow$
&regular
&
&
\\
%??P??s????
&
&$p_2(1:0:0)$
&\ctext{$p_2(0:0:1)$}$(\tilde z_2=s)\rightarrow$
&\ctext{$p_3(1:0:0)$}$\rightarrow$
&regular
\\
%??Q??s????
&
&
&
&\ctext{$q_3(0:0:1)$}$\rightarrow$
&regular
\\
%??S??s????
&
&
&
&\ctext{$r_3(1:0:-1)$}$\rightarrow$
&regular
\\
\hline
\end{tabular}
\end{center}
\label{E7E8q2first}
\end{table}%

\begin{figure}[h]
  \begin{center}
           \includegraphics[clip, width=8.6cm]{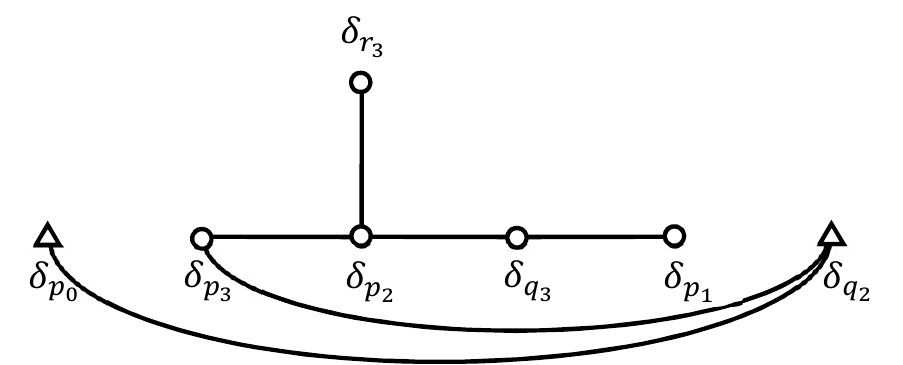}
                \caption{Incomplete intersection diagram of 
                  $E_7 \rightarrow E_8$~:~$q_2$-first ($p_0\rightarrow p_1
                            \rightarrow q_2$) case.}
    \label{Fig:E7E8q2first}
  \end{center}
%  \label{Fig:E7E8q2first}
\end{figure}

\begin{figure}[h]
  \begin{center}
         \includegraphics[clip, width=8.6cm]{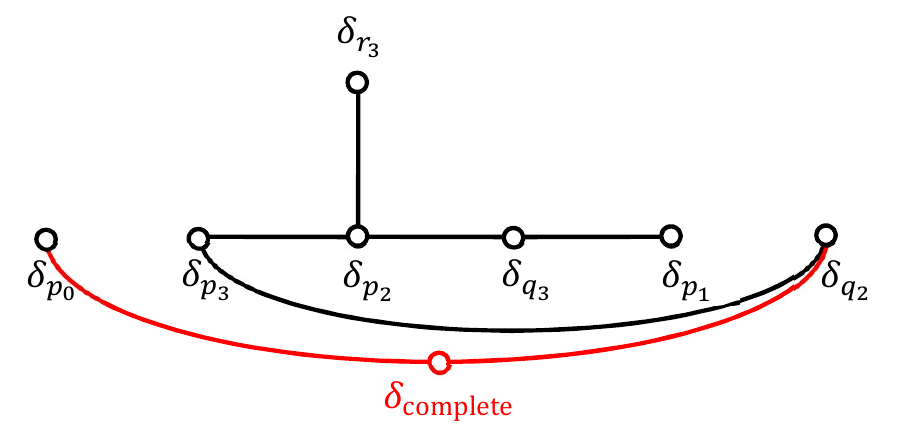}
                \caption{Complete $E_8$ intersection diagram of 
                          $E_7 \rightarrow E_8$~:~$q_2$-first ($p_0\rightarrow p_1
                            \rightarrow q_2$) case.}
    \label{Fig:E7E8q2firstcomplete}
  \end{center}
%  \label{Fig:E7E8q2firstcomplete}
\end{figure}

% Complete cace 
In the complete case, the 3rd blow up at $q_2$ does not 
end with a smooth configuration but an isolated codimension-two 
conifold singularity remains at the intersection of $\delta_{p_0}$ 
and $\delta_{q_2}$ at $s=0$ (see Table \ref{E7E8q2firstcomplete}). 
By a small resolution of this, 
the relations are modified as 
%\bes
\beqa
& {\cal C}_{p_0}=& \delta_{p_0}+\delta_{q_2}+\delta_{\rm complete}, \n %\\
& {\cal C}_{p_1}=& \delta_{p_1}, \n %\\
& {\cal C}_{q_2}=& 2\delta_{q_2}+2\delta_{p_2}+2\delta_{p_3}+\delta_{q_3}+\delta_{r_3}+\delta_{\rm complete}, \n %\\
& {\cal C}_{p_2}=& \delta_{p_2}, \n %\\
& {\cal C}_{p_3}=& \delta_{p_3}, \n %\\
& {\cal C}_{q_3}=& \delta_{q_3}, \n %\\
& {\cal C}_{r_3}=& \delta_{r_3}.
\label{CswithdsE7E8q2firstcomplete}
\eeqa
%\ees
Demanding that the intersection matrix among $\delta$'s is the minus of the proper $E_8$ Cartan matrix 
is consistent with the intersection matrix among ${\cal C}$'s.
As a result, we obtain Figure \ref{Fig:E7E8q2firstcomplete}.

%%%%%%%%%%%%%%%%%%%%%
\if0
%
\beqa
{\cal C}_1&=&\delta_1+\delta_3+\delta_{\rm complete},\n
{\cal C}_2&=&\delta_2,\n
{\cal C}_3&=&2\delta_3+2\delta_4+2\delta_{p_3}
+\delta_{q_3}+\delta_{r_3}+\delta_{\rm complete},\n
{\cal C}_4&=&\delta_4,\n
{\cal C}_{p_3}&=&\delta_{p_3},\n
{\cal C}_{q_3}&=&\delta_{q_3},\n
{\cal C}_{r_3}&=&\delta_{r_3}
\label{CswithdsE7E8q2firstcomplete}
\eeqa
%
\fi
%%%%%%%%%%%%%%%%%%%%%

% Complete case
\begin{table}[h]
\caption{$E_7\rightarrow E_8$:  Complete case when $q_2$ is blown up first ($p_0\to p_1\to q_2$).}
%\begin{center}
\vspace{0.5cm}
\hspace{0.2cm}
\begin{tabular}{|l|l|l|l|}
\hline
&1st blow up&2nd blow up&3rd  blow up\\
\hline
%??P??s????
\ctext{$p_0$}$\rightarrow$
&\ctext{$p_1(0:0:1)$}$\rightarrow$
&\ctext{$q_2(1:0:-s)$}$\rightarrow$
&{\color{red}\ctext{$p_4(1:0:0;s=0)$}$\mbox{(codim.2)}\rightarrow$}
\\
%??P??s????
&
&$p_2(1:0:0)$
&\ctext{$p_2(0:0:1)$}$(\tilde z_2=s)\rightarrow$
\\
%??Q??s????
&
&
&
\\
%??S??s????
&
&
&
\\
\hline
\end{tabular}

\hspace{9.5cm}
\begin{tabular}{|l|l|} \hline
4th blow up&5th blow up\\ \hline
{\color{red}regular} &\\
\ctext{$p_3(1:0:0)$}$\rightarrow$ &regular \\
\ctext{$q_3(0:0:1)$}$\rightarrow$ &regular \\
\ctext{$r_3(1:0:-1)$}$\rightarrow$ &regular \\ \hline
\end{tabular}
%\end{center}
\label{E7E8q2firstcomplete}
\end{table}%
%
%%%%%%%%%%%%%%%%%%%%%%%%%%
\if0
% Complete case
\begin{table}[h]
\caption{$E_7\rightarrow E_8$:  $q_2$ first, complete case.}
\begin{center}
\begin{tabular}{|l|l|l|l|l|l|}
\hline
&1st blow up&2nd blow up&3rd  blow up&4th blow up&5th blow up\\
\hline
%??P??s????
\ctext{$p_0$}$\rightarrow$
&\ctext{$p_1(0:0:1)$}$\rightarrow$
&\ctext{$q_2(1:0:-s)$}$\rightarrow$
&{\color{red}\ctext{$p_4(1:0:0;s=0)$}$\mbox{(codim.2)}\rightarrow$}
&{\color{red}regular}
&
\\
%??P??s????
&
&$p_2(1:0:0)$
&\ctext{$p_2(0:0:1)$}$(\tilde z_2=s)\rightarrow$
&\ctext{$p_3(1:0:0)$}$\rightarrow$
&regular
\\
%??Q??s????
&
&
&
&\ctext{$q_3(0:0:1)$}$\rightarrow$
&regular
\\
%??S??s????
&
&
&
&\ctext{$r_3(1:0:-1)$}$\rightarrow$
&regular
\\
\hline
\end{tabular}
\end{center}
\label{E7E8q2firstcomplete}
\end{table}%
\fi
%%%%%%%%%%%%%%%%%%%%%%%%%%%
%

\subsection{Other inequivalent orderings}

Let us summarize what other types of intersection diagrams 
are obtained for the enhancement $E_7\rightarrow E_8$ 
if we choose other orderings of the blow-ups.
So far, we have derived the intersection diagrams for blow-ups 
with orders $p_0\rightarrow p_1 \rightarrow p_2 \rightarrow q_3$ and $p_0\rightarrow p_1\rightarrow q_2$.
As we can see in the column ``3rd blow up'' in Table 
\ref{E7E8p1first}, after blowing up $p_2$, the three singular points $q_3$, $q_2$ 
and $r_3$ become an identical point $(0:0:1)$ on the $\PPsmall^2$ 
at $s=0$ (see also \eqref{eq:E7E83zxzsing}). 
Therefore, besides the case when $q_3$ is blown up after
$p_2$ as discussed in sections \ref{sec:p2firstincomp} and \ref{sec:p2firstcomp}~\footnote{$p_3$ 
is always a different point from the three and hence can be 
blown up independently at any stage.}, there are two 
other options: We can blow up either $q_2$ or $r_3$ after the blow up 
of $p_2$. 

\bigskip
\noindent 
\underline{$p_0 \rightarrow p_1 \rightarrow p_2 \rightarrow q_2$ case}
\medskip 

\begin{figure}[h]
  \begin{center}
\vspace{0.0cm}
         \includegraphics[clip, width=8.6cm]{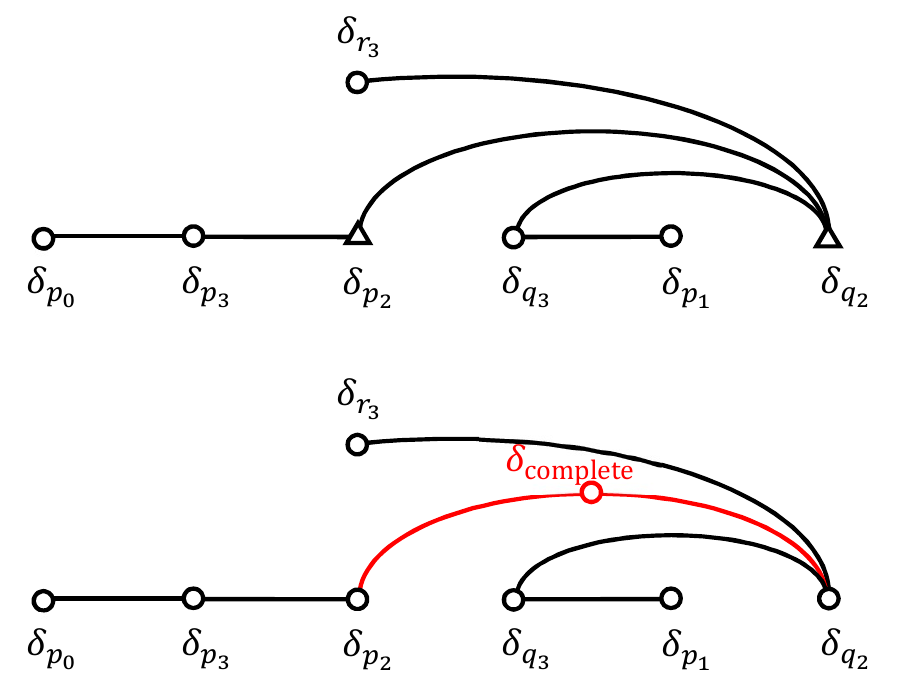}
   
                \caption{Incomplete/complete intersection diagrams of 
                            $E_7 \rightarrow E_8$~:~$p_2$-first ($p_0\rightarrow p_1
                            \rightarrow p_2 \rightarrow q_2$) case.}
    \label{}
\label{Fig:E7E8p1p2q2}
  \end{center}
%\label{Fig:E7E8p1p2q2}
\end{figure}

If we blow up $q_2$ after $p_2$, the relations between 
${\cal C}$'s and $\delta$'s are given by
%\bes
\beqa
& {\cal C}_{p_0}=& \delta_{p_0},  \n %\\
& {\cal C}_{p_1}=& \delta_{p_1}, \n %\\
& {\cal C}_{p_2} =& \delta_{p_2}+\delta_{q_2}\, (+\delta_{\scriptsize \mbox{complete}}), \n %\\
& {\cal C}_{q_2}=& 2 \delta_{q_2}+\delta_{q_3}+\delta_{r_3}\, (+\delta_{\scriptsize \mbox{complete}}), \n %\\
& {\cal C}_{q_3}=& \delta_{q_3}, \n %\\
& {\cal C}_{r_3}=& \delta_{r_3}, \n %\\
& {\cal C}_{p_3}=& \delta_{p_3}.
\label{CswithdsSO(12)E7p2q2}
\eeqa
%\ees
%
The intersection diagram of ${\cal C}$'s is the $E_7$ Dynkin diagram as before.
The intersection diagrams of  $\delta$'s for incomplete\,/\,complete cases are 
shown in Figure \ref{Fig:E7E8p1p2q2}. 
For the incomplete case, it is an $E_7$ non-Dynkin diagram with two $-\frac{3}{2}$ nodes
(the intersection among them is $\frac{1}{2}$),
while for the complete case, it is proper $E_8$ Dynkin diagram. 
Again, there is only one ${\bf 56}$ at $s=0$ as two-cycles $J$ \eqref{eq:J} with $J\cdot J=-\frac{3}{2}$.

\medskip
\noindent 
\underline{$p_0 \rightarrow p_1 \rightarrow p_2 \rightarrow r_3$ case}
\medskip 

\begin{figure}[h]
  \begin{center}
\vspace{0.2cm}
         \includegraphics[clip, width=8.6cm]{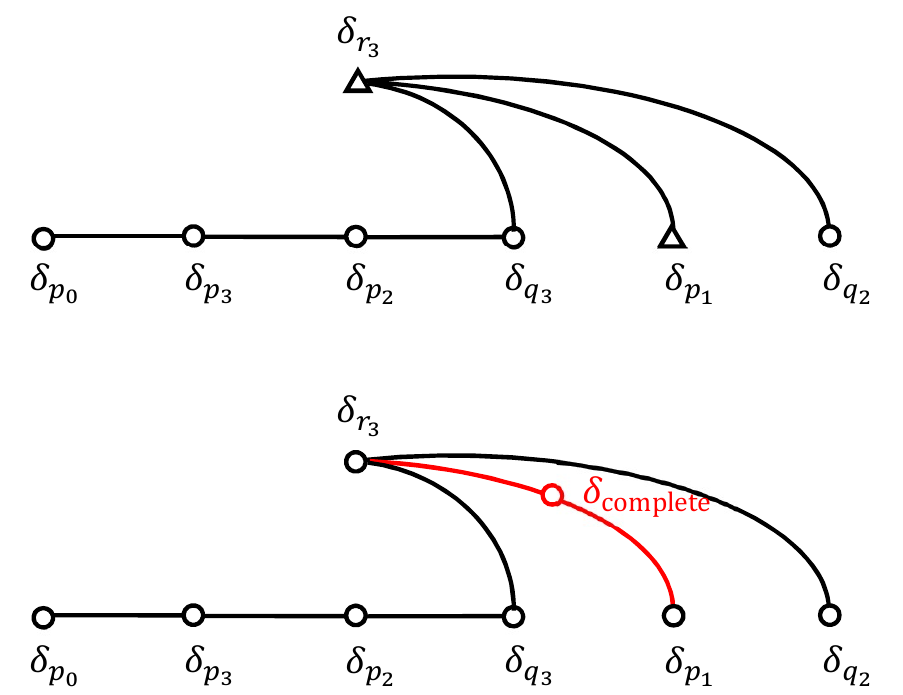}
   
                \caption{Incomplete/complete intersection diagrams of 
                            $E_7 \rightarrow E_8$~:~$p_2$-first ($p_0\rightarrow p_1
                            \rightarrow p_2 \rightarrow r_3$) case.}
    \label{}
\label{Fig:E7E8p1p2r3}
  \end{center}
%\label{Fig:E7E8p1p2q2}
\end{figure}

If we blow up $r_3$ after $p_2$, the relations between
${\cal C}$'s and $\delta$'s are given by
%\bes
\beqa
& {\cal C}_{p_0}=& \delta_{p_0}, \n % \\
& {\cal C}_{p_1}=& \delta_{p_1}+\delta_{r_3}\, (+\delta_{\scriptsize \mbox{complete}}), \n %\\
& {\cal C}_{p_2}=& \delta_{p_2} , \n %\\
& {\cal C}_{r_3}=& 2 \delta_{r_3}+\delta_{q_2}+\delta_{q_3}\, (+\delta_{\scriptsize \mbox{complete}}) , \n %\\
& {\cal C}_{q_2}=& \delta_{q_2}, \n %\\
& {\cal C}_{q_3}=& \delta_{q_3}, \n %\\ 
& {\cal C}_{p_3}=& \delta_{p_3}.
\label{CswithdsSO(12)E7p2q2}
\eeqa
%\ees
%
%The intersection matrix of ${\cal C}$'s is the same one as before.
The intersection diagrams are shown in Figure \ref{Fig:E7E8p1p2r3}. 
The diagram for the incomplete case is a $D_7$ non-Dynkin diagram with two $-\frac{3}{2}$ nodes
(their intersection is $\frac{1}{2}$).
Again, there is only one ${\bf 56}$ at $s=0$.

This exhausts all the possibilities of changing the order of 
the singularities we blow up.
We obtained four sets of incomplete\,/\,complete intersection diagrams.
Again, 
each of them corresponds to a box graph on every other row 
of Figure 44 in \cite{BoxGraphs}.
For the $p_2$-first cases, Figure \ref{Fig:E7E8} (with Figure \ref{Fig:E7E8complete}) is equivalent 
to the one in the right column:
Figure \ref{Fig:E7E8p1p2q2} and Figure \ref{Fig:E7E8p1p2r3} respectively are 
the fourth and sixth ones from the bottom in the left column.
For the $q_2$-first case, Figure \ref{Fig:E7E8q2first} (with Figure \ref{Fig:E7E8q2firstcomplete})
is the second one from the bottom in the left column.

\section{Conclusions}
We have investigated the resolutions of 
codimension-two enhanced singularities 
from $SO(12)$ to $E_7$ 
and from $E_7$ to $E_8$ in six-dimensional F-theory,  
where a half-hypermultiplet locally arises for
generic complex structures achieving them. 
A half-hypermultiplet only occurs associated with 
a Lie algebra allowing a pseudo-real representation, and 
the above are the two of the three cases in the list of 
six-dimensional F-theory compactifications \cite{BIKMSV}
that exhibit half-hypermultiplets in the massless matter 
spectrum. 
%The remaining case,  the enhancement 
%from $SU(6)$ to $E_6$, was already investigated some time 
%ago in \cite{MT}, 
As was already observed in the enhancement 
from $SU(6)$ to $E_6$ in \cite{MT}, 
we have confirmed that 
the resolution process does not generically yield 
as many number of exceptional curves as naively 
expected from the Kodaira classification of the 
codimension-one singularities. 
%This type of resolution was 
%called the incomplete resolution in \cite{MT}. 
In the present paper, we have observed similar features 
such as non-Dynkin intersection diagrams 
and half-integral %self-
intersection 
numbers of exceptional fibers. 
%The lack of the full exceptional 
%fibers may be regarded as the reason why the matter 
%arising there is a {\em half}-hyper. Even so, we have 
%confirmed that the exceptional fibers at the enhanced point form 
%extremal rays of the cone of the positive weights of the 
%relevant pseudo-real representation. 
Then we have found that the exceptional fibers at the enhanced point form 
extremal rays of the cone of the positive weights of the 
relevant pseudo-real representation,
explaining why a half-hyper multiplet arises there.

We have also found that 
a variety of different intersection diagrams of 
exceptional curves are obtained by 
altering the ordering of the singularities blown up 
in the process. 
They correspond to different ``phases'' of the 
three-dimensional M-theory.
We have obtained, for both $SO(12)\rightarrow E_7$ 
and $E_7\rightarrow E_8$, the intersection diagram 
on {\em every other row} of the figures in \cite{BoxGraphs},
but not all of them. The phases corresponding to 
the diagrams we obtained are not the ones related 
by a flop. 

We have presented detailed derivations of 
the intersection diagrams of the exceptional fibers at the 
singularity enhanced points. In particular, we have described 
how an exceptional curve is lifted up on the chart arising 
due to the subsequent blowing-up process. 
By carefully
examining whether an exceptional curve contains  
another arising afterwards as a part, 
we have obtained the intersection matrices as above. 

In the complete resolutions, where the colliding brane 
approaches the stack of branes as $O(s^2)$, we have 
obtained the full Dynkin diagrams of the group $G$ 
as the intersection diagram of the fibers at the enhanced 
point. The extra codimension-two singularity is 
always a conifold singularity, as was found in the 
previous example \cite{MT}.

%%%
Although we have studied in this paper the explicit resolutions 
of the singularities in six dimensions, the technologies we have developed 
here can also be used in more general settings such as 
codimension-three singularities in Calabi-Yau four-folds, with or 
without sections. (In the latter, one may consider 
the Jacobian fibrations. See e.g. \cite{Kimura}. )   

It would also be interesting to perform a similar analysis 
for a singularity with higher rank enhancement. 
In particular, it has been expected \cite{FFamilyUnification} 
that a codimension-three singularity enhancement from 
$SU(5)$ to $E_7$ could yield, without monodromies and with appropriate 
$G$-fluxes, the three-generation spectrum of 
the $E_7/(SU(5)\times U(1)^3)$ supersymmetric coset model 
\cite{KugoYanagida}. Such a codimension-two singularity enhancement 
in six dimensions was already studied in \cite{MizoguchiTanianomaly}.
The codimension-three case with a $\ZZsmall_2$ or a larger 
monodromy was considered in e.g.~\cite{HTV}, but was concluded 
to be not very useful for their purposes. 
The box graph analysis of \cite{BoxGraphs}, on the other hand, predicts  
the existence of the phases without monodromies, which will serve as 
basis for models of family unification in F-theory. 
We hope to come back to this issue elsewhere.

Finally, massless matter generation in F-theory may also be explained 
by string junctions stretched between various $(p,q)$ 
7-branes near the intersections. Very recently, a new pictorial method 
to keep track of non-localness of F-theory 7-branes has been developed 
by drawing  a ``dessin'' on the base of the elliptic fibration \cite{dessin1,dessin2},
which may help understanding how the difference of the resolutions is 
affected by the geometry near the enhanced point.

%\section*{Acknowledgments} 
\acknowledgments
We thank H.~Hayashi, Y.~Kimura, H.~Otsuka and S.~Schafer-Nameki 
for valuable discussions.

\appendix

\section{Symplectic Majorana-Weyl spinors, 
pseudo-real representations and 
half-hypermultiplets}\label{Appendixhalfhyper}
\subsection{Symplectic Majorana-Weyl spinors}
In six(=5+1) dimensions, consider the Dirac equation
\beqa
(i\gamma^\mu \partial_\mu-m)\psi&=&0.
\eeqa
The complex conjugate equation
\beqa
(-i\gamma^{\mu*} \partial_\mu-m)\psi^*&=&0
\eeqa
can be written in terms of the charge conjugation $\psi^c$ 
defined by
\beqa
\psi^c\equiv B\psi^*
\eeqa
with
\beqa
B\gamma^{\mu*} B^{-1}=-\gamma^{\mu} 
\label{BgammaB-1}
\eeqa 
as
\beqa
(i\gamma^\mu \partial_\mu-m)\psi^c&=&0.
\eeqa
This is the same Dirac equation as that $\psi$ obeys.
If one could impose the constraint
$
\psi^c=\psi
$
on $\psi$, one could define a Majorana spinor, 
but in six dimensions one can not as, 
if one could do so, 
\beqa
\psi=\psi^c=B\psi^*=B(B\psi^*)^*=B B^*\psi,
\eeqa
but
\beqa
BB^*&=&-1
\eeqa
in six dimensions. Alternatively, one can impose on two 
  Weyl spinors $\psi_1$, $\psi_2$
the constraint
\beqa
\psi^c_{1}=+\psi_2,~~~\psi^c_{2}=-\psi_1.
\eeqa
They are called symplectic Majorana-Weyl 
spinors. Note that in six(=5+1)-dimensions the 
charge conjugation operation does not flip the chirality 
so that one can define symplectic Majorana-{\em Weyl} spinors. 

\subsection{Pseudo-real representation and symplectic Majorana
condition}

Let us examine whether one can consistently 
impose this constraint 
on a representation space of a Lie group.
Let $G$ be a compact Lie group and $\rho$ be its 
representation on a complex $2n$-dimensional vector space 
$V$:
\beqa
\rho: G \rightarrow GL_{\CCsub}(V)~~~\mbox{homomorphism}.
\eeqa
We say $\rho$ is a pseudo-real representation 
if  there exists $P\in GL_{\CCsub}(V)$ 
such that 
\beqa
P(\rho(g))^*P^{-1}&=&\rho(g)
\eeqa
for any $g\in G$.  
${}^*$ denotes the complex conjugation.

If
\beqa
v'&=&\rho(g) v,
\label{vtransformation}
\eeqa
then
%\beqa
%v'^*&=&(\rho(g))^* v^*
%\label{v*transformation}
%\eeqa
%and
\beqa
P v'^*&=&(P(\rho(g))^*P^{-1})Pv^* \nonumber \\
&=&\rho(g)Pv^*.
\eeqa
Therefore $Pv^*$ also transforms as {\bf 2n}.

If $\rho(g)$ can be embedded in $Sp(2n)$,
we have 
\beqa
\rho(g)^T\Omega_{2n} \rho(g)&=&\Omega_{2n}, \nonumber \\
\Omega_{2n}&=&\left(
\begin{array}{cc}
0&-1_n\\
1_n&0
\end{array}
\right).
\eeqa 
Moreover, if $\rho$ is a unitary representation: 
$\rho(g)^{-1}=\rho(g)^\dagger$, we find  
\beqa
\rho(g)&=&
%\Omega_{2n}^{T-1}\rho(g)^*\Omega_{2n}^T\n&=&
\Omega_{2n}\rho(g)^*\Omega_{2n}^{-1}.
\label{rho=Omegarho*Omega-1}
\eeqa
Thus we can take
\beqa
P=\Omega_{2n}.
\eeqa

Let us construct such a vector $v$ by using symplectic Majorana(-Weyl) 
spinors.  Let $\psi_1$, $\psi_2$ be a pair of 
$n$-component column vectors consisting of $n$
 symplectic Majorana(-Weyl) spinors, and $v$ be a stack of them.  
 Then  
\beqa
v&=&\left(
\begin{array}{c}
\psi_1\\
\psi_2
\end{array}
\right)\n
&=&
\left(
\begin{array}{c}
-\psi^c_2\\
\psi^c_1
\end{array}
\right)\n
&=&
\left(
\begin{array}{cc}
0&-1_n\\
1_n&0
\end{array}
\right)
\left(
\begin{array}{c}
\psi^c_1\\
\psi^c_2
\end{array}
\right)\n
&=&
\Omega_{2n}
\left(
\begin{array}{c}
B\psi^*_1\\
B\psi^*_2
\end{array}
\right)\n
&=&
\Omega_{2n}(B\cdot 1_{2n})
v^*.
\label{v=Omegav*}
\eeqa
In general, this is not consistent as 
$v$ transforms as a ${\bf 2n}$ representation whereas 
$v^*$ as a $\overline{\bf 2n}$ representation.
However, 
for a pseudo-real representation
(\ref{rho=Omegarho*Omega-1}),
we obtain
\beqa
\rho(g)v&=&\Omega_{2n}\rho(g)^*\Omega_{2n}^{-1}
\cdot \Omega_{2n}(B\cdot 1_{2n})
v^*
\n
&=&
\Omega_{2n}(B\cdot 1_{2n})\rho(g)^*
v^*,
\eeqa
which agrees with (\ref{v=Omegav*}).
Therefore, for a pseudo-real representation 
a $2n$-dimensional representation space can be 
constructed from $n$ pairs of symplectic 
Majorana-Weyl spinors.

\subsection{$\frac{\bf 2n}2$ hypermultiplets vs. 
${\bf 2n}$ $\frac12$hypermultiplets}
In the previous section we have seen that among  
$2n$ Weyl fermions in hypermultiplets 
transforming as a pseudo-real 
${\bf 2n}$ representation one half ($=n$) of them 
can be expressed as 
the complex conjugates of the other half ($=n$).
In this section we will show that this can be viewed as
a restriction of the degrees of freedom of 
$2n$ (Weyl fermions of) hypermultiplets to 
$2n$ (``half-Weyl'' fermions of) half-hypermultiplets.

We take 
\beqa
\Gamma^0&=&\sigma_3 \otimes \sigma_3 \otimes i\sigma_2,\n
\Gamma^1&=&\sigma_3 \otimes \sigma_3 \otimes \sigma_1,\n
\Gamma^2&=&\sigma_3 \otimes \sigma_1 \otimes 1,\n
\Gamma^3&=&\sigma_3 \otimes \sigma_2 \otimes 1,\n
\Gamma^4&=&\sigma_1 \otimes 1 \otimes 1,\n
\Gamma^5&=&\sigma_2 \otimes 1 \otimes 1
\eeqa
as a realization of the gamma matrices. 
As the matrix $B$ satisfying (\ref{BgammaB-1}), we can have
\beqa
B&=&\Gamma^0\Gamma^1\Gamma^2\Gamma^4.
\eeqa
Since
\beqa
\Gamma^0\Gamma^1\Gamma^2&=&\sigma_3 \otimes \sigma_1 \otimes \sigma_3,
\eeqa
we can write $B$ in a block form as
\beqa
B&=&\left(
\begin{array}{cc}
b&0\\
0&-b
\end{array}
\right)
\left(
\begin{array}{cc}
0&{\bf 1}\\
{\bf 1}&0
\end{array}
\right),
\label{B}
\eeqa 
where
\beqa
 \sigma_1 \otimes \sigma_3&\equiv&b.
 \eeqa
{\bf 1} is the $4\times 4$ unit matrix.

Since the chirality in six dimensions is defined by
\beqa
\Gamma^\sharp~\equiv~-\Gamma^0\Gamma^1\Gamma^2\Gamma^3\Gamma^4\Gamma^5
&\equiv&(\sigma_3\otimes {\bf 1})({\bf 1}\otimes \gamma_5 ),
\eeqa
the eigenvalues of 
$\gamma_5$ and
$(\sigma_3\otimes {\bf 1})$ are correlated with each other in 
a six-dimensional spinor with a definite chirality. 
For instance, if 
$\Gamma^\sharp=+1$, $(\gamma_5,\sigma_3)=(+,+)$ or $(-,-)$.
Thus if we write  
\beqa
\psi_i&=&\left(
\begin{array}{c}
\phi_i\\
\chi_i
\end{array}
\right)~~~(i=1,2),
\eeqa
this is a decomposition with respect to the four-dimensional 
chirality.
From 
(\ref{B}), we have 
\beqa
\psi_i^c&=&B\psi_i^*\n
&=&
\left(
\begin{array}{cc}
b&0\\
0&-b
\end{array}
\right)
\left(
\begin{array}{cc}
0&{\bf 1}\\
{\bf 1}&0
\end{array}
\right)
\left(
\begin{array}{c}
\phi_i^*\\
\chi_i^*
\end{array}
\right)\n
&=&
\left(
\begin{array}{r}
b\chi_i^*\\
-b\phi_i^*
\end{array}
\right).
\eeqa
Therefore, if a collection of $n$ spinors
$\psi_2$ are written as $\psi_1^c$, the relations 
$\psi_2=\psi_1^c$ and $\psi_1=-\psi_2^c$ imply
\beqa
\chi_2=-b\phi_1^*,~~~\chi_1=+b\phi_2^*.
\eeqa
Thus the lower component of each of the $2n$ Weyl spinors 
can be expressed in terms of the upper component.

\subsection{Restriction on the complex scalars}
Let $v_i$ $(i=1,2)$ be a pair of $n$ 
 complex scalars and
\beqa
%v&=&
& &\left(
\begin{array}{c}
v_1\\
v_2
\end{array}
\right)
\eeqa
be in the ${\bf 2n}$ pseudo-real representation of $G$.
%satisfying (\ref{v=Omegav*}).
In order to similarly define $v_i^c$ such that 
\beqa
v_2=v_1^c,~~~v_1=-v_2^c
\label{v2=v1c}
\eeqa
by 
\beqa
v_i^c&\equiv&Uv^*_i
\eeqa
for some $U$ satisfying $U^*U=-1$, 
we recall that a hypermultiplet has {\em two} complex 
scalars transforming in the identical representation. 
Let $v^{(1)}_i$ and $v^{(2)}_i$ be such two scalars, 
then writing 
$v_i=\left(
\begin{array}{c}
v^{(1)}_i\\
v^{(2)}_i
\end{array}
\right)$, we define
\beqa
v_i^c&=&\left(
\begin{array}{c}
v^{(1)c}_i\\
v^{(2)c}_i
\end{array}
\right)\n
&\equiv&U\left(
\begin{array}{c}
v^{(1)*}_i\\
v^{(2)*}_i
\end{array}
\right),
\eeqa
where
\beqa
U&\equiv&\left(
\begin{array}{cc}
0&-1\\
1&0
\end{array}
\right)
\eeqa
 is a $90^\circ$ $U(1)_R$ rotation. 
Then $U^*U=-1$, and (\ref{v2=v1c}) can be imposed.
With this definition of $v_i^c$, one can 
also reduce the degrees of freedom of the complex scalars in a 
hypermultiplet.

\section{Summary of $SU(6)\rightarrow E_6$}
\label{AppendixSU(6)E6}

In this appendix we summarize the results of the 
analysis \cite{MT} of the codimension-two 
enhancement $SU(6)\rightarrow E_6$. 
Consider
\beqa
\Phi&=&-y^2+x^3+(t_r^2+3 t_r z)x^2 
+(2 t_r z^2 + 3z^3)x + z^4.
\label{PhiSU6E6}
\eeqa
One can verify that $\Phi$ has an $SU(6)$ singularity at $(x,y,z)$ 
for fixed $t_r\neq 0$, and this is enhanced to an $E_6$ singularity 
for $t_r=0$. $\Phi$ (\ref{PhiSU6E6}) can be obtained by putting
$h_{n+2-r}=\frac1{\sqrt{3}}$, $H_{n+4-r}=-\frac12$, 
$u_{r+4}=\frac12$, $f_{n+8-r}=f_8=g_{12}=0$ in 
(\ref{fSU(6)}) and (\ref{gSU(6)}) and making a 
change of variables.

If we take $t_r=s$, we are led to the incomplete resolution.
The process is summarized in Table \ref{SU(6)E6}.
After the codimension-one blow up at the singularity $p_0$ 
of $\Phi$ (\ref{PhiSU6E6}), 
we obtain two exceptional curves 
${\cal C}_{1\pm}$ 
at fixed $s\neq 0$, 
which come on top of each other into a single curve $\delta_1$ 
at $s=0$. These two curves intersect at a singularity $p_1$, 
which forms a singular line along the $s$ direction. 
We perform a codimension-one blow up along this line 
to find, again,  two exceptional curves ${\cal C}_{2\pm}$ 
at $s\neq 0$, which becomes $\delta_{2\pm}$ at $s=0$. 
The intersection of ${\cal C}_{2\pm}$ is again another 
singularity $p_2$. We then blow up the singularity line 
$p_2$ to get a regular curve ${\cal C}_3$ at $s\neq 0$, 
which splits into two curves $\delta_{3\pm}$ at $s=0$. 

\begin{figure}[h]
  \begin{center}
\vspace{0.2cm}
         \includegraphics[clip, width=7.4cm]{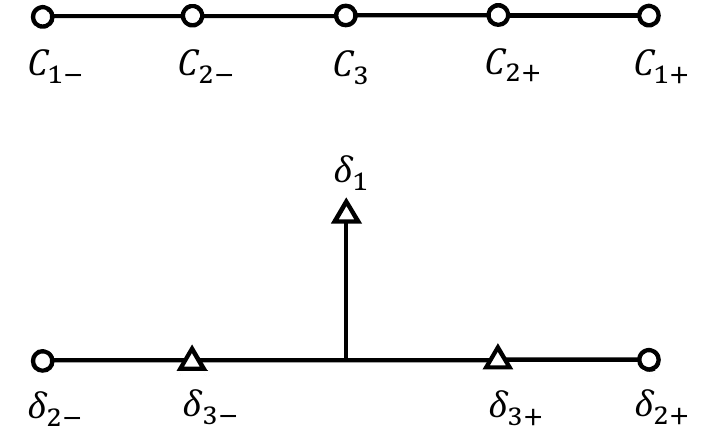}
   
                \caption{Generic intersection diagram at $s\neq 0$ (upper) and 
                        incomplete intersection diagram at $s=0$ (lower) of $SU(6) \rightarrow E_6$.}
    \label{}
\label{Fig:SU(6)E6incomp}
  \end{center}
%\label{Fig:SO(12)E7p1q1}
\end{figure}

In the incomplete case, this is the end. The relations among 
${\cal C}$'s and $\delta$'s are given by
%
%\bes
\beqa
& & {\cal C}_{1\pm}= \delta_1+\delta_{3\pm},  \n %\\
& & {\cal C}_{2\pm}= \delta_{2\pm}, \n %\\
& & {\cal C}_3= \delta_{3+}+\delta_{3-}.
\label{CswithdsSU(6)E6}
\eeqa
%\ees
One can derive these relations in the same way as explained in section \ref{sec:matrixp1}.
Their intersection diagrams are shown in Figure \ref{Fig:SU(6)E6incomp}.
\begin{table}[h]
\caption{$SU(6)\rightarrow E_6$: Incomplete case. }
\begin{center}
\begin{tabular}{|l|l|l|l|}
\hline
&1st blow up&2nd blow up&3rd blow up\\
\hline
\ctext{$p_0$}$\rightarrow$
&\ctext{$p_1(0:0:1)$}$\rightarrow$
&\ctext{$p_2(1:0:-s)$}$\rightarrow$
&regular
\\
\hline
\end{tabular}
\end{center}
\label{SU(6)E6}
\end{table}%

For the complete resolution, we take $t_r=s^2$.
In this case, 
as is shown in Table \ref{SU(6)E6complete}, we have an additional codimension-two 
isolated conifold singularity after the 3rd blow up.

\begin{figure}[h]
  \begin{center}
\vspace{0.2cm}
         \includegraphics[clip, width=7.4cm]{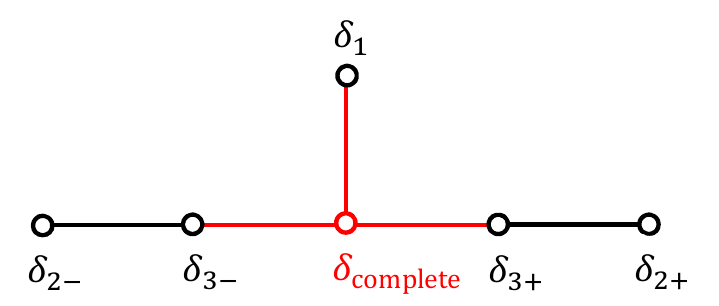}
   
                \caption{Complete $E_6$ intersection diagram of $SU(6) \rightarrow E_6$.}
    \label{}
\label{Fig:SU(6)E6comp}
  \end{center}
\end{figure}

The relations (\ref{CswithdsSU(6)E6}) are modified to
%\bes
\beqa
& & {\cal C}_{1\pm}=\delta_1+\delta_{3\pm}+\delta_{\rm complete}, \n %\\
& & {\cal C}_{2\pm}=\delta_{2\pm}, \n %\\
& & {\cal C}_3=\delta_{3+}+\delta_{3-}+\delta_{\rm complete}.
\label{CswithdsSU(6)E6modified}
\eeqa
%\ees
This result is obtained by following the analysis explained in section \ref{sec:conifoldlimit}. 
The intersection diagram of these six $\delta$'s is the proper $E_6$ Dynkin diagram as in 
Figure \ref{Fig:SU(6)E6comp}.
\begin{table}[h]
\caption{$SU(6)\rightarrow E_6$: Complete case. }
\begin{center}
\begin{tabular}{|l|l|l|l|l|}
\hline
&1st blow up&2nd blow up&3rd blow up&4th blow up\\
\hline
\ctext{$p_0$}$\rightarrow$
&\ctext{$p_1(0:0:1)$}$\rightarrow$
&\ctext{$p_2(1:0:-w)$}$\rightarrow$
&{\color{red}\ctext{$p_3(1:0:0;s=0)$}$\mbox{(codim.2)}\rightarrow$}
&{\color{red}regular}
\\
\hline
\end{tabular}
\end{center}
\label{SU(6)E6complete}
\end{table}%

\section{Small resolution of a conifold singularity}
\label{sec:smallresolution}

In this appendix, we give a brief review of the small resolution (see, {\it e.g.}, \cite{CD}). 
A conifold ${\cal M}$ is a three dimensional space given by the polynomial in $\CCsmall^4$
\beq
  X_1 X_4-X_2 X_3 = 0.
\label{eq:conifold}
\eeq
It has a singularity at $(X_1,X_2,X_3,X_4) = (0,0,0,0)$, which is called the conifold singularity.
In general, an isolated singularity on a hypersurface in $\CCsmall^4$ can be resolved by 
inserting $\PPsmall^3$ in the ambient space $\CCsmall^4$.
For the conifold singularity, it is equivalent to inserting $\PPsmall^1 \times \PPsmall^1$ on 
the conifold ${\cal M}$.
However, it is sufficient to insert only one $\PPsmall^1$ to resolve the conifold singularity.
This process is called the small resolution.
Inserting $\PPsmall^1$ at the origin of ${\cal M}$ is given by
\beq
 \breve{\cal M}=\left\{   (X_1,X_2,X_3,X_4) \times (y_1:y_2) \hspace{-0.1cm}
                       \begin{array}{r|} 
                          \, \\
                          \, \\
                       \end{array}   
                \left[ \begin{array}{cc} 
                       X_1 & X_2 \\
                       X_3 & X_4 
                      \end{array}
                \right]
                \left[ \begin{array}{c} 
                         y_1 \\
                         y_2  
                      \end{array}
                \right]
             =  \left[ \begin{array}{c} 
                         0 \\
                         0  
                      \end{array}
                \right]
                \right\}.
\label{eq:smallresolution}
\eeq
Let us write 
$
M={\small \left[ \begin{array}{rl} 
                       X_1 & X_2 \\
                       X_3 & X_4 
                      \end{array}
                \right]}.
$
Since $(y_1:y_2) \neq (0,0)$, $\mbox{det}M = 0$.
This gives the defining equation of the conifold \eqref{eq:conifold}.
Furthermore, if
\bes
(X_1,X_2,X_3,X_4) 
\left\{ 
\begin{array}{l}
\hspace{-0.1cm}\neq (0,0,0,0) \Rightarrow 
     \mbox{rank}M = 1 \Rightarrow (y_1:y_2) \mbox{ is determined,}  \\
\hspace{-0.1cm}= (0,0,0,0) \Rightarrow \mbox{rank}M = 0 \Rightarrow (y_1:y_2) \mbox{ is undetermined,}
\end{array}
 \right. 
\nonumber
\ees
which means that $\PPsmall^1 = (y_1:y_2)$ is inserted only at the origin.
As a result, $\breve{\cal M}$ is regular, since
$\partial_{X_1}(X_1y_1+X_2y_2) = y_1 = 0$ and $\partial_{X_2}(X_1y_1+X_2y_2) = y_2 = 0$ 
are not compatible.

$\breve{\cal M}$ is covered by two local coordinate patches.
If $y_1 \neq 0$, we can solve \eqref{eq:smallresolution} as 
\beq
 X_1 = -\frac{y_2}{y_1}X_2 \,\, , \,\, X_3 = -\frac{y_2}{y_1}X_4,
\eeq
and hence
\beq
(X_1,X_2,X_3,X_4) = (-\lambda X_2,X_2,-\lambda X_4,X_4) \quad \mbox{with} \,\, \lambda \equiv \frac{y_2}{y_1}.
\label{eq:H+}
\eeq
We call this patch $H_+$. Local coordinates of $H_+$ are $(X_2,X_4,\lambda)$ and the inserted $
\PPsmall^1=(y_1:y_2)$ is located at
\beq
 (X_2,X_4,\lambda) = (0,0,\lambda).
\label{eq:H+delta}
\eeq
If $y_2 \neq 0$, we can solve \eqref{eq:smallresolution} as 
\beq
 X_2 = -\frac{y_1}{y_2}X_1 \,\, , \,\, X_4 = -\frac{y_1}{y_2}X_3,
\eeq
and hence
\beq
(X_1,X_2,X_3,X_4) = (X_1,-\mu X_1,X_3,-\mu X_3) \quad \mbox{with} \,\, \mu \equiv \frac{y_1}{y_2}.
\label{eq:H-}
\eeq
We call this patch $H_-$. Local coordinates of $H_-$ are $(X_1,X_3,\mu)$ and the inserted $\PPsmall^1$ is 
\beq
 (X_1,X_3,\mu) = (0,0,\mu).
\label{eq:H-delta}
\eeq

%\clearpage
%%%%%%%%%%%%%%%%%%%%%%%%%%%%%%%%%%%%%%%%%%%%%%%%%%%%%%%%
%%   References
%%%%%%%%%%%%%%%%%%%%%%%%%%%%%%%%%%%%%%%%%%%%%%%%%%%%%%%%

\end{document}